\documentclass[a4paper,fleqn,usenatbib]{mnras}

\usepackage{newtxtext,newtxmath}

\usepackage[T1]{fontenc}
\usepackage{ae,aecompl}

\usepackage{graphicx}	
\usepackage{amsmath}	
\usepackage{amssymb}	
\usepackage{gensymb}	
\usepackage[compatibility=false]{caption}
\usepackage{subcaption}
\usepackage[online,flushleft,para]{threeparttable}

\makeatletter
\let\TPT@hookin\@gobble
\let\TPT@hookarg\@gobble
\makeatother

\title[Dust distributions in debris discs]{Dust size and spatial distributions in debris discs: predictions for exozodiacal dust dragged in from an exo-Kuiper belt}

\author[J. K. Rigley et al.]{
Jessica K. Rigley,$^{1}$\thanks{E-mail: jkr40@ast.cam.ac.uk}
and Mark C. Wyatt,$^{1}$
\\
$^{1}$Institute of Astronomy, University of Cambridge, Madingley Road, Cambridge, CB3 0HA, UK\\
}

\date{Accepted XXX. Received YYY; in original form ZZZ}

\pubyear{2020}

\begin{document}
\label{firstpage}
\pagerange{\pageref{firstpage}--\pageref{lastpage}}
\maketitle
%
\begin{abstract}
The SEDs of some nearby stars show mid-infrared excesses from warm habitable zone dust, known as exozodiacal dust. This dust may originate in collisions in a planetesimal belt before being dragged inwards. This paper presents an analytical model for the size distribution of particles at different radial locations in such a scenario, considering evolution due to destructive collisions and Poynting-Robertson (P-R) drag. Results from more accurate but computationally expensive numerical simulations of this process are used to validate the model and fit its free parameters. The model predicts $11~\mu\text{m}$ excesses ($R_{11}$) for discs with a range of dust masses and planetesimal belt radii using realistic grain properties. We show that P-R drag should produce exozodiacal dust levels detectable with the Large Binocular Telescope Interferometer (LBTI) ($R_{11} > 0.1\%$) in systems with known outer belts; non-detection may indicate dust depletion, e.g. by an intervening planet. We also find that LBTI could detect exozodiacal dust dragged in from a belt too faint to detect at far-infrared wavelengths, with fractional luminosity $f\sim10^{-7}$ and radius $\sim 10-80$~au. Application to systems observed with LBTI shows that P-R drag can likely explain most (5/9) of the exozodiacal dust detections in systems with known outer belts; two systems ($\beta$~Uma and $\eta$~Corvi) with bright exozodi may be due to exocomets. We suggest that the three systems with exozodiacal dust detections but no known belt may have cold planetesimal belts too faint to be detectable in the far-infrared. Even systems without outer belt detections could have exozodiacal dust levels $R_{11} > 0.04\%$ which are problematic for exo-Earth imaging.
\end{abstract} 


\defcitealias{vLieshout14}{V14}

\section{Introduction}
About $20 \%$ of main sequence stars show far-infrared (IR) excesses, indicating the presence of circumstellar dust orbiting at tens or hundreds of au, known as debris discs \citep{Wyatt08,Krivov07,Matthews14}. For example, surveys have been conducted at $24~\mu$m and $70~\mu$m with Spitzer/MIPS \citep[e.g.][]{Rieke05,Meyer06,Su06,Hillenbrand08} and $70 - 160~\mu$m with Herchel/PACS \citep{Eiroa13,Thureau14,Sibthorpe18} in order to detect cold debris discs. These observations are explained by large planetesimals in a belt colliding and grinding down to produce a collisional cascade of particles covering a range of sizes. Poynting-Robertson (P-R) drag causes particles to lose angular momentum, leading to inward migration as their semimajor axes decrease and their orbits become circularised. However, \citet{Wyatt05} showed that for the debris discs that have been detected, the majority of the dust does not manage to migrate in by P-R drag, since these belts are so dense that the dust is destroyed on a much shorter timescale in mutual collisions. \par 
Nevertheless, observations of stars in the mid-infrared show the presence of warm ($\sim 300~$K) dust in many systems, at closer proximity to the star than a standard debris disc. These clouds of warm dust are referred to as `exozodi', with exozodiacal dust in analogy to our solar system's zodiacal cloud. The term 'exozodiacal dust' can also be used to refer to the hot circumstellar dust which produces near-infrared excesses \citep[originally detected by][]{Absil06}. While this is important to the study of the very innermost regions of planetary systems within a few 0.1~au of the star, it requires additional physics which is not considered here, such as sublimation. The focus of this paper is on warm exozodiacal dust, which produces mid-infrared excesses. For observations of hot exozodi, see \citet{Absil13} and \citet{Ertel14}. \par
The first observations of warm exozodiacal dust were done using mid-infrared photometry, such as with the Infrared Astronomical Satellite \citep[IRAS,][]{Gaidos99}, the Infrared Space Observatory \citep[ISO,][]{Laureijs02}, Spitzer \citep{Lawler09}, and WISE \citep{Kennedy13}, all finding less than $2\%$ of observed stars had mid-infrared excesses at the given sensitivities. However, the sensitivity of photometry is limited, as only the brightest exozodi can be detected above the stellar photosphere, with mid-infrared fluxes in excess of 10$\%$ of the stellar flux required for a detection. Interferometry gives much better spatial resolution, allowing the dust emission to be separated from that of the star such that much fainter excesses can be detected, at levels $\lesssim 1\%$. Previously the Keck Interferometer Nuller \citep[KIN,][]{Colavita09,Serabyn12} studied 47 nearby main-sequence stars \citep{Mennesson14,Millan-Gabet11}. These studies found that five stars had an 8-9~$\mu$m excess at a sensitivity of 150 zodis (where the unit of zodi refers to dust with the same optical depth at 1~au as the zodiacal cloud), equivalent to an excess of $\sim 1\%$. NASA's Large Binocular Telescope Interferometer \citep[LBTI,][]{Defrere16,Hinz14} is a nulling interferometer which is sensitive to warm dust down to the level of a few zodis, equivalent to a null excess of $\sim 0.05\%$. Recently the HOSTS survey \citep{Ertel18_hosts,Ertel20} was conducted in the N band to search for levels of exozodiacal dust around 38 nearby main sequence stars using LBTI, with a detection rate of $26 \%$. \par
The origin of exozodiacal dust is still not well understood. In a cold planetesimal belt at tens of au, km-sized bodies can survive for Gyr timescales, but in exozodiacal clouds at just a few au, collisional lifetimes of such planetesimals are much shorter \citep{Dominik03,Wyatt07_zodi}. While some exozodi can be explained by an in situ planetesimal belt  \citep{Geiler17}, particularly for those found in young systems, this is not always the case \citep[see, e.g.][]{Wyatt07_zodi,Lebreton13}. For example, \citet{Kennedy13} concluded that another component is needed in addition to in situ belts to explain the bright exozodi detected in WISE $12~\mu$m observations. Some of these exozodi could be explained by transient phenomena, such as a dynamical instability similar to the Late Heavy Bombardment (LHB) of the solar system, that produces a short-lived enhancement of dust in the inner regions \citep{Booth09,Bonsor13}. Rare, bright exozodi may also be explained by recent collisions of large planetesimals similar to the Moon-forming impact \citep{Jackson12,Kral15}. However, such events cannot explain a phenomenon as common as 26\%.\par
Another possibility is that exozodiacal dust is produced in a cold outer planetesimal belt, and then transported to the inner regions of the planetary system, either by P-R drag \citep{Kennedy15} or comet delivery \citep{Bonsor12_zodi,Marboeuf16,Faramaz17}. Indeed, it is believed that the zodiacal cloud is primarily sustained by Jupiter-family comets delivering material to the inner solar system \citep{Nesvorny10,Nesvorny11_ZC,Rowan-Robinson13,Ueda17}, along with dust from the asteroid belt being dragged in towards the Sun by P-R drag. Both the results of KIN \citep{Mennesson14} and HOSTS \citep{Ertel20} found a higher occurrence rate of exozodiacal dust around stars with known far-infrared excesses which imply the presence of a cold debris disc. \par
Given the potential correlation of the presence of cold and warm dust, it is important to explore the viability of transport of dust from an outer planetesimal belt as a source of the observed exozodiacal dust. The full distribution of grains created in a planetesimal belt, including all grain sizes and distances to the star, can be studied numerically. For example, ACE \citep{Krivov05,Krivov06,Krivov08,Reidemeister11_drag} finds the distribution of particles in phase space based on the gain and loss of particles to collisions and drag, with simplified dynamics. Similarly, \citet[hereafter V14]{vLieshout14} produced a numerical model of the evolution of particles in a debris disc, including the effects of collisions, P-R drag, and sublimation. However, numerical methods such as these models are computationally expensive, meaning they are less straightforward to implement than a simple analytical model and are more time-consuming. \par
A simple analytical prescription for the process of P-R drag transporting dust inward from a planetesimal belt exists, but is only approximate. For example, \citet{Mennesson14} showed consistency between observations with KIN and the simple analytical prescription of \citet{Wyatt05} for the interplay of collisions and P-R drag. A modified version of the \citet{Wyatt05} model was used by \citet{Kennedy15} to predict the levels of dust transported inwards from Kuiper belt analogues by P-R drag. They found that LBTI, which probes lower excess levels, should be able to detect this component of dust brought inwards by P-R drag for systems with known Kuiper belt analogues, and that it may detect such dust for some systems with no detectable parent belt. Nevertheless, this model was still inaccurate, and only considered a single grain size, moreover assuming the grains were black bodies. \par
This provides the motivation for this study, which aims to produce an analytical model that considers all particle sizes, rather than only grains just above the blowout size, to give a distribution of dust for a system with an outer planetesimal belt evolving via collisional evolution and P-R drag. The size distribution is described in terms of geometrical optical depth, defined such that $\tau(D, r)dD$ is the cross-sectional area surface density in particles of size $D \rightarrow D + dD$ at radius $r$. Combining the approach of \citet{Wyatt11} for the size distribution of a planetesimal belt at a single radial distance with \citet{Wyatt05} for the radial profile of a given particle size produces a two-dimensional distribution. Results from the analytical model are validated against the numerical model of \citetalias{vLieshout14} to show how it can predict two-dimensional distributions in debris discs, and to find the limitations of the analytical model. This gives predictions for the levels of dust transported to the inner regions of a planetary system. Realistic grain properties can be applied to find the corresponding flux, and so the model can be used to assess whether this scenario could explain the observed mid-infrared excesses of exozodi, such as those found by LBTI. \par
We briefly summarise the numerical model of \citetalias{vLieshout14} and our numerical results in Section~\ref{sec:numerical}, then describe our analytical model in Section~\ref{sec:analytical}. In Section~\ref{sec:comparison} we compare the predictions of our analytical model with results from the numerical model in order to fit the model parameters before exploring the parameter space in Section~\ref{sec:parspace}. Applications of the model to the zodiacal cloud and exozodi are presented in Section~\ref{sec:obs}, and our conclusions are given in Section~\ref{sec:conc}. \par

\section{Numerical Model}
\label{sec:numerical}
The numerical model of \citetalias{vLieshout14} considers the evolution of a belt of planetesimals and the debris created when they are destroyed in mutual collisions. It self-consistently takes into account the effects of collisions, P-R drag, and sublimation on these particles. This model uses a statistical method based on that of \citet{Krivov05, Krivov06}, applying kinetic theory to obtain the spatial and size distribution of particles in a phase space of orbital elements and particle masses. The phase space is over orbital eccentricity $e$, periastron distance $q$, and particle mass $m$; other orbital elements are averaged over under the assumption that the disc is axisymmetric. It is assumed that there is a uniform distribution of particles over inclination. \par 
The continuity equation is solved numerically to find the number of particles in each phase space bin at successive times. P-R drag and sublimation act as diffusion terms which shift particles to adjacent phase space bins. For these processes orbit-averaged equations are used, assuming that the relevant timescales are longer than an orbital period. \par
Collision rates are calculated between pairs of phase-space bins according to analytical equations from \citet{Krivov06}, including the number densities of the particles, their relative velocity, collisional cross-section, and effective volume of interaction. The outcome of collisions is determined by the impact energy per unit mass: if this exceeds the critical specific energy, $Q_\mathrm{D}^\star$, the collisions are destructive such that the largest fragment has at most half the mass of the target particle. Cratering collisions, which have specific energy below $Q_\mathrm{D}^\star$, are not considered by the model.  When two particles collide, their mass is redistributed amongst the bins according to a redistribution function, which is a power law $n_\mathrm{r}(D) \propto D^{-\alpha_\mathrm{r}}$, where $D$ is particle diameter. Integrating this redistribution function gives the number of particles which go into each bin, up to a maximum mass determined by the specific energy. The orbital elements of collision fragments are determined based on conservation of momentum and the effects of radiation pressure. If a particle has a mass below the lowest mass bin, it is considered lost due to blowout. \par
The strength of radiation pressure acting on particles of a given size is determined by the ratio of radiation pressure to gravity acting on a particle:
\begin{equation}
\label{eq:beta}
\beta = \frac{F_{\mathrm{rad}}}{F_\mathrm{g}} = \frac{3L_\star Q_{\mathrm{pr}}}{8\pi GM_\star cD\rho},
\end{equation}
where $L_\star$ is the stellar luminosity, $M_\star$ is stellar mass, $\rho$ is particle density, and $c$ is the speed of light. $Q_{\mathrm{pr}}$ is the radiation pressure efficiency averaged over the stellar spectrum, which is a function of particle size; this can be found numerically given assumptions about the dust composition and stellar spectrum (see Section~\ref{subsec:optprops}). Grains released from parent bodies on circular orbits will have eccentricities $e~=~\beta /(1~-~\beta)$, and will be blown out of the system on hyperbolic orbits when $\beta \geq 0.5$. From this, we can estimate the smallest particle size which can remain bound, under the assumption of $Q_\mathrm{pr} = 1$, such that grains are perfect absorbers, as
\begin{equation}
\label{eq:Dbl}
D_\mathrm{bl} = \frac{3 L_\star}{4\pi G M_\star c \rho}.
\end{equation}
\par
It would be computationally expensive to model the entire collisional cascade from km-sized planetesimals down to sub-micron grains, so only particles $0.1~\mu\text{m} < D < 2~\text{cm}$ are modelled. The largest bodies will remain confined to the planetesimal belt, with negligible P-R drag, producing dust via collisions. This is taken into account with a source of dust in the belt, which mimics the production of grains by larger bodies, replenishing the dust in the belt each time step. The model is run from an initially empty disc until steady state is reached, such that the distribution changes by less than 1\% in a logarithmic time step. For a very massive, collisional disc it takes $10~$Gyr to reach steady state due to the time taken by the largest particles to migrate inwards from the belt via P-R drag, as the migration timescale is proportional to particle size. However, the smallest, barely-bound grains will be in steady state after $\sim$10~Myr, and dominate the optical depth. The time taken to reach steady state increases as disc mass is decreased, and the least massive disc considered here takes $10^{14}$~yr to reach steady state at $<1$~au. The optical depth of the lowest mass discs is dominated by larger grains, which take longer to evolve inwards. While this is an unrealistically long time, the time taken to reach steady state would likely be shorter with alternative initial conditions. The starting conditions chosen here assume an initially empty disc. It may be more realistic for the disc to start with dust spread throughout the system and so closer to steady state, which would lead to shorter convergence times. In terms of computing time, it takes about a week on a standard desktop. Only particles at < 0.02~au are affected by sublimation around a Sun-like star, and the minimum distance used in this paper is 0.03~au. Sublimation is therefore ignored throughout this paper, as the focus is on the overall distribution, rather than the innermost edges of the disc. Sublimation would, however, be very important when studying hot exozodiacal dust, which gives rise to near-infrared excesses. Our model is only aimed at explaining warm exozodi, for which sublimation is less important. The output of the numerical model is a steady-state distribution of particles in the phase space of orbital elements and particle masses, which can be converted to a distribution over radial distance and particle size via Haug's integral \citep{Haug58}. \par

\subsection{Model inputs}
\label{subsec:numinputs}
The inputs to the model include the radius of the parent belt $r_0$, stellar mass $M_\star$ and luminosity $L_\star$, the semi-opening angle of the disc $\epsilon$, and the slope of the redistribution function $\alpha_\mathrm{r}$. The critical specific energy for catastrophic collisions follows a combination of power laws to represent the strength and gravity regimes as $Q_\mathrm{D}^\star = Q_\mathrm{a}D^{-a} + Q_\mathrm{b}D^b$, with the parameters of the power laws being inputs. The overall level of dust in the source belt is set with an input parameter which is the mass supply rate of dust in the belt from collisions of large bodies, $\dot{M}_\mathrm{in}$. The mass from the break-up of the largest bodies is distributed according to the redistribution function of collisional fragments $n_\mathrm{r}(D)$ across the range of sizes considered, down to the blowout size. The model also takes the grain density $\rho$, as well as values of $\beta$ for different particles sizes (see Section~\ref{subsec:optprops}). \par 
Typically it is assumed that the slope of the redistribution function is in the range $3 < \alpha_\mathrm{r} < 4$, so $\alpha_\mathrm{r} = 3.5$ is used. The effect of varying $\alpha_\mathrm{r}$ is studied in Section~\ref{subsec:alpha_r}. All particles which are modelled are small enough such that they are in the strength regime of $Q_\mathrm{D}^\star$. Laboratory experiments with high-velocity collisions of small particles find a constant value of $Q_\mathrm{D}^\star = 10^7~\text{erg g}^{-1}$ \citep{Flynn04}, which is used for ease of comparison between models. Section~\ref{subsec:QD} investigates the effect of using a power law for $Q_\mathrm{D}^\star$. Most of the simulations are for a Sun-like star, $M_\star = \mathrm{M}_{\sun}$, $L_\star = \mathrm{L}_{\sun}$, and a disc semi-opening angle (equivalent to the maximum orbital inclination of particles) of $\epsilon = 8.5 \degree$. The fiducial value for the belt radius is $r_0 = 30~$au, and for the mass input rate is $\dot{M}_\mathrm{in} = 10^{-15} \mathrm{M}_{\earth}~\text{yr}^{-1}$, though a range of values is considered for each. \par
A similar phase space grid is used to that of \citetalias{vLieshout14}. For eccentricity, the grid has ten logarithmically spaced bins for $0 \leq e \leq 1$, for which the lowest bin is at $e = 10^{-4}$, with two linear bins each for hyperbolic orbits ($1 \leq e \leq 2$) and the anomalous hyperbolic orbits of $\beta > 1$ grains ($-2 \leq e \leq -1$). The periastron grid has 60 logarithmic bins from $q=0.03~$au to $q = 100~$au for the fiducial model, which has a belt at $r_0 = 30~$au. The mass grid has logarithmic bins, with higher resolution for the smallest particles. There are 45 high resolution bins from $D = 0.1~\mu$m to $D = 20~\mu$m, with 18 low resolution bins going up to the maximum size $D = 2~$cm.

\subsection{Optical properties}
\label{subsec:optprops}
Optical properties of the grains are calculated using the same method as \citet{Wyatt02}, with compositions from the core-mantle model of \citet{Li97}, first used for interstellar dust, which can also be used for dust in debris discs \citep{Li98,Augereau99}. This model assumes fluffy aggregates with a silicate core and organic refractory mantle. Grains are nominally assumed to be asteroidal, such that they have volume fractions of $1/3$ amorphous silicate and $2/3$ organic refractory material with zero porosity, which gives a dust density of $\rho = 2.37 \text{g cm}^{-3}$ Alternative compositions are considered later in the paper. The radiation pressure efficiency $Q_\mathrm{pr}$ and absorption efficiencies $Q_\mathrm{abs}(\lambda, D)$ are calculated using Mie Theory \citep{Bohren83}, Rayleigh-Gans theory, or geometric optics, depending on the wavelength \citep{Laor93}. Given $Q_\mathrm{pr}$, realistic values of $\beta$ can be found for use in the numerical model. For the assumed asteroidal composition, the blowout size (i.e. that for which $\beta = 0.5$) is found to be $D_\mathrm{bl} \sim 1.5~\mu$m around a Sun-like star. Grain temperatures $T(D, r)$ and absorption efficiencies $Q_\mathrm{abs}(\lambda, D)$ are used when predicting fluxes from distributions of dust (Section~\ref{subsec:applyexo}). \par

\begin{figure*}
	\centering
	\includegraphics[width=0.85\textwidth]{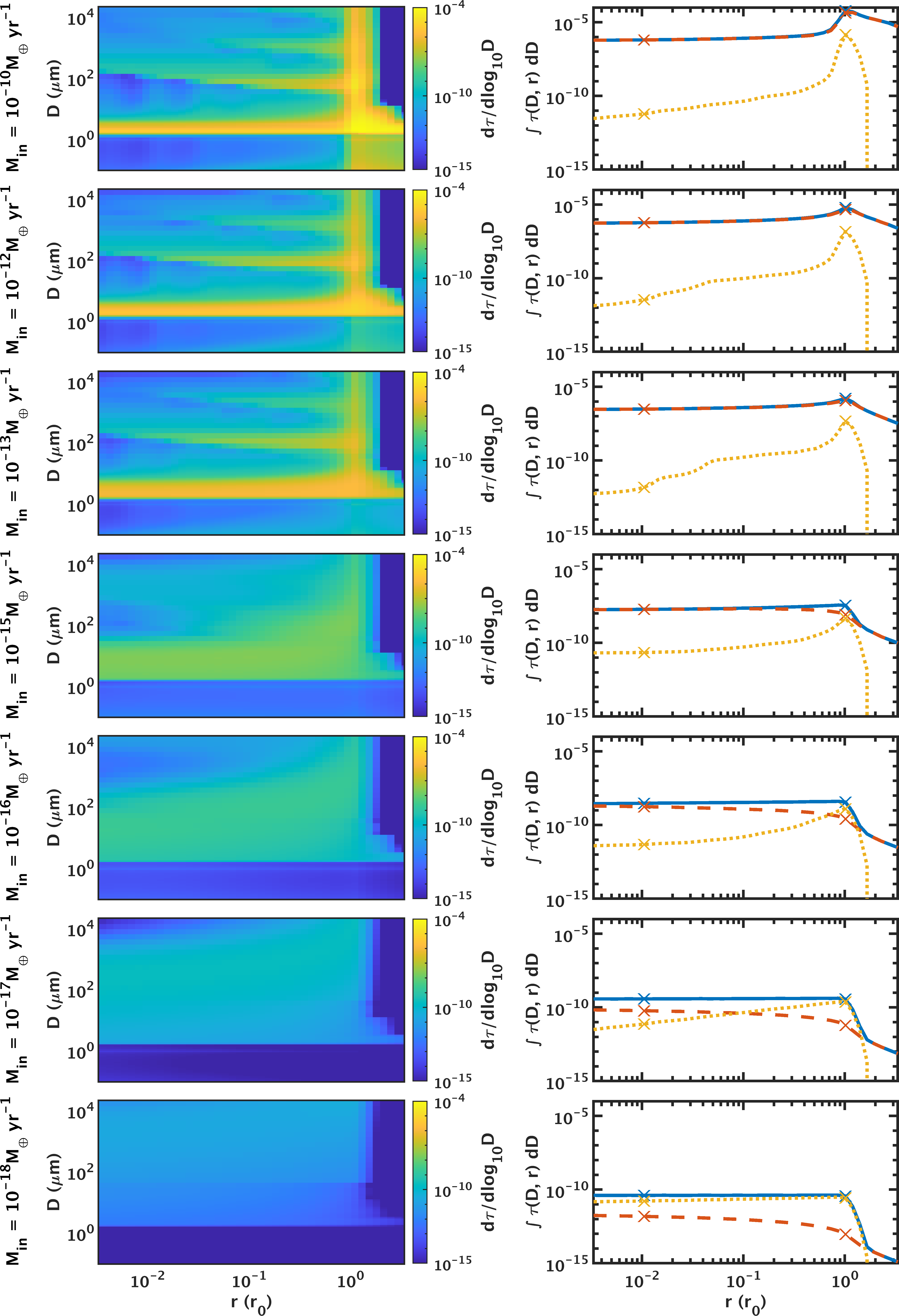}
	\caption{Left: two-dimensional size distribution of particles over size and radial distance from the numerical model of \citetalias{vLieshout14} for discs with different dust mass input rates and a belt radius of 30~au. The colour scale gives the optical depth per unit size decade. Right: radial distribution of optical depth integrated over different size ranges: all particles (blue), barely bound grains (orange, dashed), and cm-sized particles (yellow, dotted). Markers show the values close in to the star ($0.01 r_0$) and in the belt, which are used later to characterise the behaviour of the model.}
	\label{fig:num_Min}
\end{figure*}

\subsection{Numerical results}
\label{subsec:numres}
The left column of Figure~\ref{fig:num_Min} shows the results of simulations for belts with the standard values of Sections~\ref{subsec:numinputs} and \ref{subsec:optprops}, but with mass input rates varying from $10^{-18} \mathrm{M}_{\earth}~\text{yr}^{-1}$ to $10^{-10} \mathrm{M}_{\earth}~\text{yr}^{-1}$. The colour scale shows the geometrical optical depth per unit size decade, $\mathrm{d}\tau / \mathrm{d}\log_{10}D$, as a function of both particle size $D$ and radial distance relative to the belt radius $r_0$. \par
Some characteristics are seen at all disc masses, such as a concentration of dust at $r_0$ with a drop off outside of the belt, while the smallest grains are put onto highly eccentric orbits by radiation pressure, forming a halo at larger radii.  Plots are truncated at a radius of $3r_0$, as this study is not focussed on modelling the halo. In all cases grains below the blowout size, $D_\mathrm{bl} \sim 1.5\mu\text{m}$, are blown away by radiation pressure such that their contribution is negligible. Grains below the blowout size are present in the belt, where they are produced, and in the halo as they are blown out, with a density which is proportional to the mass input rate. However, these grains do not contribute significantly to the optical depth overall as they are orders of magnitude lower than other grain sizes in the belt. As disc mass is decreased, the discs evolve from being collisional to being dominated by P-R drag, which is reflected by a flattening of the radial distribution. \par
Where particles are collisional, they are destroyed by collisions before they have a chance to migrate in towards the star, such that their optical depth is heavily depleted inwards of the belt. A characteristic wavy pattern is seen in collisional discs. This is a well-known phenomenon in the size distributions of collisional cascades \citep[see, e.g.][]{CampoBagatin94,Durda97,Thebault03,Krivov06} which is caused by the truncation of the size distribution below $D_\mathrm{bl}$, where particles get blown away by stellar radiation. The lack of particles just below $D_\mathrm{bl}$ means that particles just above the cutoff are not destroyed by collisions due to a lack of impactors, causing an increase in their numbers. This increased number of particles just above $D_\mathrm{bl}$ then breaks up larger particles faster, causing them to be more depleted than in an infinite cascade, and so on. Within the belt the size distribution follows the standard result for a collisional cascade \citep{Dohnanyi69}, such that $n(D) \propto D^{-\alpha}$, with a slope of $\alpha = 3.5$. The wavy pattern is then superposed on this size distribution, and the pattern also extends inwards of the belt since collisions continue to operate on the dust as it is dragged into the inner regions. The barely bound grains just above $D_\mathrm{bl}$ dominate the optical depth and have a flatter radial profile than larger grains. \par

For the lowest disc masses, P-R drag is the dominant loss mechanism, and particles migrate inwards while suffering very few collisions. This gives radial profiles which are almost flat, although the largest particles still see a small amount of depletion due to collisions. In these cases, the particle size which dominates the optical depth is the largest particle size which is not significantly depleted by collisions. As expected, the overall number and mass of particles also decreases as mass input rate decreases, leading to lower total cross-section. \par
The waves seen in the distributions cannot be modelled analytically, so to better compare the numerical and analytical models we also consider the optical depth integrated over decades of particle size to smooth over the waves. The right hand column of Figure~\ref{fig:num_Min} shows the radial distribution of integrated optical depth. This is shown for three different size ranges: the total optical depth, small grains which are barely bound ($D_\mathrm{bl} < D < 20 \mu\text{m}$), and the largest particles ($2~\text{mm} < D < 2~\text{cm}$). For the most massive discs, the total optical depth is very close to that from the smallest particles, which is due to grains just above the blowout size dominating the cross-section. Larger particles are very depleted, and contribute much less. As disc mass is decreased, the relative contribution of barely bound grains decreases, and the largest particles contribute more. For the least massive disc, most of the cross-section is in the largest grains. \par

\section{Analytical Model}
\label{sec:analytical}
Here we present a model which predicts the size distribution of the disc at different radii by considering the balance between collisional evolution and migration due to P-R drag. First we consider the size distribution expected at the location of the planetesimal belt, then apply a model for how these particles evolve inwards of the belt. \par

The size distribution within the planetesimal belt can be found using the model of \citet{Wyatt11}, which determines the size distribution in a planetesimal belt at a single radius undergoing catastrophic collisions with loss processes acting. Particles are considered lost from the belt when P-R drag causes them to migrate past the belt's inner edge.
To find how the distribution of a given particle size evolves radially we use the model of \citet{Wyatt05}, which found the radial optical depth profile for a population of single-sized particles evolving via destructive collisions and P-R drag. The shape depends on the ratio of the P-R drag timescale to the collision time. \par 
Combining the models of \citet{Wyatt11} and \citet{Wyatt05} gives the size distribution of a debris disc at different radial locations, taking into account the collisional evolution of particles and P-R drag. 

\subsection{Parent belt size distribution}
\label{subsec:beltdist}
Consider a belt of planetesimals at a radius $r_0$ from the star, collisionally evolving to produce smaller grains. The method of \citet{Wyatt11} can be used to find the one-dimensional size distribution of particles in the belt, which will extend up to some maximum particle size $D_\mathrm{max}$. The lower end of the distribution is determined by the blowout size $D_\mathrm{bl}$ (equation~\ref{eq:Dbl}). The size distribution is approximated by a series of broken power laws; the precise power laws depend on the collision timescales, as collisions move material down the collisional cascade.  We calculate the collision rate between particles in the disc with the particle-in-a-box approach. The rate of impacts onto a particle of size $D$ from impactors of sizes $D_{\mathrm{im}} \rightarrow D_{\mathrm{im}} + dD_{\mathrm{im}}$ is
\begin{equation}
\label{eq:Rcoll}
R_{\mathrm{coll}}(D,~D_{\mathrm{im}})dD_{\mathrm{im}} = \frac{n(D_{\mathrm{im}})dD_{\mathrm{im}}}{V}\frac{\pi}{4}(D + D_{\mathrm{im}})^2v_{\mathrm{rel}}.
\end{equation}
Here $V$ is the disc volume, $n(D_{\mathrm{im}})$ is the number of particles per unit diameter, and $v_{\mathrm{rel}}$ is the relative velocity of the collisions. As in the numerical model (Section~\ref{sec:numerical}), only catastrophic collisions are considered; these are destructive such that the largest fragment has less than half the mass of the target particle, producing fragments according to some redistribution function $n_\mathrm{r}(D) \propto~ D^{-\alpha_\mathrm{r}}$. Catastrophic collisions require the impact energy per unit target mass to be above some critical dispersal value $Q_{\mathrm{D}}^\star$, so destructive collisions only occur with impactors of a diameter greater than $X_\mathrm{C}D$, where
\begin{equation}
\label{eq:XC}
X_{\mathrm{C}} = \left(\frac{2Q_{\mathrm{D}}^\star}{v_{\mathrm{rel}}^2}\right)^{\frac{1}{3}}.
\end{equation}
The critical specific energy for dispersal in the strength regime is parameterised as $Q_\mathrm{D}^\star = Q_\mathrm{a}D^{-a}$. It is assumed that the velocity of collisions is related to the Keplerian velocity by the maximum inclination, $I_\mathrm{max}$, as
\begin{equation}
\label{eq:vel}
v_{\mathrm{rel}} = I_\mathrm{max}v_{\mathrm{k}} = I_\mathrm{max}\sqrt{\frac{GM_\star}{r}},
\end{equation}
where the semi-opening angle of the disc $\epsilon$ in the \citetalias{vLieshout14} model would correspond to the maximum inclination. This assumes that the relative velocity of collisions is dominated by the vertical motion perpendicular to the plane of the disc. The volume of a disc of width $dr$  and radius $r$ can be approximated as 
\begin{equation}
\label{eq:vol}
V = 4\pi r^3\left(\frac{dr}{r}\right)I_\mathrm{max}.
\end{equation}
Within the belt we assume that the number of particles per unit diameter follows a power law
\begin{equation}
\label{eq:nD}
n(D) = KD^{-\alpha},
\end{equation}
such that $n(D)dD$ is the number of particles with diameters $D~\rightarrow~D~+~dD$, and the classical power law has $\alpha = 3.5$ when $Q_{\mathrm{D}}^\star$ is independent of particle size \citep{Dohnanyi69,Tanaka96}. Integrating over all possible impactors, we find that the rate of catastrophic collisions for a particle of size $D$ is 
\begin{align}
\label{eq:Rcc}
R_{\mathrm{cc}}(D) &= \int_{X_{\mathrm{C}}D}^{D_{\mathrm{max}}}R_{\mathrm{coll}}(D, D_{\mathrm{im}})dD_{\mathrm{im}} \nonumber \\
&\approx \frac{\pi}{4(\alpha - 1)}\frac{Kv_{\mathrm{rel}}}{V}X_{\mathrm{C}}^{1-\alpha}D^{3-\alpha},
\end{align}
where we assume that $\alpha > 3$ and that $X_{\mathrm{C}} \ll 1$ to find the most relevant term in the collision rate. Therefore, the collision timescale for particles of size $D$ is
\begin{equation}
\label{eq:tcoll}
t_{\mathrm{coll}}(D) \approx \frac{4(\alpha-1)}{\pi}\frac{V}{Kv_{\mathrm{rel}}}X_{\mathrm{C}}^{\alpha - 1}D^{\alpha - 3}.
\end{equation}
Note that this may have additional size dependence via $X_\mathrm{C}$ when $Q_\mathrm{D}^\star$ is a power law with size. The normalisation of the size distribution in equation~\ref{eq:nD} can be found by
\begin{equation}
\label{eq:KD}
K = \frac{6(4-\alpha)}{\pi \rho}D_{\mathrm{max}}^{\alpha - 4} M_{\mathrm{dust}},
\end{equation}
where $\rho$ is the density of particles, $D_{\mathrm{max}}$ is the maximum particle diameter, and $M_{\mathrm{dust}}$ is the total mass of dust particles, under the assumption that $\alpha < 4$ such that the mass distribution is dominated by the largest particles. \par
The timescale for a particle on a circular orbit to migrate in to the star via P-R drag from a radius $r$ is \citep{Wyatt50,Burns79}
\begin{equation}
\label{eq:tPR}
t_{\mathrm{PR}}(r) = \frac{cr^2}{4GM_\star\beta},
\end{equation}
where $M_\star$ is the stellar mass, $c$ is the speed of light, and $\beta$ the ratio of radiation pressure to gravity acting on a particle. 

\label{subsec:PR}
The balance between collisions and P-R drag is described by the ratio of their timescales for particles in the belt,
\begin{equation}
\label{eq:eta0}
\eta_0(D) = \frac{t_{\mathrm{PR}}(D, r_0)}{t_{\mathrm{coll}}(D, r_0)}. 
\end{equation}
Both timescales are dependent on particle size, so the relative strength of collisions and P-R drag is a function of particle size. For low mass discs there is a critical particle diameter $D_{\mathrm{pr}}$ such that the P-R drag and collisional timescales are equal,
\begin{equation}
\label{eq:tDpr}
\eta_0(D_{\mathrm{pr}}) = 1, \qquad t_{\mathrm{coll}}(D_{\mathrm{pr}}, r_0) = t_{\mathrm{PR}}(D_{\mathrm{pr}}, r_0).
\end{equation}
\citet{Wyatt11} showed that the size distribution in a planetesimal belt can be approximated by two power laws of different slope, with a transition at the critical particle size $D_\mathrm{pr}$. Particles larger than $D_\mathrm{pr}$ ($\eta_0 > 1$) are dominated by destructive collisions, and they follow the classical size distribution given in equation~\ref{eq:nD}, with a slope determined by the power law of $Q_\mathrm{D}^\star$ as 
\begin{equation}
\label{eq:alphaQD}
\alpha = \frac{7 - a/3}{2 - a/3}.
\end{equation}
However, \citet{Wyatt11} found a turnover in the slope of the size distribution for particles smaller than $D_\mathrm{pr}$ such that $\eta_0 < 1$, for which P-R drag is the dominant loss mechanism. The new slope is given by $\alpha_\mathrm{pr} = \alpha_\mathrm{r} - 1$:
\begin{equation}
\label{eq:nPR}
n(D) = K_{\mathrm{pr}}D^{1-\alpha_\mathrm{r}},\qquad D\leq D_{\mathrm{pr}},
\end{equation}
where continuity of the size distribution at $D_{\mathrm{pr}}$ gives that
\begin{equation}
\label{eq:kPR}
K_{\mathrm{pr}} = KD_{\mathrm{pr}}^{\alpha_\mathrm{r} -\alpha - 1}.
\end{equation}
Once again integrating equation~\ref{eq:Rcoll} over the size distribution of impactors for particles smaller than $D_\mathrm{pr}$ and assuming that $\alpha > 3$ and $3 < \alpha_\mathrm{r} < 4$, for particles such that $D \ll D_{\mathrm{pr}}$ and $X_\mathrm{C} \ll 1$, the dominant term in the collision timescale is
\begin{equation}
\label{eq:tcollPR}
t_{\mathrm{coll}}(D<D_{\mathrm{pr, eff}}) \approx \frac{4(\alpha_\mathrm{r}-2)}{\pi}\frac{V}{Kv_{\mathrm{rel}}}D_{\mathrm{pr}}^{\alpha - \alpha_\mathrm{r}+1}X_\mathrm{C}^{\alpha_\mathrm{r}-2}D^{-(4-\alpha_\mathrm{r})}. 
\end{equation}
This is equivalent to replacing $\alpha$ by $\alpha_\mathrm{r}- 1$ and $K$ by $K_\mathrm{pr}$ in equation~\ref{eq:tcoll}. So under the given assumptions, equation~\ref{eq:tcoll} applies to both regimes of the size distribution, with different parameters $\alpha$ and $K$. \par
While the size distribution is continuous, with two regimes which match at $D_\mathrm{pr}$, generally $t_{\mathrm{coll}}$ and thus $\eta_0$ are discontinuous at $D_{\mathrm{pr}}$, motivating the introduction of an effective critical size $D_{\mathrm{pr,eff}}$ at which they are continuous. Then $t_\mathrm{coll}$ and $\eta_0$ have two power laws which join at $D_\mathrm{pr, eff}$. It is to be expected that particles slightly bigger than $D_\mathrm{pr}$ will be affected by the turnover of the size distribution, as it affects the number of impactors that can catastrophically destroy them. The size at which $t_\mathrm{coll}$ and $\eta_0$ are continuous is close to the particle size at which $X_\mathrm{C}D = D_{\mathrm{pr}}$ such that the smallest impactors are of size $D_{\mathrm{pr}}$, so that 
\begin{align}
\label{eq:DPReff}
D_{\mathrm{pr, eff}} &= \left(\frac{\alpha_\mathrm{r} - 2}{\alpha - 1}\right)^{\frac{1}{1+\alpha - \alpha_\mathrm{r}}}X_\mathrm{C}^{-1}D_{\mathrm{pr}}\\
&= \left(\frac{\alpha_\mathrm{r} - 2}{\alpha - 1}\right)^{\frac{3}{(3-\alpha)(1+\alpha -\alpha_\mathrm{r})}}\left(\frac{v_{\mathrm{rel}}^2}{2Q_\mathrm{a}}\right)^{\frac{1}{3-\alpha}}D_{\mathrm{pr}}^{\frac{3}{3-\alpha}}.	
\end{align}

\subsection{Two-dimensional distribution}
\label{subsec:2ddist}
Once the size distribution of the parent belt has been found, each particle size is assumed to evolve independently inwards of the belt. The radial profile of a given particle size can be found using the model of \citet{Wyatt05}, which takes into account P-R drag and mutual collisions for a single particle size. The collision timescales from Section~\ref{subsec:beltdist} (equations \ref{eq:tcoll} and \ref{eq:tcollPR}) and the P-R drag timescales (equation~\ref{eq:tPR}) are used to calculate the values of $\eta_0$ for each size (equation~\ref{eq:eta0}), which determines the shape of the radial profile. This gives a two-dimensional size distribution over particle size $D$ and radius from the star $r$. We express the distribution of particles in terms of vertical geometrical optical depth, which is the surface density of cross-sectional area. \par
For an annulus of particles all of the same size at $r \rightarrow r + dr$, the geometrical optical depth is given by
\begin{equation}
\label{eq:tau0r}
\tau(r) = \frac{\sigma n(r)dr}{2\pi rdr} = \frac{\sigma n(r)}{2\pi r},
\end{equation}
where $n(r)$ is the number density of particles per unit radius and $\sigma$ is the cross-section of a particle. \citet{Wyatt05} solved the continuity equation for $n(r)$ to show that the optical depth due to particles inwards of a belt at radius $r_0$ is given by
\begin{equation}
\label{eq:taur}
\tau(r) = \frac{\tau(r_0)}{1 + 4\eta_0(1 - \sqrt{\frac{r}{r_0}})}.
\end{equation}
The shape of the profile depends on the balance between collisions and P-R drag via the parameter $\eta_0$. For massive debris discs, such as those which are currently detectable, collisions dominate and there is a sharp depletion of particles inwards of the belt, as grains are destroyed before they have a chance to migrate inwards. For less massive discs, however, the dominant process is migration via P-R drag; with negligible collisions the surface density becomes constant throughout the disc. \par
Now considering a distribution of particle sizes, if $\tau(D)dD$ is the cross-sectional area surface density in particles of size $D \rightarrow D + dD$ at radius~$r$, the optical depth in a belt with size distribution $n(D)$ is given by
\begin{equation}
\label{eq:tau0D}
\tau_0(D) = \frac{n(D)\pi \frac{D^2}{4}}{2\pi r_0 dr} = \frac{n(D) D^2}{8r_0^2\left(\frac{dr}{r}\right)}.
\end{equation}
As stated above, the model of \citet{Wyatt05} is applied to each particle size independently. By only considering a single size of particle, this does not take into account the gain of smaller particles due to the fragmentation of larger grains or the overall size distribution. While particles will interact and be destroyed in collisions with particles of different sizes, we assume that the collision rate of grains of a given size scales with the number of similarly sized particles. This approximation will be corrected for in Section~\ref{subsec:applmodel}. Let $\tau(D, r)dD$ be the cross-sectional area surface density in particles of size $D \rightarrow D + dD$ at radius r, then applying equation~\ref{eq:taur} to each size $D$, the two-dimensional distribution is
\begin{equation}
\label{eq:tauDr}
\tau(D, r) = \frac{\tau_0(D)}{1+4\eta_0(D)\left(1-\sqrt{\frac{r}{r_0}}\right)}.
\end{equation}

\subsection{Application of model}
\label{subsec:applmodel}
The analytical model makes many assumptions, such as that all particles are on circular orbits. It also assumes that inwards of the belt, particles are only destroyed by similarly sized particles, or at a rate which scales with the local density of similarly sized particles. To take into account approximations in the model, we follow \citet{Kennedy15} in introducing a factor $k$ which modifies the previously derived collisional timescales, affecting how collisional particles are such that
\begin{equation}
\label{eq:eta0k}
\eta_0(D) = \frac{t_{\mathrm{PR}}(D, r_0)}{kt_{\mathrm{coll}}(D, r_0)}.
\end{equation}
Section~\ref{subsec:fitting} shows that it is necessary for this to be dependent on size, which is implemented as
\begin{equation}
\label{eq:kD}
k = k_0 \left(\frac{D}{D_\mathrm{bl}}\right)^{-\gamma},
\end{equation}  
for some parameters $k_0$ and $\gamma$ which are to be fitted by comparison with the numerical model (Section~\ref{sec:numerical}), where these approximations were not made. \par
Our model can be used to find the two-dimensional size distribution for a disc, given the belt radius $r_0$, the grain density $\rho$, the stellar mass and luminosity $M_\star$ and $L_\star$, the maximum inclination $I_\mathrm{max}$, the dust mass $M_\mathrm{dust}$, and the disc fractional width $\frac{dr}{r}$. These parameters can be input to find the disc volume (equation~\ref{eq:vol}), relative velocity (equation~\ref{eq:vel}), the critical impactor sizes (equation~\ref{eq:XC}), and the $\beta$ values (equation~\ref{eq:beta}). Then the P-R drag timescale can be found with equation~\ref{eq:tPR}, and equations~\ref{eq:tcoll} and~\ref{eq:tcollPR} give the collisional timescales in the two regimes. The ratio of these timescales gives $\eta_0$, and $D_\mathrm{pr}$ is the particle size for which $\eta_0 = 1$. Then $D_\mathrm{pr, eff}$ can be calculated to find where the two regimes for the collision timescales apply. The $k$ factor should be applied to $\eta_0$ after finding $D_\mathrm{pr}$. The size distribution within the belt is
\begin{equation}
\label{eq:nD_all}
n(D) = \begin{cases}
KD_\mathrm{pr}^{\alpha_\mathrm{r} - \alpha - 1}D^{1 - \alpha_\mathrm{r}}, & D \leq D_\mathrm{pr} \\
KD^{-\alpha}, & D \geq D_\mathrm{pr},
\end{cases}
\end{equation}
where the normalisation $K$ is given by equation~\ref{eq:KD}. This size distribution can be used to find the optical depth in the belt (equation~\ref{eq:tau0D}), which should be applied at radii $r_0 \rightarrow r_0 + dr$ to take into account the finite extent of the belt. Inwards of the belt, the size distribution can be found using equation~\ref{eq:tauDr}. The size distribution is cut off for particles smaller than $D_\mathrm{bl}$. \par
The numerical model uses the rate at which dust mass is introduced to the system by collisions of larger parent bodies, $\dot{M}_{\mathrm{in}}$, as an input. We assume that the disc is in steady state, such that this mass input is balanced by the loss of the largest particles to collisions or P-R drag. Since the largest particles should dominate the mass distribution given the assumption that $\alpha < 4$, the belt mass can be estimated using
\begin{equation}
\label{eq:Min}
M_\mathrm{belt} \approx \dot{M}_\mathrm{in} \min(t_{\mathrm{coll}}(D_{\mathrm{max}}), t_{\mathrm{PR}}(D_{\mathrm{max}})).
\end{equation} \par
As in the numerical model, the fiducial model has a belt location of $r_0 = 30~$au and a maximum particle inclination of $I_\mathrm{max} = 8.5\degree$. The particles included in the model are in the strength regime, with a maximum size $D_\mathrm{max} = 2~$cm, so the critical dispersal threshold is taken to be constant at $Q_\mathrm{D}^\star = 10^7~\mathrm{erg~g}^{-1}$. The slope of the size distribution of particles has the standard value when $Q_\mathrm{D}^\star$ is independent of particle size, $\alpha = 3.5$. As for the numerical model, a value of $\alpha_\mathrm{r} = 3.5$ is used. The fiducial model considers a Sun-like star with $M_\star = \mathrm{M}_{\sun}$, $L_\star = \mathrm{L}_{\sun}$, and the mass input rate is varied from $\dot{M}_\mathrm{in}~=~10^{-18}~\mathrm{M}_{\earth}~\text{yr}^{-1}$ to $\dot{M}_\mathrm{in} = 10^{-12}~\mathrm{M}_{\earth}~\text{yr}^{-1}$.

\section{Comparison of numerical and analytical models}
\label{sec:comparison}
Comparison of results from the analytical and numerical models was used to find the best fit parameterisation of $k$ and test predictions of parameter space trends. \par 

\begin{figure*}
	\centering
	\includegraphics[width=0.85\textwidth]{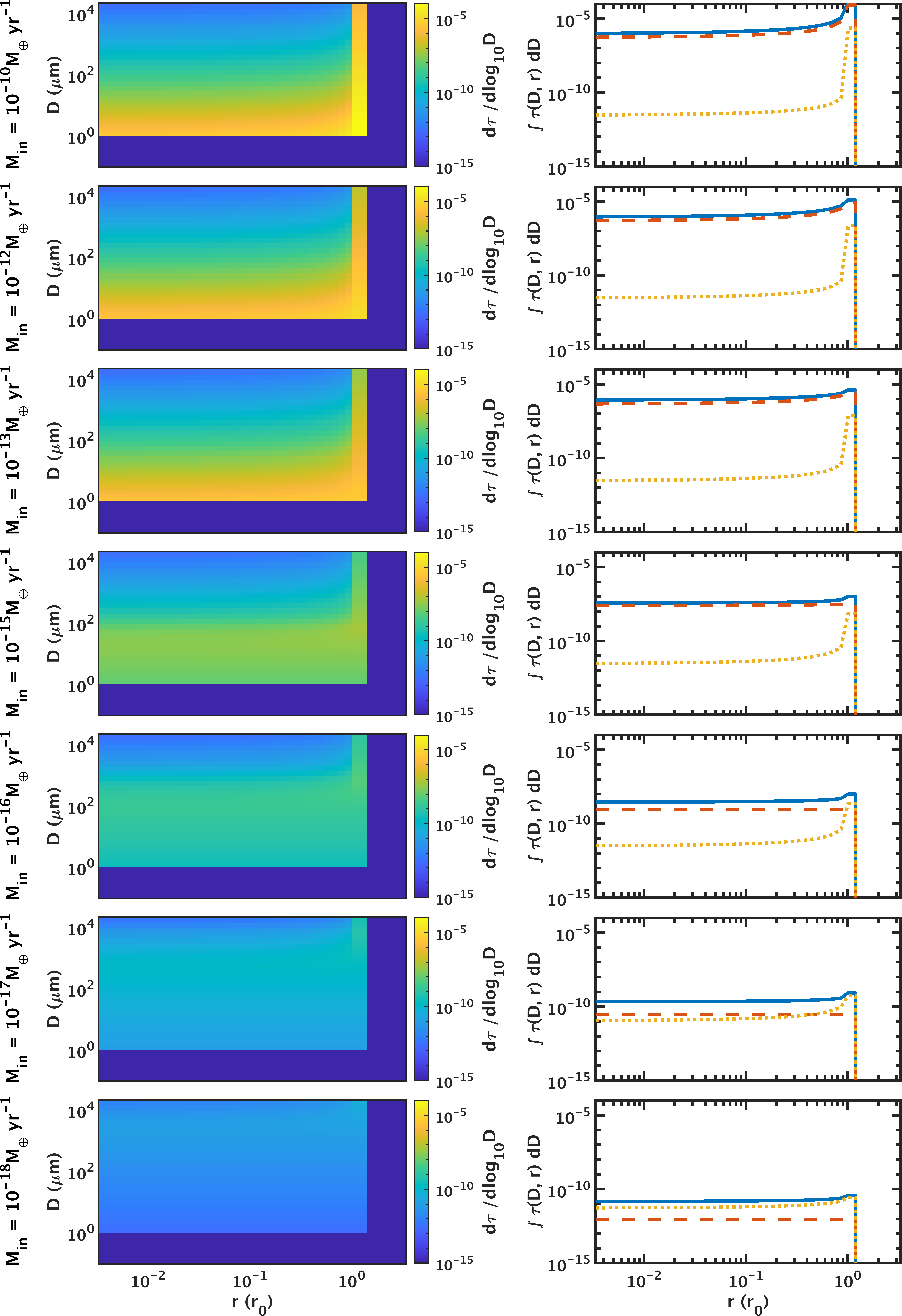}
	\caption{Left: two-dimensional size distribution of particles over size and radial distance for discs with different dust mass input rates and a belt radius $r_0 = 30~$au as predicted by our analytical model, with our best fit of the factor $k$. The colour scale gives the optical depth per unit size decade. Right: radial distribution of integrated optical depth for all particles (blue), barely bound grains (orange, dashed), and cm-sized particles (yellow, dotted). Markers show values at $0.01 r_0$ and $r_0$; ratios of these integrated optical depths are used to characterise the behaviour of the model.}
	\label{fig:an_Min}
\end{figure*}

\subsection{Disc mass} 
\label{subsec:Min}
The left column of Figure~\ref{fig:an_Min} shows the optical depth per unit size decade $d\tau / d\log_{10}D$ as a two-dimensional distribution in grain size $D$ and radius $r$ for the best fit of $k$, to be compared with the numerical model in Figure~\ref{fig:num_Min}. In terms of the analytical model, this is equivalent to 
\begin{equation}
\label{eq:dlogD}
\frac{d\tau}{d\log_{10}D} = D\log{10}\times\tau(D, r).
\end{equation} \par
Regardless of the value of $k$, our model captures the broad trends with disc mass seen in the numerical model, going from particles being collisional to drag-dominated as disc mass is decreased. Low mass discs are in the drag-dominated regime, and the radial profiles are close to flat. The highest mass discs are in the collisional regime, showing a rapid decrease in optical depth inwards of the belt. Intermediate disc masses such as $\dot{M}_\mathrm{in}~=~10^{-15}~\mathrm{M}_{\earth}~\text{yr}^{-1}$ show a transition between being drag-dominated for the smallest particles, while large particles are collisionally depleted. The transition between these two regimes for a given disc mass occurs at a similar particle size in both models. There is an abrupt drop-off in optical depth outside of the belt, as the model has not been constructed to include consideration of the halo, instead focussing on structure within and interior to the belt. There is also a drop-off at particle sizes below $D_\mathrm{bl}$, the lower limit of our size distribution, as grains below this size will be blown out by stellar radiation pressure. The analytics (equation~\ref{eq:Dbl}) give a value of $D_\mathrm{bl}~=~0.98~\mu$m, slightly smaller than $1.5~\mu$m which is found numerically. The main issue is the wavy patterns seen in the most massive discs for the numerical model (Figure~\ref{fig:num_Min}). As noted in Section~\ref{subsec:numres}, to avoid the waves biasing the comparison between models, the behaviour of the models is characterised by integrating the size distribution over particle size. \par
The right hand column of Figure~\ref{fig:an_Min} shows the radial distribution of optical depth integrated over particle size to better compare the behaviour of the two models. Radial profiles are given for the total optical depth, the smallest particles ($D_\mathrm{bl}~<~D~<~20~\mu\text{m}$), and the largest particles ($2~\text{mm}~<~D~<~2~\text{cm}$). Similar trends are seen to the numerical model for the relative contributions of grain sizes. Most of the optical depth for massive discs is in barely bound grains, with the large particles heavily depleted by collisions. As disc mass decreases, the contribution of barely bound grains also decreases. In the lowest mass disc, most of the optical depth comes from the largest particles. The radial profiles of large particles are much flatter in the analytical model, while in the numerical model the profiles are less uniform due to the effect of collisions, and the aforementioned waviness. \par

\begin{figure}
	\centering
	\includegraphics[width=\linewidth]{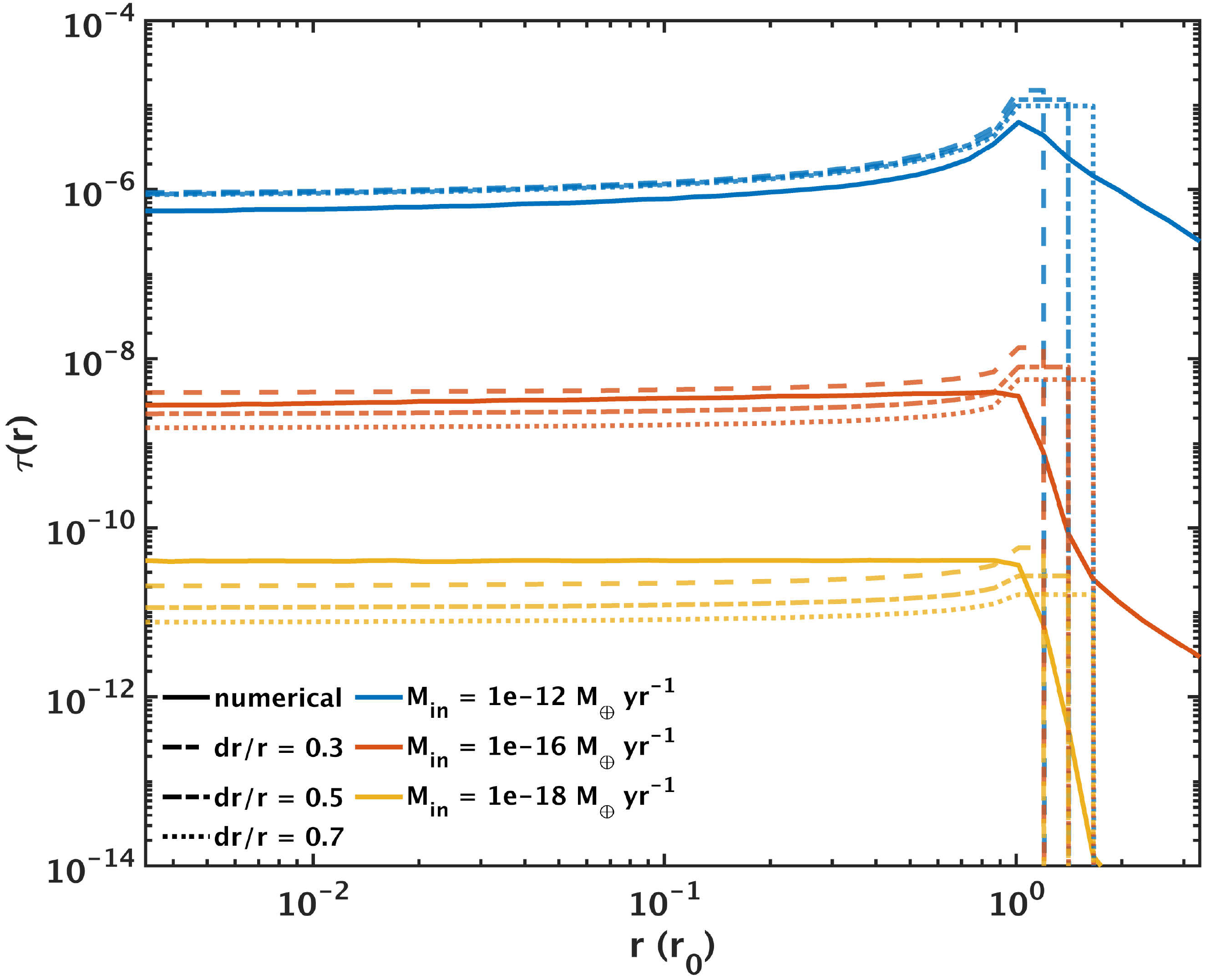}
	\caption{Radial optical depth profiles for discs with mass input rates $\dot{M}_\mathrm{in}$ of $10^{-12}~\mathrm{M}_{\earth}~\text{yr}^{-1}$ (blue), $10^{-16}~\mathrm{M}_{\earth}~\text{yr}^{-1}$ (orange), and $10^{-18}~\mathrm{M}_{\earth}~\text{yr}^{-1}$ (yellow). Profiles from the numerical model are shown with solid lines, while the analytical model is shown for three values of $dr/r$: 0.3 (dashed), 0.5 (dash-dotted), and 0.7 (dotted). The analytical model is shown with the values $k_0 = 10$ and $\gamma = 0.7$.}
	\label{fig:dr_r}
\end{figure}

\begin{figure*}
	\centering
	\begin{subfigure}[b]{0.48\textwidth}
		\centering
		\includegraphics[width=\textwidth]{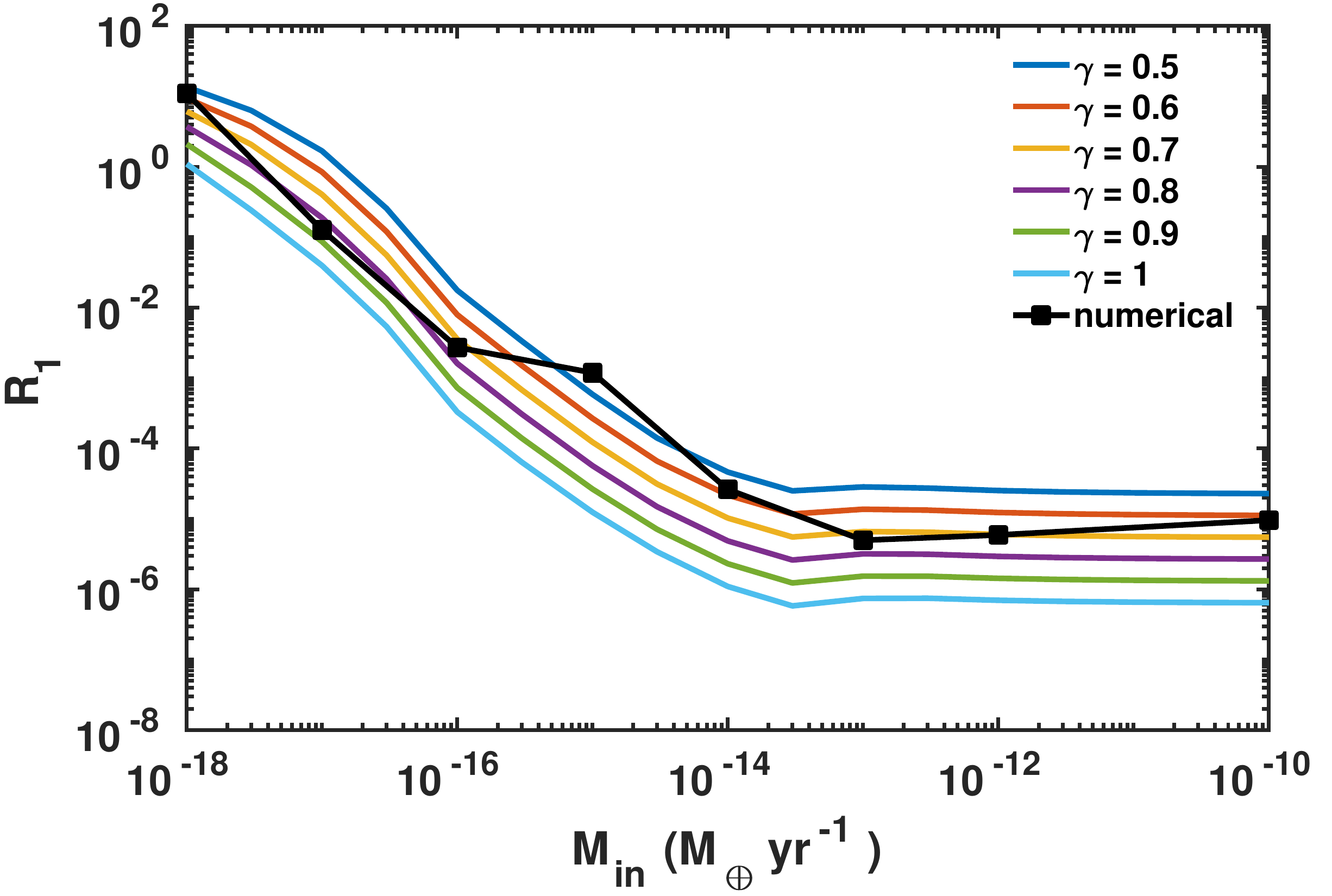}
		\caption{$R_1$ with varying $\gamma$ for fixed $k_0 = 4.2$.}
		\label{fig:R1_gamma}
	\end{subfigure}
	\hfill
	\begin{subfigure}[b]{0.48\textwidth}
		\centering
		\includegraphics[width=\textwidth]{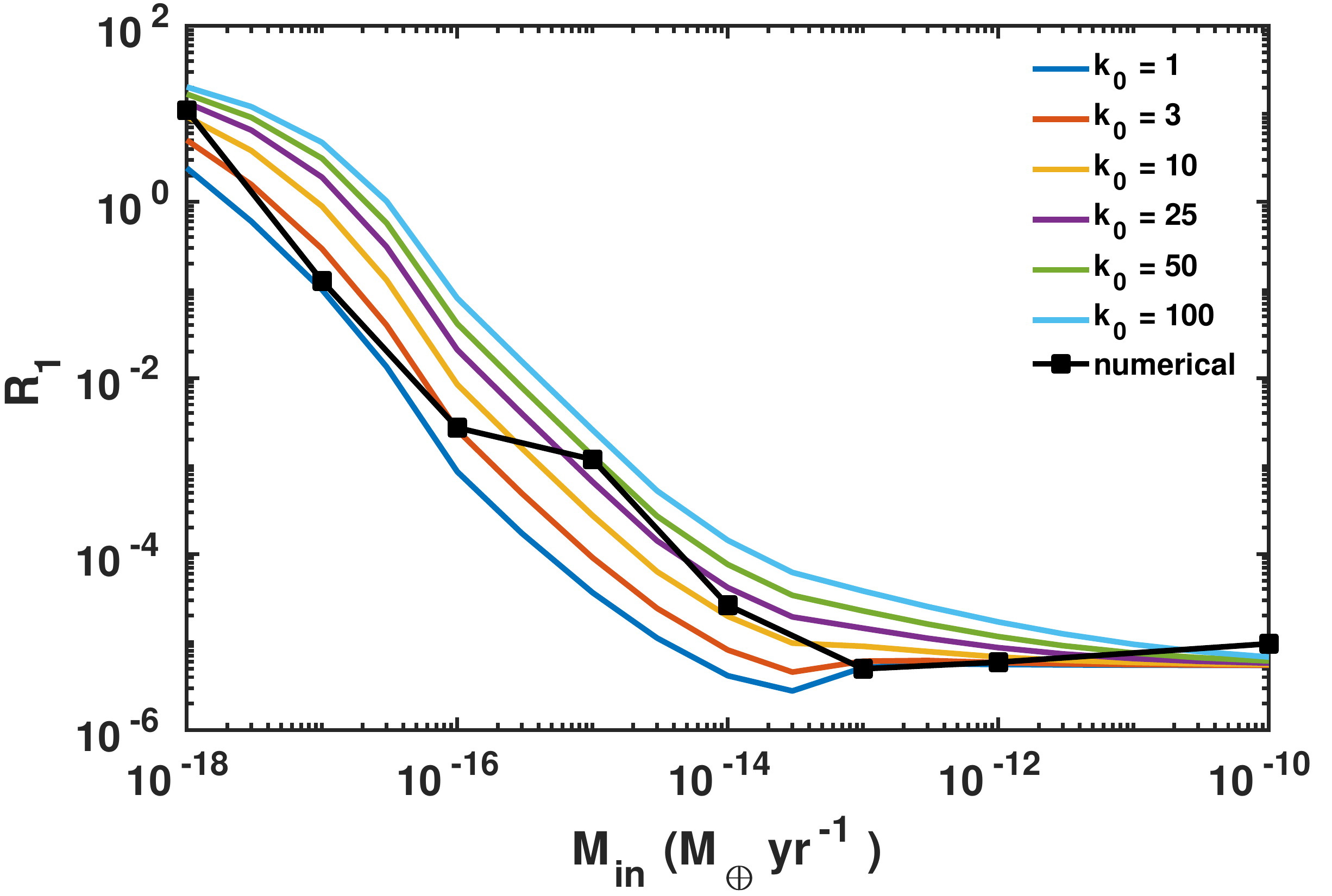}
		\caption{$R_1$ with varying $k_0$ for fixed $\gamma = 0.7$.}
		\label{fig:R1_k0}
	\end{subfigure}
	
	\begin{subfigure}[b]{0.48\textwidth}
		\centering
		\includegraphics[width=\textwidth]{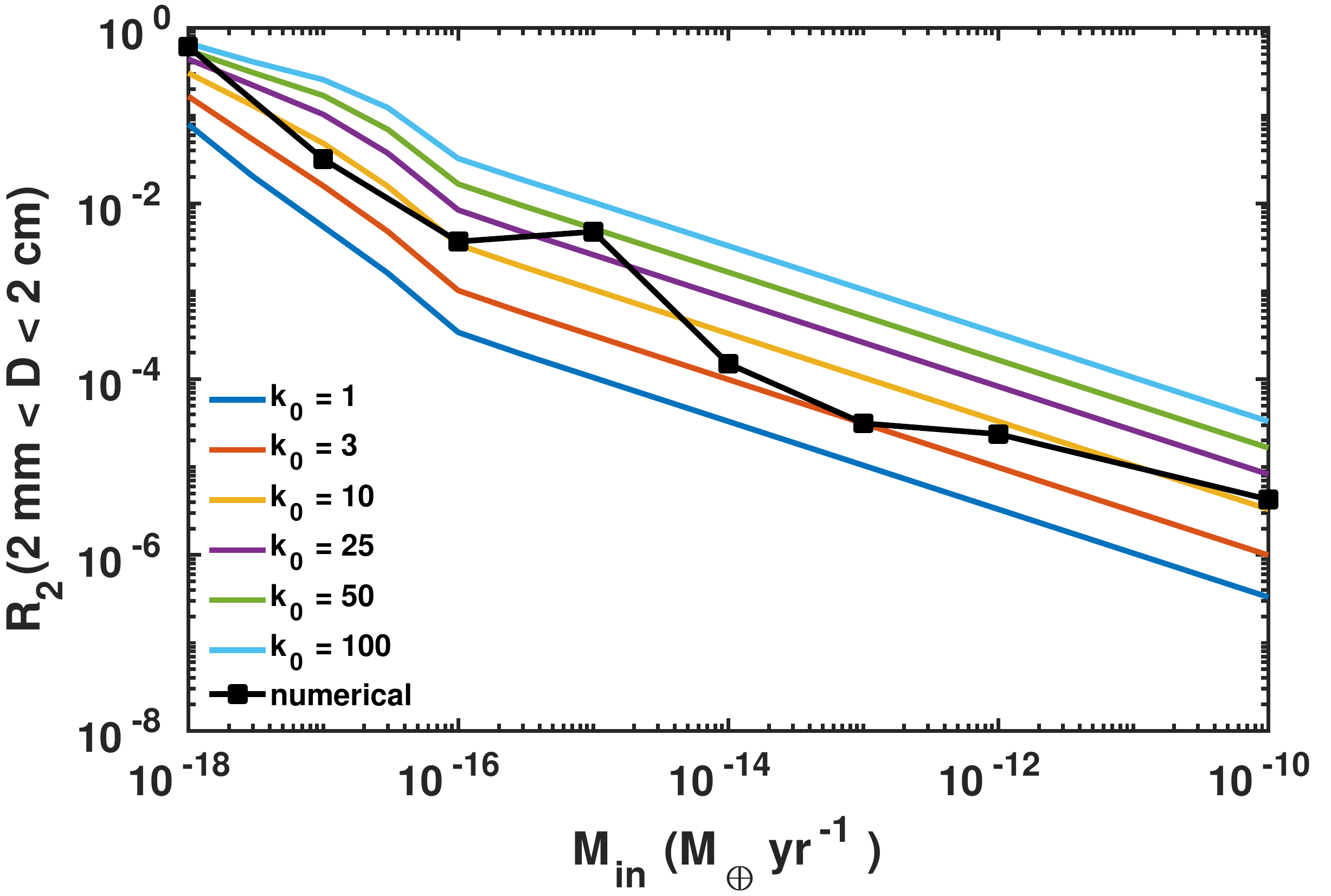}
		\caption{$R_2$ for large particles, $2 \mathrm{mm} < D < 2 \mathrm{cm}$, with $\gamma = 0.7$ and varying $k_0$.}
		\label{fig:R2_cm}
	\end{subfigure}
	\hfill
	\begin{subfigure}[b]{0.48\textwidth}
		\centering
		\includegraphics[width=\textwidth]{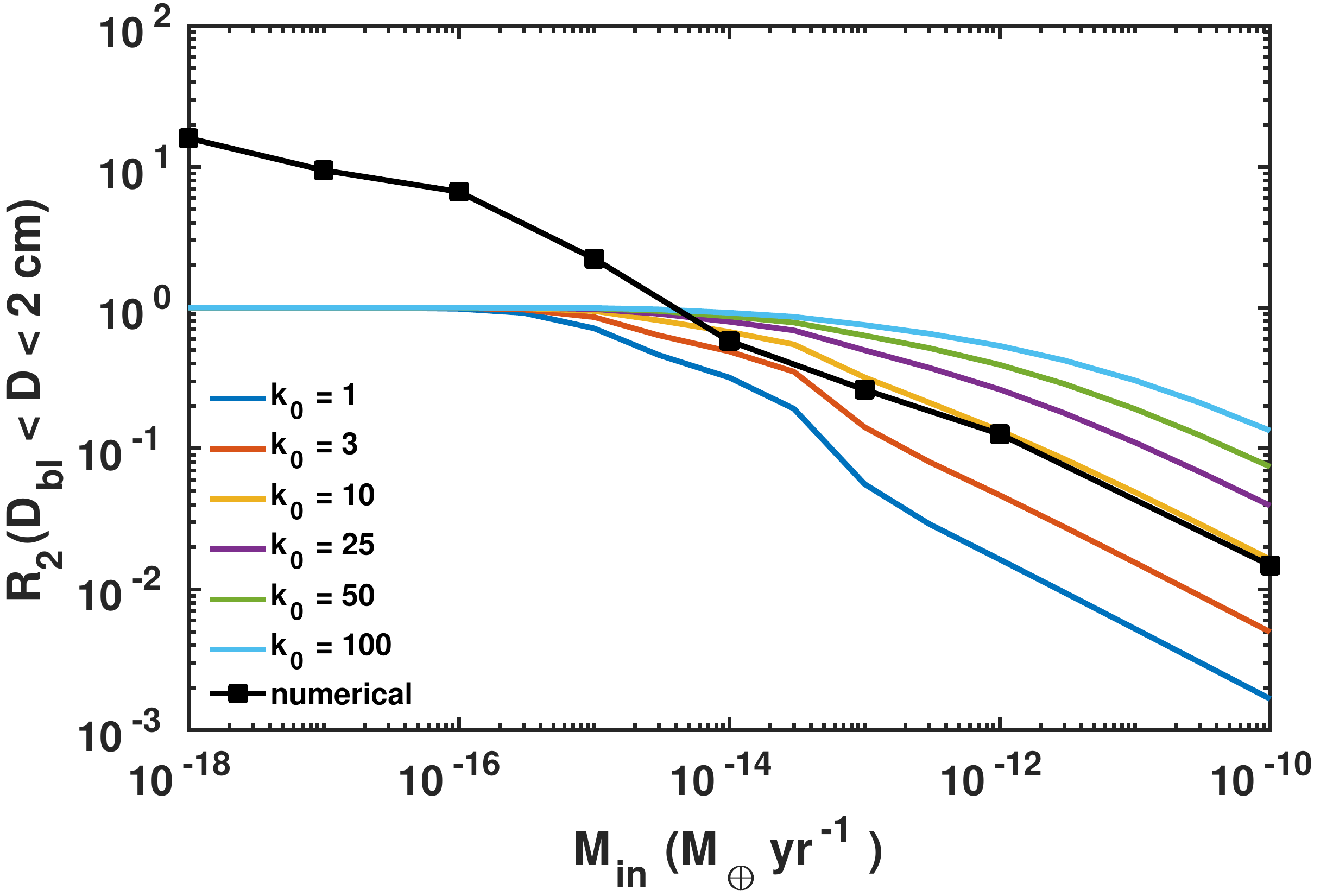}
		\caption{$R_2$ for the smallest particles, $D_\mathrm{bl} < D < 20~\mu\mathrm{m}$, with $\gamma = 0.7$ and varying $k_0$.}
		\label{fig:R2_bl}
	\end{subfigure}
	
	\caption{Ratios used to fit the parameters of the factor $k$ (equation~\ref{eq:kD}), which matches the analytical model to the numerical one. The top panels show $R_1$ (equation~\ref{eq:R1}), the ratio of integrated optical depth in large particles to small particles when close in to the star at $0.01~r_0$, as a function of mass input rate $\dot{M}_\mathrm{in}$, which is related to disc mass. The bottom panels show $R_2$ (equation~\ref{eq:R2cm}), the ratio of integrated optical depth close in to the star to that in the belt as a function of mass input rate $\dot{M}_\mathrm{in}$ for two different particle sizes. In all cases the numerical model (black) is compared with the analytical model for various values of the parameters.}
	\label{fig:ratios}
\end{figure*}

\subsection{Model fitting}
\label{subsec:fitting}
The model has three parameters to fit: the belt width $\frac{dr}{r}$, and the two parameters $k_0$ and $\gamma$ which determine the collisional factor $k$. While $\frac{dr}{r}$ is not a direct input parameter of the numerical model, the belt width can be altered either by changing the eccentricity of the parent bodies in the belt, which mostly affects large grains, or changing the range of initial periastra. This was found to have little effect on the distribution inwards of the belt, only affecting the breadth of the belt. Therefore, the same initial conditions are used throughout for the numerical model. It is assumed in the analytical model that particles are on circular orbits, with no eccentricity inherited from the parent bodies. Varying $\frac{dr}{r}$ in the analytical model affects the breadth of the belt, but also has an effect on the optical depth profile due to the dependence of belt volume and area on its width. The belt width was fitted using the radial optical depth profiles simultaneously along with $k_0$ and $\gamma$, which were fitted using metrics described below. \par 
Figure~\ref{fig:dr_r} shows the effect of varying the belt width on the radial optical depth profile. Smaller belt widths cause the optical depth of the belt to be higher (equation~\ref{eq:tau0D}). Similarly, the collision timescale (equation~\ref{eq:tcoll}) is proportional to the disc volume, which depends on the belt width, such that smaller belt widths have shorter collisional timescales, meaning they have profiles which are more depleted. The best fit of $\frac{dr}{r}$ is different for discs with different masses. For example, for the most collisional discs (e.g. $10^{-12}~\mathrm{M}_{\earth}~\text{yr}^{-1}$), no value is a very good match as the analytical model overestimates the optical depth. For intermediate mass discs (e.g. $10^{-16}~\mathrm{M}_{\earth}~\text{yr}^{-1}$), the best fit is $\frac{dr}{r}=0.5$. For low mass discs (e.g. $10^{-18}~\mathrm{M}_{\earth}~\text{yr}^{-1}$), the optical depth is underestimated, but the best fit would be $\frac{dr}{r}=0.3$. \par
The best fit of the factor $k$ was found by considering the model's behaviour in two dimensions, comparing ratios of optical depths for different particle sizes and locations. The right hand columns of Figures~\ref{fig:num_Min} and \ref{fig:an_Min} show markers for which points are compared in the ratios. The first ratio, 
\begin{equation}
\label{eq:R1}
R_1 = \frac{\int_{2 \mathrm{mm}}^{2 \mathrm{cm}} \tau(0.01 r_0, D)dD}{\int_{D_{\mathrm{bl}}}^{20~\mu\mathrm{m}} \tau(0.01 r_0, D)dD},
\end{equation}
compares the relative contributions of large particles and small particles to the optical depth when close in to the star. In highly collisional discs, $\eta_0 \gg 1$, the optical depth (equation~\ref{eq:tauDr}) close in to the star tends towards a value 
\begin{equation}
\label{eq:anlimits}
\tau(D, 0.01 r_0) \approx \frac{\tau_0(D)}{3.6 \eta_0(D)} \propto \frac{k(D)D^{2 - \alpha}}{D^{4 - \alpha}} = k(D)D^{-2}.
\end{equation}
This means that the ratio of optical depths in two different sizes just depends on the values of $D$ and $k$ as
\begin{equation}
\label{eq:R1_an}
\frac{\tau(D_1, 0.01 r_0)}{\tau(D_2, 0.01 r_0)} = \frac{k(D_1)}{k(D_2)} \left(\frac{D_1}{D_2}\right)^{-2} = \left(\frac{D_1}{D_2}\right)^{-2-\gamma}.
\end{equation}
Therefore at large disc masses, $R_1$ should tend to a constant value depending only on the chosen particles sizes to be compared and $\gamma$. \par

Figure~\ref{fig:R1_gamma} shows $R_1$ plotted as a function of mass input rate, which determines disc mass. As predicted, a plateau is seen at the largest disc masses, where the disc is collisional. When $k$ is a constant with particle size, this plateau cannot be fitted with the analytical model. However, when $k$ becomes a function of particle size, the value of the analytical model can be shifted to match the numerical one. This explains the choice of the prescription of equation~\ref{eq:kD} with $k$ a function of particle size, for which $k_0$ and $\gamma$ are parameters to be fitted to the numerical model. Figure~\ref{fig:R1_gamma} shows how varying $\gamma$ shifts the value of the plateau, with a best fit expected to be close to $\gamma = 0.7$.\par
The effect on $R_1$ of varying $k_0$ with fixed $\gamma = 0.7$ is shown in Figure~\ref{fig:R1_k0}. A broad range of values are feasible for $k_0$, but the most consistent value is close to $k_0 = 3$ or 10. The other ratio used to fit $k$ is the ratio of optical depth close in to the star to that in the belt, given by 
\begin{equation}
\label{eq:R2cm}
R_2 = \frac{\int_{\mathrm{D_{lower}}}^{\mathrm{D_{upper}}} \tau(0.01 r_0,D)dD}{\int_{\mathrm{D_{lower}}}^{\mathrm{D_{upper}}} \tau(r_0, D)dD}.
\end{equation}
This varies for different ranges of particle size $\mathrm{D_{lower}}$ to $ \mathrm{D_{upper}}$. When considering the largest particles, as in Figure~\ref{fig:R2_cm}, which shows $R_2$ for $2~\text{mm}~<~D~<~2~\text{cm}$, there is some fluctuation, but broadly the best fit is expected to be close to $k_0 = 10$. However, when considering the smallest particles as in Figure~\ref{fig:R2_bl} ($D_{\mathrm{bl}}~<~D~<~20~\mu\text{m}$), our analytical model cannot fit to the drag-dominated regime (low disc masses) for any values of the parameters. The integrated optical depths in Figure~\ref{fig:num_Min} show that in low mass discs the optical depth of barely bound grains in the numerical model decreases with radius, causing the ratio plotted in Figure~\ref{fig:R2_bl} to exceed 1. By construction, the radial profile of a single particle size in the analytical model must either be flat, or increase with radius, meaning that this ratio cannot exceed 1. In the drag-dominated regime, a flat radial profile is predicted, giving a ratio of 1. \par 
One reason for this discrepancy is the assumption in the analytical model that particles of different sizes evolve independently inwards of the belt. However, the breakup of large particles will act as a source of smaller grains, such that optical depth of small particles actually increases in towards the star, rather than being flat, for the numerical model. \par
Fitting of the three parameters was done by minimising $\chi^2$ for the logarithms of the ratios $R_1$ (equation~\ref{eq:R1}) and $R_2$ for large particles (equation~\ref{eq:R2cm}), along with the radial profile of optical depth $\tau(r)$. For example, the $\chi^2$ for $R_1$ would be given by
\begin{equation}
\label{eq:chisq}
\chi^2(R_1) = \sum_{\dot{M}_{in}} \left(\log R_{1\mathrm{, analytical}} - \log R_{1\mathrm{, numerical}} \right)^2.
\end{equation}
The ratios $R_1$ and $R_2$ have a single point for each mass input rate, while the radial optical depth profile is sampled at $n_{\mathrm{r}}$ different points for each mass input rate. Therefore, the $\chi^2$ for $\tau(r)$ is weighted by $1/n_{\mathrm{r}}$ such that each of these factors contributes equally to the fit. While a range of values for each parameter gives reasonable results, combined minimisation gives the best fit values $\frac{dr}{r} = 0.4$, $\gamma = 0.7$, and $k_0 = 4.2$. \par

\subsection{Thermal emission}
\label{subsec:SEDs}
Thermal emission from dust grains can be seen at infrared wavelengths in the spectral energy distributions (SEDs) of stars as excess flux above the stellar photosphere. This can be described as a fractional excess, $R_\nu~=~F_{\nu~\mathrm{disc}}/F_{\nu\star}$. Once the size and spatial distribution of dust in a disc has been determined using a model, infrared excesses of the disc can be predicted by applying realistic grain properties (Section~\ref{subsec:optprops}) for the absorption efficiencies $Q_\mathrm{abs}(\lambda, D)$ and temperatures $T(D, r)$ of grains. The disc flux in Jy at a given wavelength can be found from the model by summing the emission from different radii and particle sizes as
\begin{equation}
\label{eq:Fnu}
F_\nu = 2.35\times10^{-11} d^{-2} \iint Q_\mathrm{abs}(\lambda, D) B_\nu[T(D, r)] 2\pi r \tau(D, r) dDdr,
\end{equation}
where d is the distance from the star in pc, radius $r$ is in au, and $B_\nu$ is the spectral radiance in Jy~$\text{sr}^{-1}$ \citep{Wyatt99}. \par
\begin{figure}
	\centering
	\includegraphics[width=\linewidth]{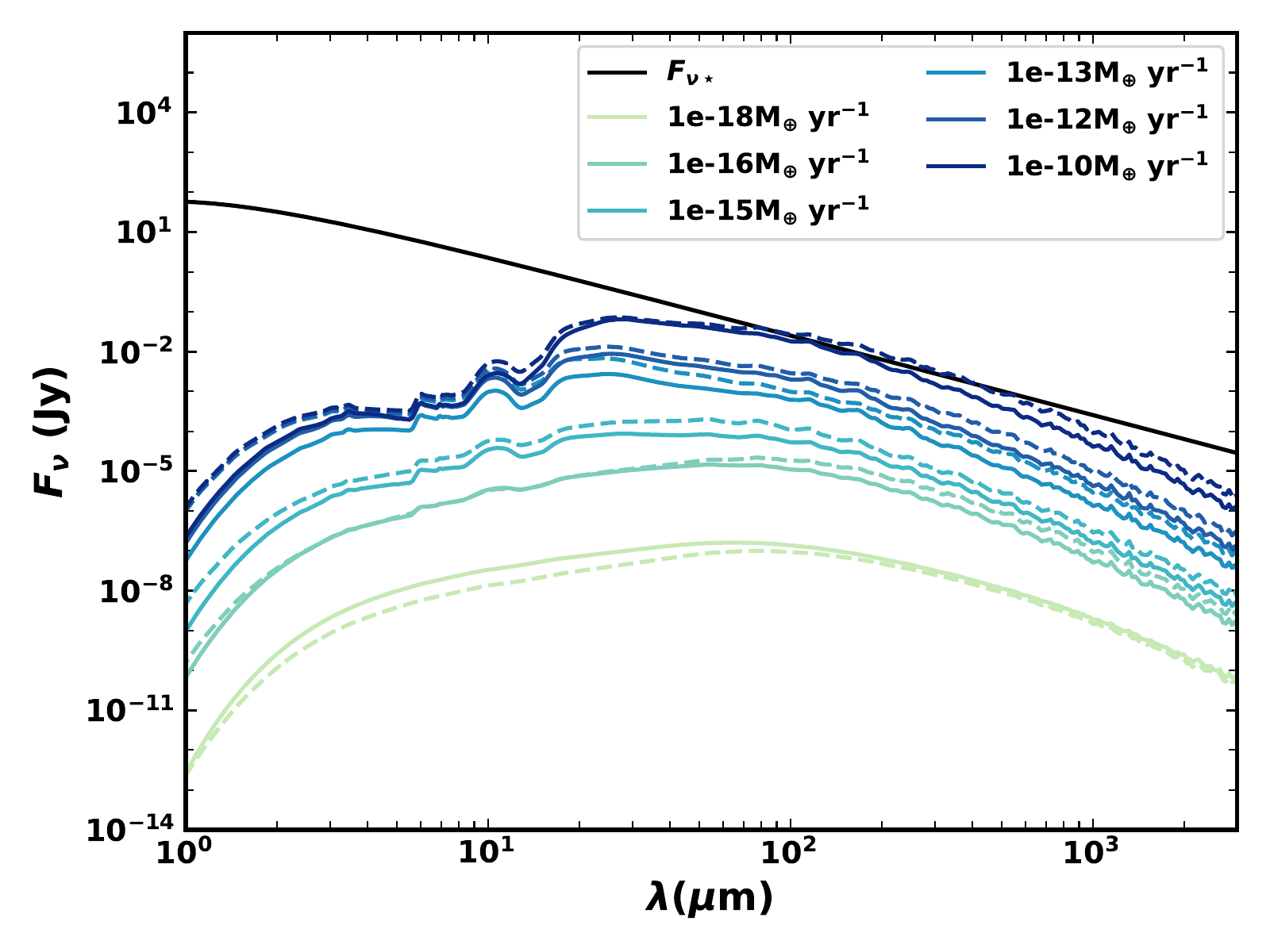}
	\caption{Comparison of SEDs based on the numerical (solid lines) and analytical (dashed lines) models for the discs considered in Sections~\ref{subsec:numres} and~\ref{subsec:Min}. Realistic optical properties were used, assuming asteroidal grains.}
	\label{fig:SED_comp}
\end{figure}

As an example, Figure~\ref{fig:SED_comp} shows a comparison of the SEDs resulting from the numerical and analytical models for the discs considered in Sections~\ref{subsec:numres} and~\ref{subsec:Min}. An inner cut-off at a radius of 0.1~au is used, such that the SEDs will not be fully accurate at wavelengths below $\sim6~\mu$m. Only thermal emission has been included, while scattered light will also contribute below $\sim5~\mu$m. Realistic optical properties for asteroidal grains are used for both models, so both models include similar features in the SED. A similar pattern is seen as when fitting the optical depth in Figure~\ref{fig:dr_r}. The analytical model fits very well at intermediate disc masses, but slightly overestimates the optical depth for the most massive discs, and slightly underestimates the optical depth of the least massive discs. For the disc with a mass input rate of $10^{-10}~\mathrm{M}_{\earth}~\text{yr}^{-1}$, the fractional 11~$\mu$m excess is overestimated by a factor 2.0, while for the disc with a mass input rate of $10^{-18}~\mathrm{M}_{\earth}~\text{yr}^{-1}$, the 11~$\mu$m flux is underestimated by a factor of 2.5. This shows that while there are differences between the two models, the 11~$\mu$m excess should not differ by more than a factor $\sim$3 for discs of masses similar to those considered here.

\section{Parameter space}
\label{sec:parspace}
Once the model has been fitted for the fiducial planetesimal belt properties, its ability to predict parameter space trends can be tested by further comparison with the numerical model.The dependence of size distributions on individual input parameters is investigated in the following subsections.

\begin{table}
	\centering
	\caption{Stellar types used in comparison of the models (Figure~\ref{fig:vary_Ms}).}
	\label{tab:stars}
	\begin{tabular}{|c|c|c|c|}
		\hline
		Stellar type & Mass & Luminosity & Temperature \\
		& $\mathrm{M}_{\sun}$ & $\mathrm{L}_{\sun}$ & K\\
		\hline
		M0 & 0.5 & 0.074 & 3822 \\
		K2 & 0.75 & 0.31 & 4958 \\
		G2 & 1.0 & 1.1 & 5868 \\
		F7 & 1.25 & 2.4 & 6264 \\
		A5 & 2.0 & 14.0 & 8204 \\
		A0 & 3.0 & 61 & 9722 \\
		\hline
	\end{tabular}
\end{table}

\begin{figure*}
	\centering
	\includegraphics[width=0.85\textwidth]{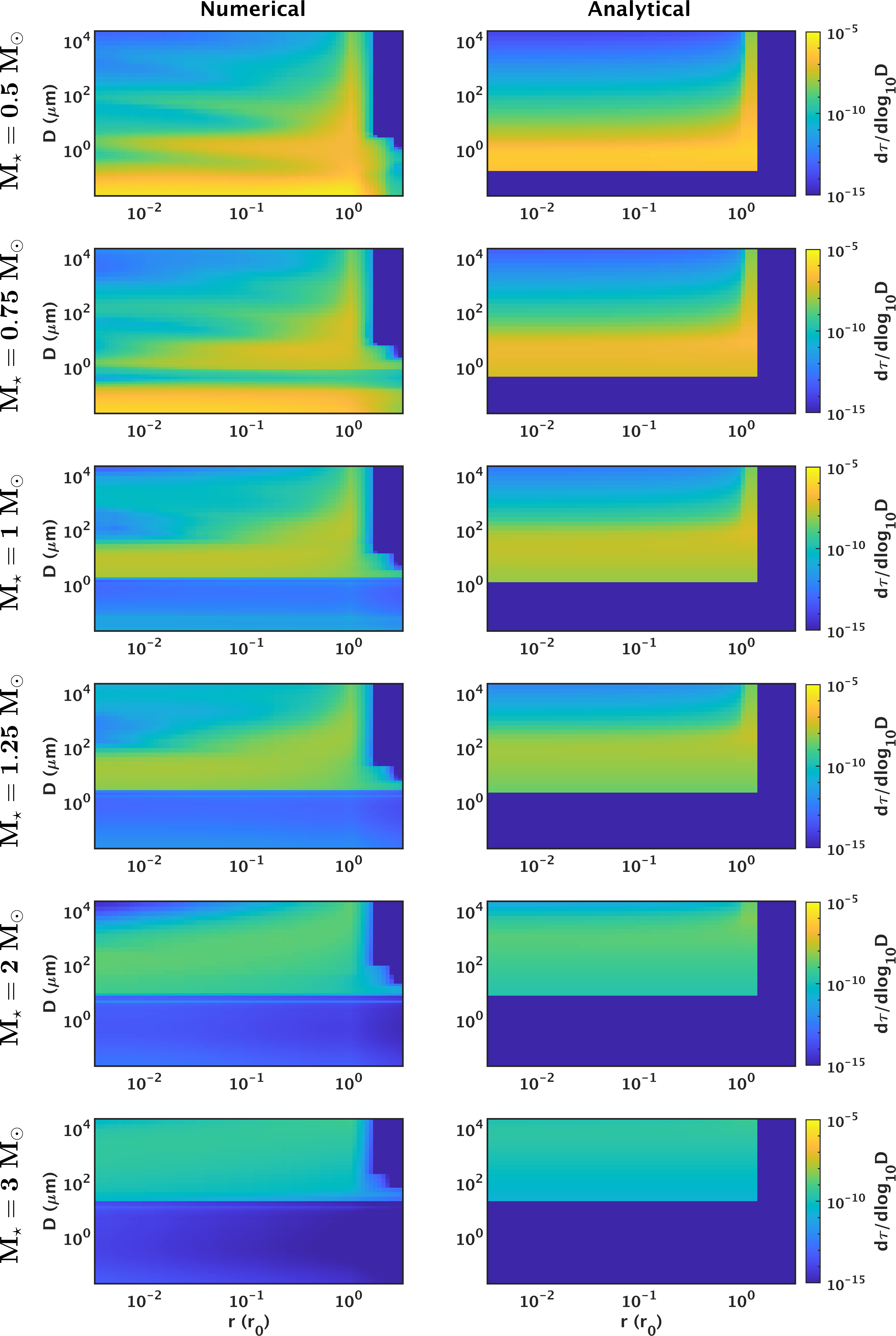}
	\caption{Two-dimensional size distribution of particles over size and radial distance for discs with different stellar types, as produced by the numerical model of \citetalias{vLieshout14} (left) and our analytical model (right). The colour scale gives the optical depth per unit size decade. The mass input rate is fixed throughout at $\dot{M}_\mathrm{in} = 10^{-15}~\mathrm{M}_{\earth}~\text{yr}^{-1}$, and the planetesimal belt has a radius of $r_0 = 30~$au.}
	\label{fig:vary_Ms}
\end{figure*}

\begin{figure}
	\centering
	\includegraphics[width=\linewidth]{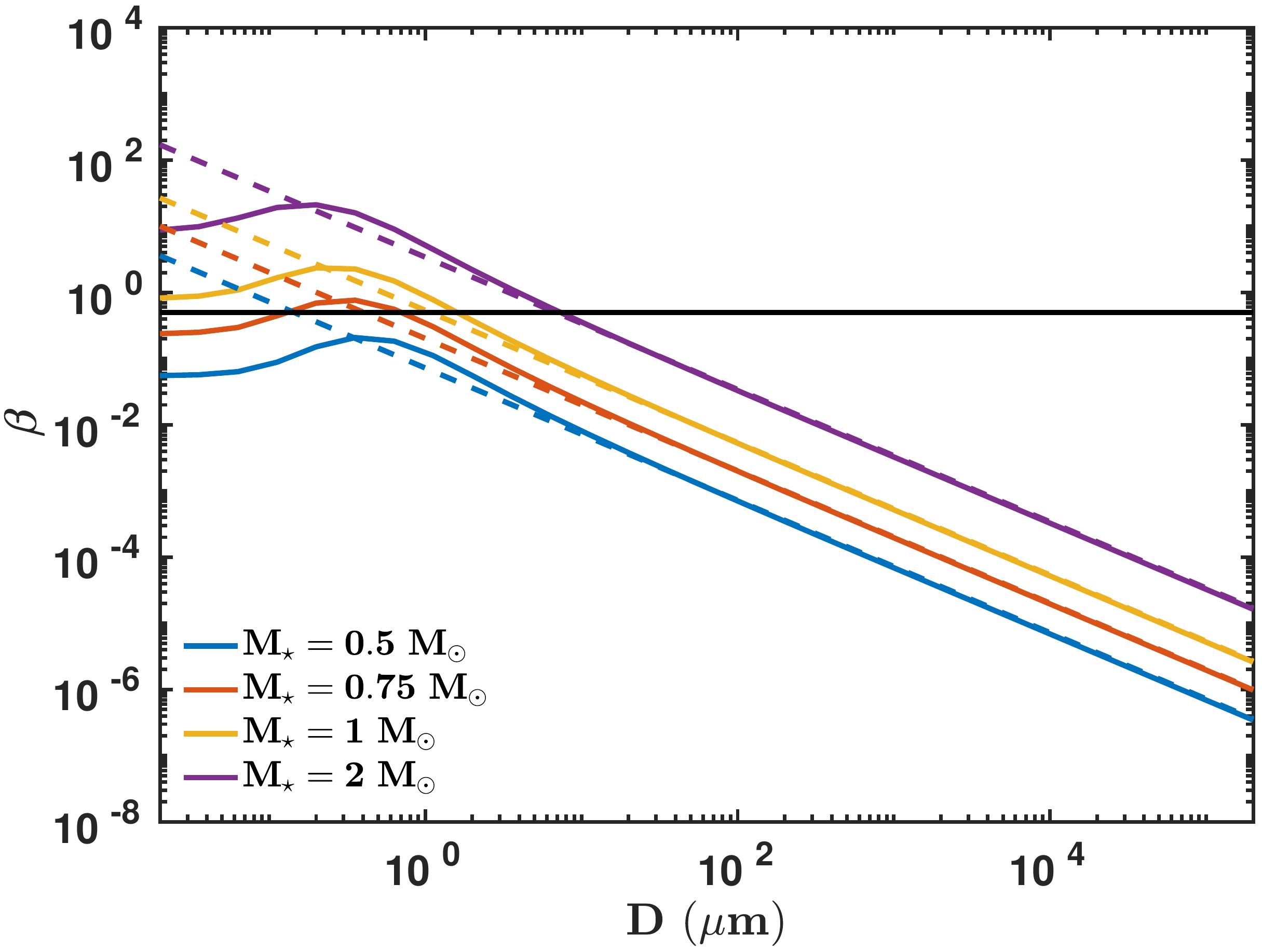}
	\caption{$\beta$ as a function of particle diameter, $D$, for grains around stars of different masses. Solid lines show numerical values, while dashed lines show the analytical predictions from equation~\ref{eq:beta}, assuming $Q_\mathrm{pr} = 1$. The black line shows $\beta = 0.5$, the limit above which particles should be blown out of the system on unbound orbits by stellar radiation pressure.}
	\label{fig:beta}
\end{figure}

\begin{figure}
	\centering
	\includegraphics[width=\linewidth]{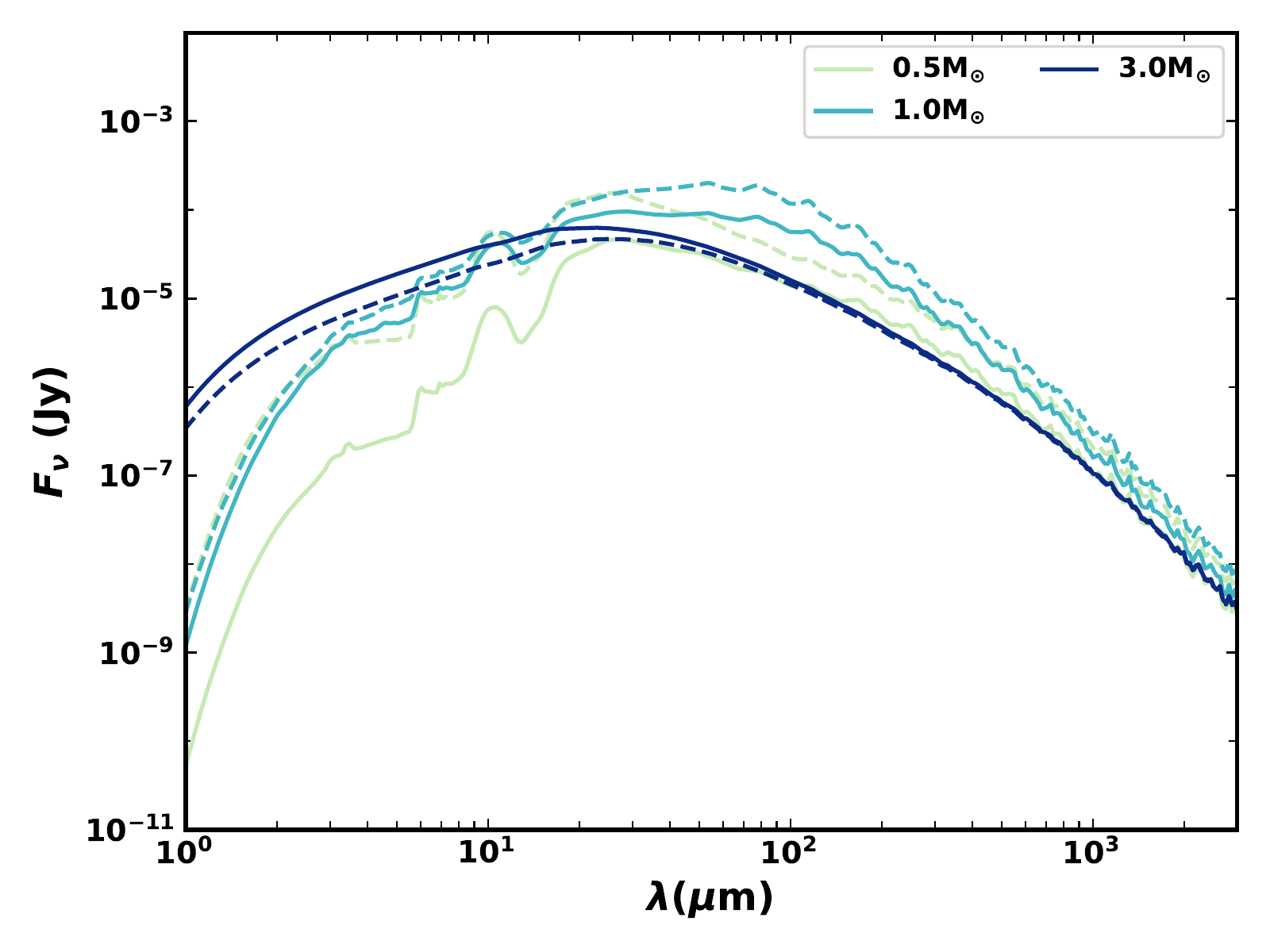}
	\caption{Comparison of SEDs from the numerical (solid lines) and analytical (dashed lines) models for discs with a mass input rate of $10^{-15}~\mathrm{M}_{\earth}~\text{yr}^{-1}$ and different stellar masses, assuming asteroidal grains.}
	\label{fig:SED_Ms}
\end{figure}

\subsection{Stellar type}
\label{subsec:star}
To ascertain the effect of stellar mass and luminosity, the model was applied to stars of different types, whose parameters are given in Table~\ref{tab:stars}. The optical properties for dust around different stars were calculated for the numerical model using the method described in Section~\ref{subsec:optprops}. Equation~\ref{eq:beta} was used for the analytical model, assuming $Q_\mathrm{pr} = 1$. A comparison of the size distributions from the numerical and analytical models is given in Figure~\ref{fig:vary_Ms} with fixed mass input rate $\dot{M}_\mathrm{in} = 10^{-15} \mathrm{M}_{\earth}~\text{yr}^{-1}$ and a belt of radius $r_0 = 30~$au. As stellar mass is increased, the blowout size increases due to higher stellar luminosity and therefore stronger radiation pressure. Both models broadly follow the same trends without having to change the factor $k$, although for the lowest mass stars sub-micron grains are present in the numerical results, whereas such grains are assumed to be removed by radiation pressure in the analytical model. \par
The $\beta$ profiles for grains around stars of different masses are plotted in Figure~\ref{fig:beta}, calculated numerically using the approach in Section~\ref{subsec:optprops}, and analytically using equation~\ref{eq:beta}. The standard shape of the profile of $\beta$ is that it is inversely proportional to particle size, with a turnover at the smallest particle sizes. Since $\beta \propto L_\star/M_\star$, low mass stars have lower values of $\beta$, such that either when the profile turns over it drops back below $0.5$ (as seen for $0.75\mathrm{M}_{\sun}$), or $\beta$ never actually exceeds $0.5$ (as seen for $0.5\mathrm{M}_{\sun}$). Sub-micron grains are therefore present around low mass stars in contradiction to the simple analytical prescription suggesting that they should be blown out on hyperbolic orbits. A limitation of the analytical model is that it cannot faithfully reproduce the distribution of small grains for late-type stars, for which radiation pressure is weaker, such that it is not always possible to blow particles out \citep[e.g.][]{Sheret04}. Furthermore, drag forces around late-type stars are significantly enhanced by stellar winds \citep{Plavchan05}, which have not been considered. It may be possible to incorporate the effects of stellar wind drag by modification of $\beta$ if the magnitude of the stellar wind is known, as described in \citetalias{vLieshout14}, but that is beyond the scope of this work. \par
The effect of stellar type on the SED is shown in Figure~\ref{fig:SED_Ms} for discs with a mass input rate of $10^{-15}~\mathrm{M}_{\earth}~\text{yr}^{-1}$. Overall the shapes of the SEDs are similar between the two models, with slight differences in the magnitude of the flux. For an A star (3~$\mathrm{M}_{\sun}$), the 11~$\mu$m excess is underestimated by a factor of 1.6, while the Sun-like star (1~$\mathrm{M}_{\sun}$) is overestimated by a factor 1.3. For the M dwarf, the analytical model overestimated the 11~$\mu$m flux by a factor of 7. Therefore the model is most applicable to Sun-like and A stars. As concluded previously from the optical depth, there is a much poorer fit for M stars due to the discrepancy with $\beta$. However, how well the model fits will also vary with the mass of the disc. \par

\subsection{Belt radius}
\label{subsec:r0}
\begin{figure}
	\centering
	\includegraphics[width=\linewidth]{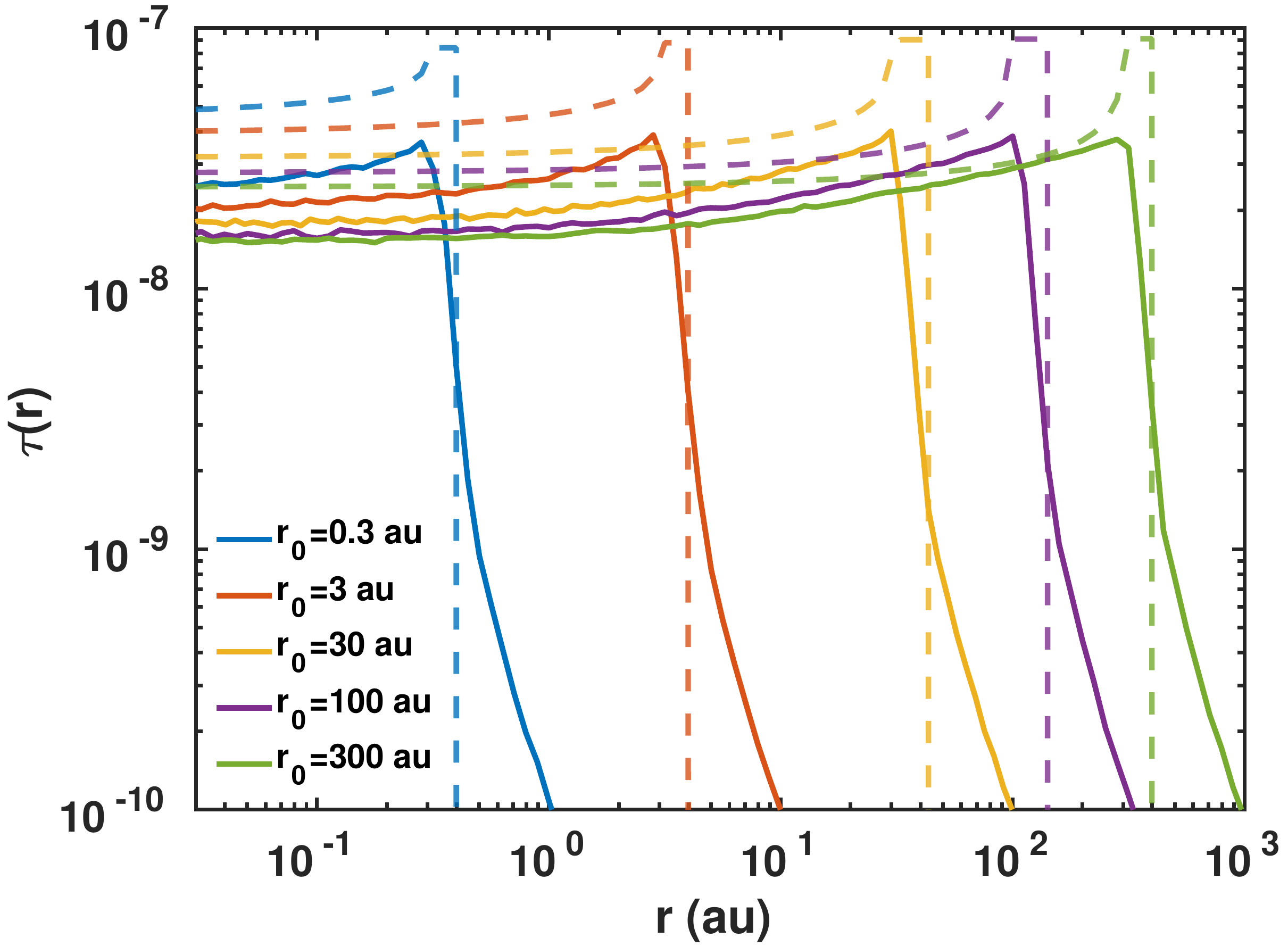}
	\caption{Total geometrical optical depth as a function of radius for discs with parent belts at different radii and a mass input rate of $\dot{M}_{\mathrm{in}}~=~10^{-15}~\mathrm{M}_{\earth}~\text{yr}^{-1}$. Results from the numerical model are solid lines, while the dashed lines show the analytical model.}
	\label{fig:tau_r0}
\end{figure}

Another parameter of the model which can be varied is the distance of the planetesimal belt from the star, $r_0$. Simulations were run with belt radii in the range $0.3-300~$au. Figure~\ref{fig:tau_r0} shows the radial profile of optical depth as $r_0$ is varied at constant mass input rate. As was seen for the fiducial model in Figure~\ref{fig:an_Min}, the radial profiles predicted analytically are quite flat, with either a sharp drop inwards of the belt before becoming flat for collisional particles, or a completely flat profile for drag-dominated particles. The numerical radial profiles are moderately flat, but less so than the analytical ones. In the analytical model, the optical depth in the belt has a weak dependence on belt radius for fixed mass input rate, so varying the radius over a few orders of magnitude causes little variation. The numerical results also show similar optical depth levels in the belt regardless of $r_0$. As the belt location moves outwards there is a slight decrease in the level the optical depth flattens out to in the inner regions of the disc. Comparison of the radial profiles shows that both models follow similar trends with radius, but the analytical profiles overestimate the optical depth by a factor of $\sim 2$. However, this offset does not significantly affect the predicted size distribution, and is expected due to the compromise needed when choosing $\frac{dr}{r}$ to fit the model over a large range of $\dot{M}_\mathrm{in}$ (see Figure~\ref{fig:dr_r}). \par

\subsection{Dispersal threshold of particles}
\label{subsec:QD}
So far the critical dispersal energy, $Q_\mathrm{D}^\star$, has been taken to be independent of particle size. However, this is not a realistic prescription, and many attempts have been made to characterise its dependence on particle size. For example, \citet{Benz99} used SPH to simulate collisions, while \citet{Housen99} used laboratory experiments to measure the outcomes of collisions between small particles. \citet{Durda97} and \citet{Durda98} ran numerical collisional evolution models, and fit these to observations of the main belt asteroid size distribution to find the best fit of $Q_\mathrm{D}^\star$. In general, it is assumed that $Q_\mathrm{D}^\star$ can be approximated by a sum of two power laws, representing the strength and gravity regimes. All particles considered in this paper are small enough that material strength determines their critical dispersal threshold, which can be described by a single power law, $Q_\mathrm{D}^\star = Q_\mathrm{a}D^{-a}$. \citet{Holsapple94} found a dependence in the strength regime of $Q_\mathrm{D}^\star \propto D^{-0.33}$, while  \citet{Housen90} found $Q_\mathrm{D}^\star \propto D^{-0.24}$. \citet{Benz99} considered both basalt and ice, and found that depending on the material and impact velocity, the dependence varied between $a = 0.36 - 0.45$. Here we choose to use the same values as \citet{Lohne08}, which in our notation gives $Q_\mathrm{a} = 2.45 \times 10^7 \text{erg g}^{-1}$ and $a = 0.3$, with $D$ given in cm. \par
\begin{figure}
	\centering
	\includegraphics[width=\linewidth]{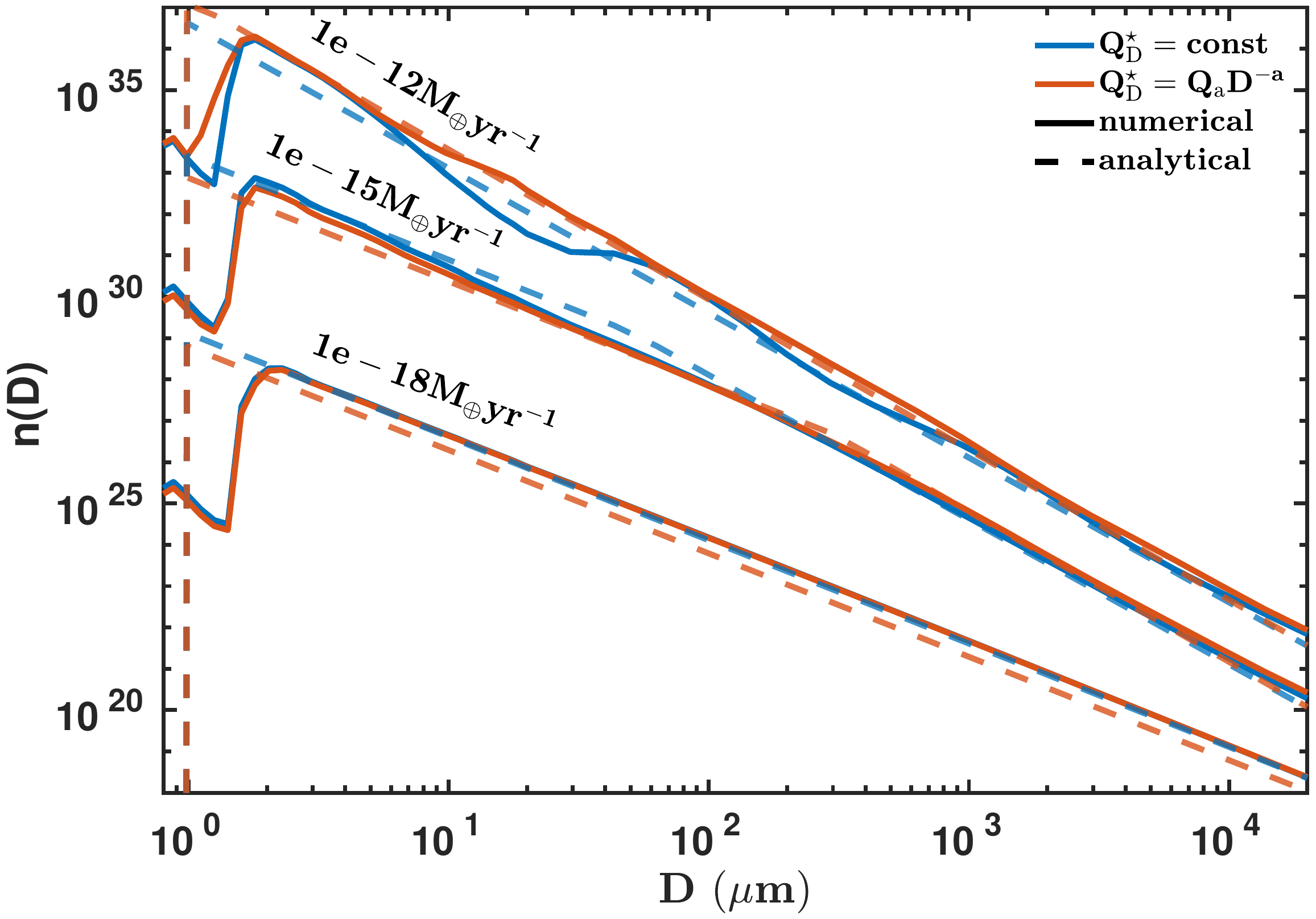}
	
	\caption{The size distribution $n(D)$ of particles within the belt for discs of different mass input rates, both with $Q_\mathrm{D}^\star = $ constant, and $Q_\mathrm{D}^\star = Q_\mathrm{a}D^{-a}$. Results from the numerical model are shown with solid lines, and the analytical predictions are shown with dashed lines.}
	\label{fig:QD}
\end{figure}

 
As shown in Figure~\ref{fig:QD}, using a realistic prescription for $Q_\mathrm{D}^\star$ only had a minor effect on the size distribution of particles in the belt. For the numerical model, particles are assumed to be within the parent belt if they are between radii $30 \leq r_0 \leq 45~$au. The most significant effect is that for collisional discs, such as $\dot{M}_\mathrm{in}~=~10^{-12}~\mathrm{M}_{\earth}~\text{yr}^{-1}$, using a power law for $Q_\mathrm{D}^\star$ dampens the collisional waves which are due to truncation of the distribution at the blowout size. For the chosen values, $Q_\mathrm{D}^\star$ also has a greater value when a power law is used, so this may be an effect of using a greater $Q_\mathrm{D}^\star$ value. The slope of the size distribution in the collisional regime should also be affected. \citet{Wyatt11} derived that the slope of the steady state size distribution depends on the power law of $Q_\mathrm{D}^\star$ as in equation~\ref{eq:alphaQD}, which gives a slope of -3.63 for the chosen prescription of $Q_\mathrm{D}^\star$. Comparison of the distributions for $\dot{M}_\mathrm{in}~=~10^{-12}~\mathrm{M}_{\earth}~\text{yr}^{-1}$ shows that with a power law, the slope of the distribution becomes steeper, going from -3.52 to -3.57. The increased $Q_\mathrm{D}^\star$ for the realistic prescription also causes the collisional waves in size distributions to move closer together, due to particles being destroyed by other particles which are a larger fraction of their own size (as shown by equation~\ref{eq:XC}). This change in spacing of the waves is most evident in the inner regions of the disc, where the waves are more significant. \par 
The discs with $\dot{M}_\mathrm{in}~=~10^{-15}~\mathrm{M}_{\earth}~\text{yr}^{-1}$ show a transition between regimes, with large particles being collisional, while smaller particles are drag-dominated, with a shallower slope. The analytical model predicts that the location of the turnover, $D_\mathrm{pr}$, should be smaller with constant $Q_\mathrm{D}^\star$ due to its lower value. The numerical model similarly has a change in slope of the size distribution at a larger particle size for the realistic $Q_\mathrm{D}^\star$. Low mass discs, such as $\dot{M}_\mathrm{in}~=~10^{-18}~\mathrm{M}_{\earth}~\text{yr}^{-1}$, are drag-dominated, such that the slope of their size distribution depends only on the redistribution function. Therefore, no difference is seen in the size distribution when $Q_\mathrm{D}^\star$ is changed, with a slope of -2.5 in both cases agreeing with the analytical prediction of $1 - \alpha_\mathrm{r}$. A minor difference is seen in the analytic model due to the different values of $D_\mathrm{pr}$ obtained with the different $Q_\mathrm{D}^\star$ values. The normalisation of the size distribution of the drag-dominated regime, $K_\mathrm{pr}$, depends on $D_\mathrm{pr}$ as in equation~\ref{eq:nPR}. \par
Collision timescales increase with $Q_\mathrm{D}^\star$ (equation~\ref{eq:tcoll}), so the analytical model predicts that discs will be less collisional with the realistic prescription. This will cause the radial optical depth profiles to become flatter. However, very minimal differences are seen in the radial profiles of the numerical model, as they are dominated by barely bound grains, which are mostly flat. Thus these profiles are not shown. The grains which are most affected by $Q_\mathrm{D}^\star$ are cm-sized, but these grains contribute less to the overall optical depth profile of the disc. \par

\subsection{Redistribution function}
\label{subsec:alpha_r}
\begin{figure}
	\centering
	\includegraphics[width=\linewidth]{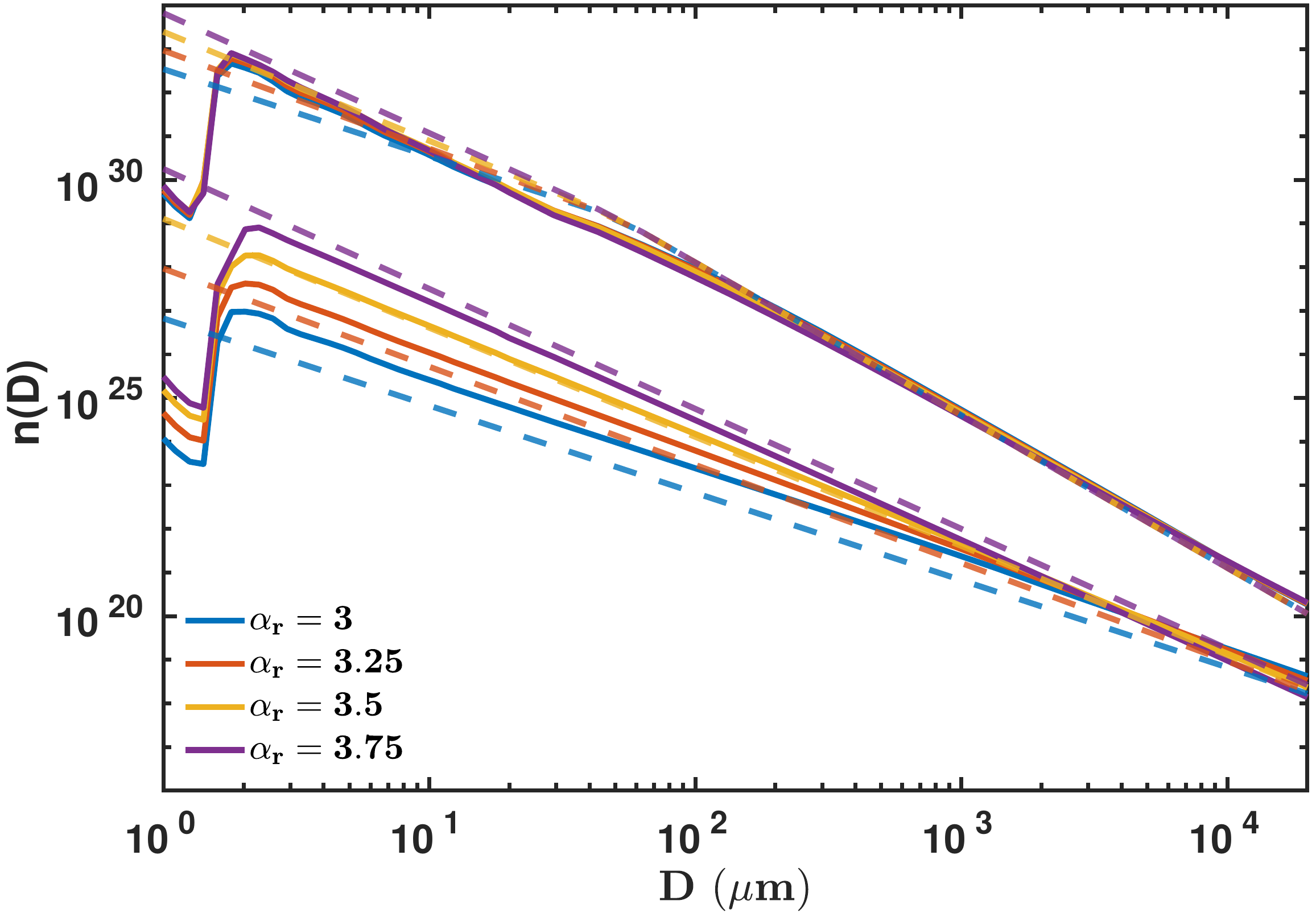}

	\caption{The size distribution $n(D)$ of particles in the belt for a disc with mass input rate $\dot{M}_\mathrm{in} = 10^{-15}~\mathrm{M}_{\earth}~\text{yr}^{-1}$ (top) and $10^{-18}~\mathrm{M}_{\earth}~\text{yr}^{-1}$ (bottom). The slope of the redistribution function, $\alpha_\mathrm{r}$, is varied between $3$ and $3.75$ in both the numerical model (solid lines) and the analytical model (dashed lines) to show the corresponding change in slope of the size distribution for drag-dominated particles.}
	\label{fig:alphar}
\end{figure}

As derived in \citet{Wyatt11}, the redistribution function of collisions should determine the slope of the size distribution for particles which are dominated by P-R drag, such that their size distribution has a slope $1 - \alpha_\mathrm{r}$. It is expected that the size distribution should change slope at the critical particle size $D_\mathrm{pr}$. Figure~\ref{fig:alphar} compares the size distribution of particles in the belt from the numerical and analytical models as $\alpha_\mathrm{r}$ is varied. These distributions show the expected decrease in steepness of the slope as $\alpha_\mathrm{r}$ is decreased. \par
Figure~\ref{fig:alphar} shows the size distribution for a disc with $\dot{M}_\mathrm{in}~=~10^{-15}~\mathrm{M}_{\earth}~\text{yr}^{-1}$. This is a disc for which the largest particles are collisional, while the smaller particles are drag-dominated. For the larger, collisional particles, the slope is the same for all redistribution functions and agrees with the analytical model. Below $100~\mu$m there is a change in slope as predicted, though the slope does not perfectly match the analytical prediction. One possible explanation for the numerical model having a steeper slope is the inclusion of small particles which are put onto eccentric orbits by radiation pressure, forming the halo. \par 
The size distribution of a lower mass disc is given at the bottom of Figure~\ref{fig:alphar}, which shows a drag-dominated disc with $\dot{M}_\mathrm{in} = 10^{-18}~\mathrm{M}_{\earth}~\text{yr}^{-1}$. Since this disc is more fully in a P-R drag dominated regime, the slopes in the numerical model match better with the analytical predictions, and once more a slope change is seen when varying $\alpha_\mathrm{r}$. The slope of the numerical model varies between $-2.7$ for $\alpha_\mathrm{r} = 3.75$ and $-2.02$ for $\alpha_\mathrm{r} = 3$, while the analytical predictions of the slope are $-2.75$ and $-2$ respectively. 

\subsection{Limitations of model}
\label{subsec:limitations}
The model has been fitted over a large range of mass input rates, with the highest mass input rate, $10^{-10}~\mathrm{M}_{\earth}~\text{yr}^{-1}$, corresponding to a dust mass of $3\times10^{-4}~\mathrm{M}_{\earth}$ for a belt with a radius of 30~au, while the lowest disc mass the model was fitted to is $3\times10^{-8}~\mathrm{M}_{\earth}$. Some of the discs this model is applied to may have higher dust masses, such that they are outside the range the model has been fitted to. Further, the model was fitted for a Sun-like star, and may slightly underestimate the flux for A stars (see Section~\ref{subsec:star}). It is advised not to apply the model to M dwarfs, where the $\beta$ profile differs significantly from the analytical prediction, and stellar winds need to be considered. It would be possible to include stellar winds by adjusting $\beta$ for a specific system, if the magnitude of the stellar wind were known. \par
Further, real debris discs can either be very broad or very narrow, while our model assumes discs to have a width $dr/r = 0.4$, and does not predict the correct dependence when varying disc width. Therefore it is difficult to apply the analytical model to discs which are very broad, such as $\tau$~Ceti, which has inner and outer radii of 6 and 52~au respectively. While the model has been fitted to a range of disc masses, it is difficult to constrain the best value of $k$, and a range of parameters can produce acceptable results. For a different stellar type or belt radius, it may be that slightly different values of $k_0$ and $\gamma$ fit better than those chosen here. Overall for the discs considered in this paper, the model should fit to within a factor $\sim3$, except for M stars.

\section{Applications}
\label{sec:obs}
\subsection{Solar system}
\label{subsec:applySS}
The model can be used as a simple way to predict the distribution of particles within the inner solar system. As a toy model, we assume a belt with a radius of $r_0 = 3~$au, and vary its mass to best fit the optical depth of the zodiacal cloud at 1~au. This is not meant to provide a more accurate description of the zodiacal cloud than currently available models. Rather it is used to serve as a quick illustration of the model in a situation where there are observational constraints on its predictions before it is applied to systems with fewer constraints. The source of dust at 3~au could either be from collisions of asteroids, or delivery of material from comets. For a face-on optical depth at 1~au of $\tau(1~\text{au}) = 7.12\times10^{-8}$ \citep{Kelsall98}, fitting the radial optical depth profile to this value gives a dust mass of $M_{\mathrm{dust}}~=~{6.62\times10^{-9}}~\mathrm{M}_{\earth}$, including all particle sizes up to $D~=~2~$cm. This agrees with the predicted mass of the inner zodiacal cloud from \citet{Nesvorny11_ZC}, who predicted a mass of $\sim6.6\times10^{-9}~\mathrm{M}_{\earth}$ within the inner 5~au, assuming a single grain size of $D~=~100~\mu$m and grain density $\rho~=~2~\text{g~cm}^{-3}$, though this estimate is dependent upon the model parameters and chosen grain size. This low value of the dust mass means that P-R drag is significant, such that the toy model predicts only a modest drop (factor of $\sim2$) in optical depth inwards of the source region due to collisions, giving a relatively flat radial profile. \par
\begin{figure}
	\centering
	\begin{subfigure}[b]{\linewidth}
		\centering
		\includegraphics[width=\linewidth]{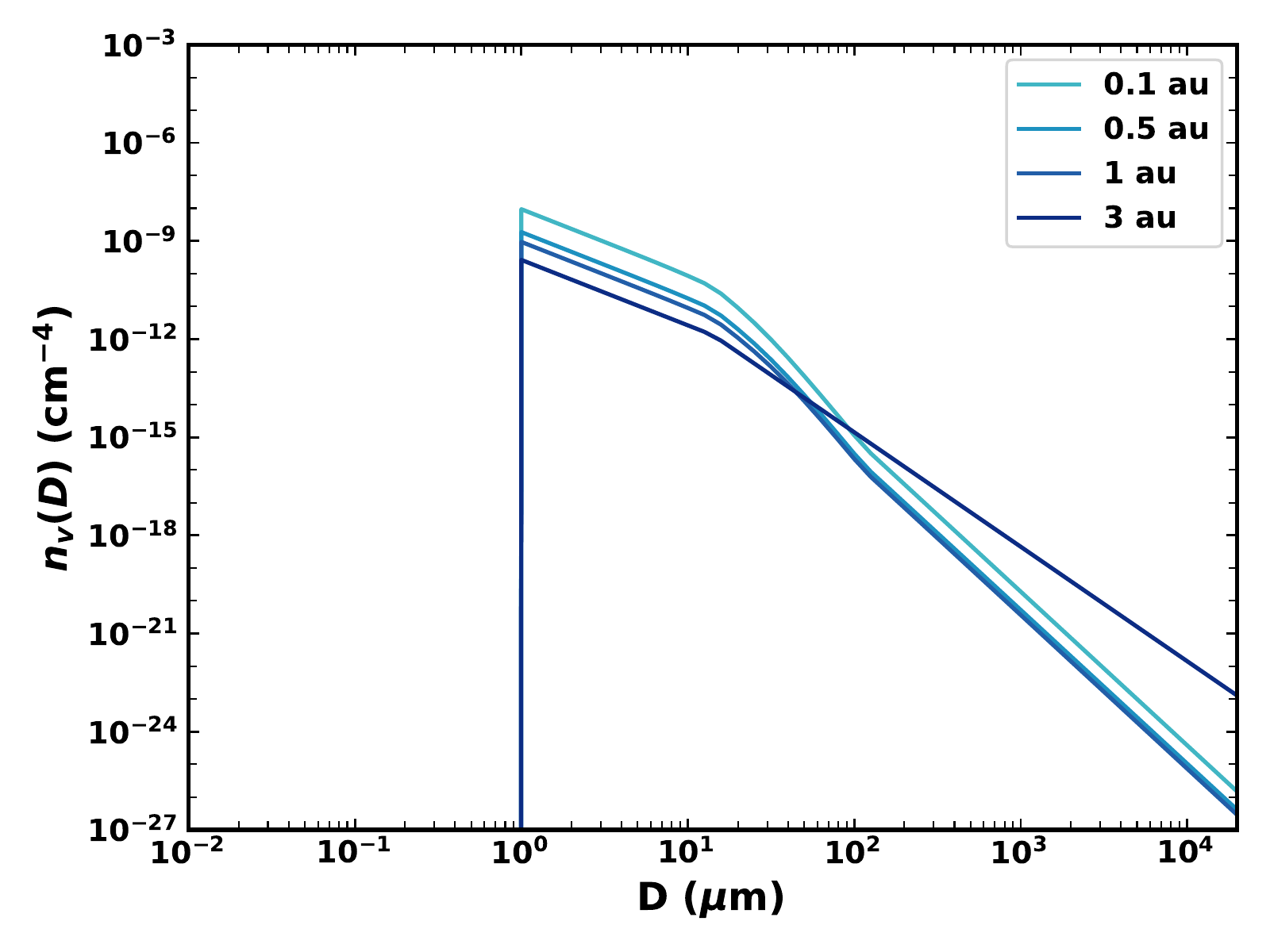}
	\end{subfigure}
	
	\begin{subfigure}[b]{\linewidth}
		\centering
		\includegraphics[width=\linewidth]{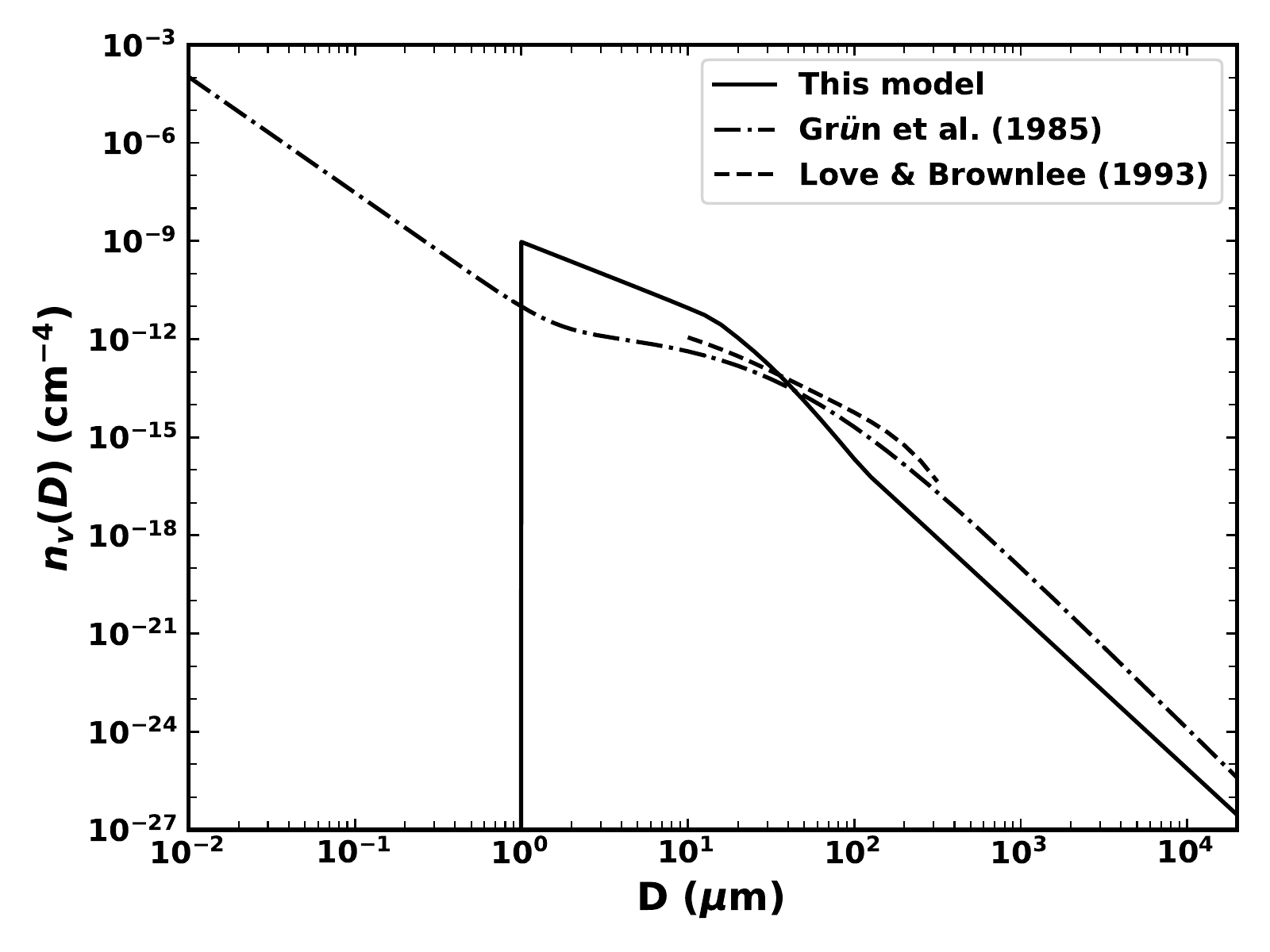}
	\end{subfigure}
	\caption{Size distribution, shown as differential number density, for a belt with radius 3~au fitted to the optical depth of the inner solar system at 1~au as a toy model for the zodiacal cloud. Top: the size distribution at different radii in the disc. Bottom: the size distribution at 1~au for the model (solid), and the empirical models of \citet{Grun85} (dash-dotted) and \citet{Love93} (dashed).}
	\label{fig:SS}
\end{figure}

  
The predicted size distribution at different radii from the Sun is given in Figure~\ref{fig:SS} (top) in terms of the differential number density of particles. Number density is used to better compare with observations; the number density at a given radius can be found from optical depth as
\begin{equation}
\label{eq:nV_D}
n_\mathrm{v}(D) = \frac{\tau(D, r)}{h\sigma},
\end{equation}
where $h = 2r\sin\epsilon$ is the height of the disc, and $\sigma = \pi D^2/4$ is the cross-sectional area of a grain with a given size. At 3~au the size distribution is as described in Section~\ref{subsec:beltdist}, with a turnover to a shallower slope below $D_\mathrm{pr}~=~27.5~\mu$m. Small grains have a flat radial optical depth profile, and converting number per cross-sectional area to number density requires dividing by r, such that closer to the star the number density of small grains increases. Particles larger than $D_\mathrm{pr}$ are depleted by collisions inwards of the source belt, causing the slope of the size distribution to be steeper inwards of 3~au. The model predicts a steep slope of -4.7 for large particles inwards of 3~au, with a shallower slope for small grains.\par 
Also shown in Figure~\ref{fig:SS} (bottom) is a comparison of the model with two empirical models for the size distribution of dust at 1~au in the solar system which were obtained by fitting to measurements of interplanetary dust particles (IDPs). \citet{Grun85} developed an empirical model for the interplanetary meteoroid flux at 1~au based on data from the lunar crater size distribution for large meteoroids ($m \gtrsim 10^{-6}~$g, or $D \gtrsim~91~\mu$m), and in situ measurements from micrometeoroid detectors on board the Highly Eccentric Orbit Satellite 2 (HEOS-2) and the Pegasus satellite for small meteoroids ($m \lesssim 10^{-9}~$g, or $D \lesssim 9.1~\mu$m). \citet{Love93} determined the mass flux distribution of meteoroids in the mass range $10^{-9} \leq m \leq 10^{-4}$~g accreted onto Earth using hypervelocity impact craters on the Long Duration Exposure Facility (LDEF) satellite. This is equivalent to a size range of $9.1 \leq D \leq 424~\mu$m. The size distribution shown for LDEF has taken into account the effect of gravitational focussing, as the Earth's gravity will increase the flux of particles onto the Earth. Fluxes were converted to number densities using equation 3 from \citet{Grun85}, assuming an isotropic flux. \par 
For large particle sizes, which will be the most collisional, the slope of the model from \citet{Grun85} is -4.9, in good agreement with the analytical model. For collisional particles, the analytical size distribution close in will be $\propto n_0(D)/\eta_0(D)$, such that the size dependence of the factor $k$ affects the slope of the size distribution inside of the belt. Without the size dependence of $k$, the slope of the size distribution would be shallower, and have a poorer fit to that of the \citet{Grun85} model, providing further justification for the size dependence of $k$. \par
Despite the good agreement for large particles, there are some differences between the analytical and empirical models at small sizes. While all the models include a turnover to a shallower slope at smaller particle sizes, the turnover is smoother in the empirical models, whereas the simpler analytical prescription necessitates a sharper change. The analytical model turns over at a value of $D_\mathrm{pr} = 27.5~\mu$m, however the empirical models suggest that this should be slightly larger, perhaps closer to $D_\mathrm{pr} = 100~\mu$m. \citet{Wyatt11} showed that the size distribution of drag-dominated particles can be indicative of the redistribution function. Generally the redistribution function power law is assumed to lie in the range $3 \leq \alpha_\mathrm{r} \leq 4$. The empirical size distributions have shallower slopes than the analytical model at small particle sizes, so a value of $\alpha_\mathrm{r} = 3$ is chosen to better match the empirical models. This fits the slope of the LDEF model, and is the smallest value that would typically be expected for $\alpha_\mathrm{r}$. In order to best fit the \citet{Grun85} model, a value of $\alpha_\mathrm{r} \sim 2$ would be needed. Another discrepancy between the models is that \citet{Grun85} suggests there is a high density of submicron grains. However, such small grains are not included in this paper as it is expected that they will be blown out by stellar radiation pressure. Despite these discrepancies, the analytical model captures most of the main features of the empirical models, which are also not completely accurate, and is good for a rough approximation of the size distribution at 1~au. This provides confidence that our model gives reasonable predictions for the exozodi properties in other systems, which are described in the following sections. \par

\subsection{Thermal emission}
\label{subsec:applyexo}
As discussed in Section~\ref{subsec:SEDs}, SEDs can be found for a given distribution by finding realistic grain temperatures and absorption efficiencies, and integrating the optical depth over grain size and radius. \par
\begin{figure}
	\centering
	\includegraphics[width=\linewidth]{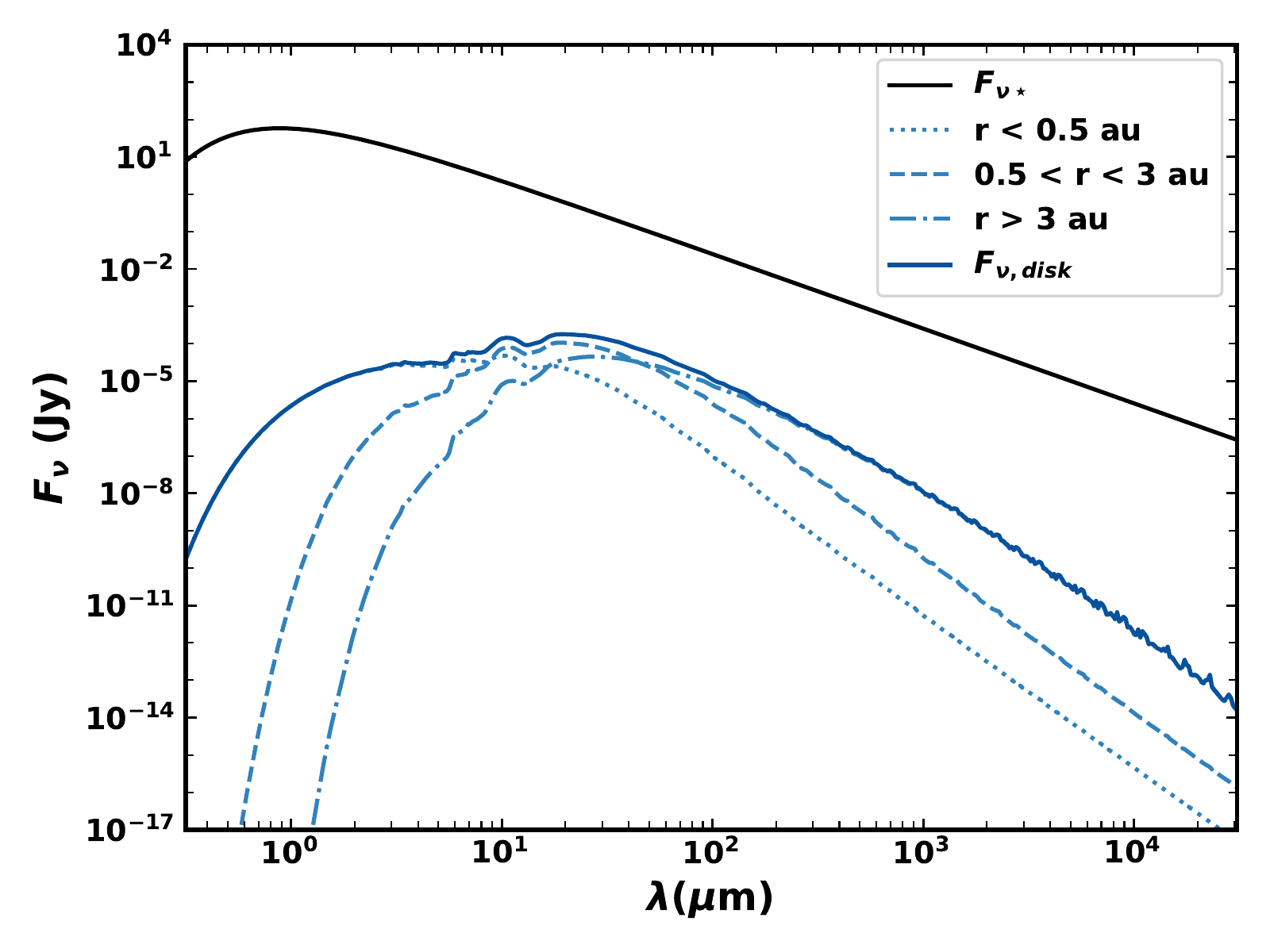}
	\caption{SED for the toy model for the zodiacal cloud from Section~\ref{subsec:applySS}, showing the contributions from different radii ranges, as well as the total disc emission and that of the Sun, as observed at a distance of 10~pc.}
	\label{fig:SS_SED}
\end{figure}

For example, Figure~\ref{fig:SS_SED} shows the resulting SED for the disc used in the toy model of the zodiacal cloud in Section~\ref{subsec:applySS}, as well as the contributions from different radii. This disc produces small excesses at all wavelengths relative to the stellar flux. The fractional excess at $11~\mu$m is $7.6\times10^{-5}$, and the emission peaks at $19~\mu$m, where the fractional excess is $2.9\times10^{-4}$. As would be expected, at the shortest wavelengths the emission is dominated by the warmest dust, which is close to the star. Habitable zone dust dominates the mid-infrared, and the colder dust further out dominates far-infrared emission.
\par

\begin{figure*}
	\centering
	\begin{subfigure}[t]{0.49\textwidth}
		\centering
		\includegraphics[width=\textwidth]{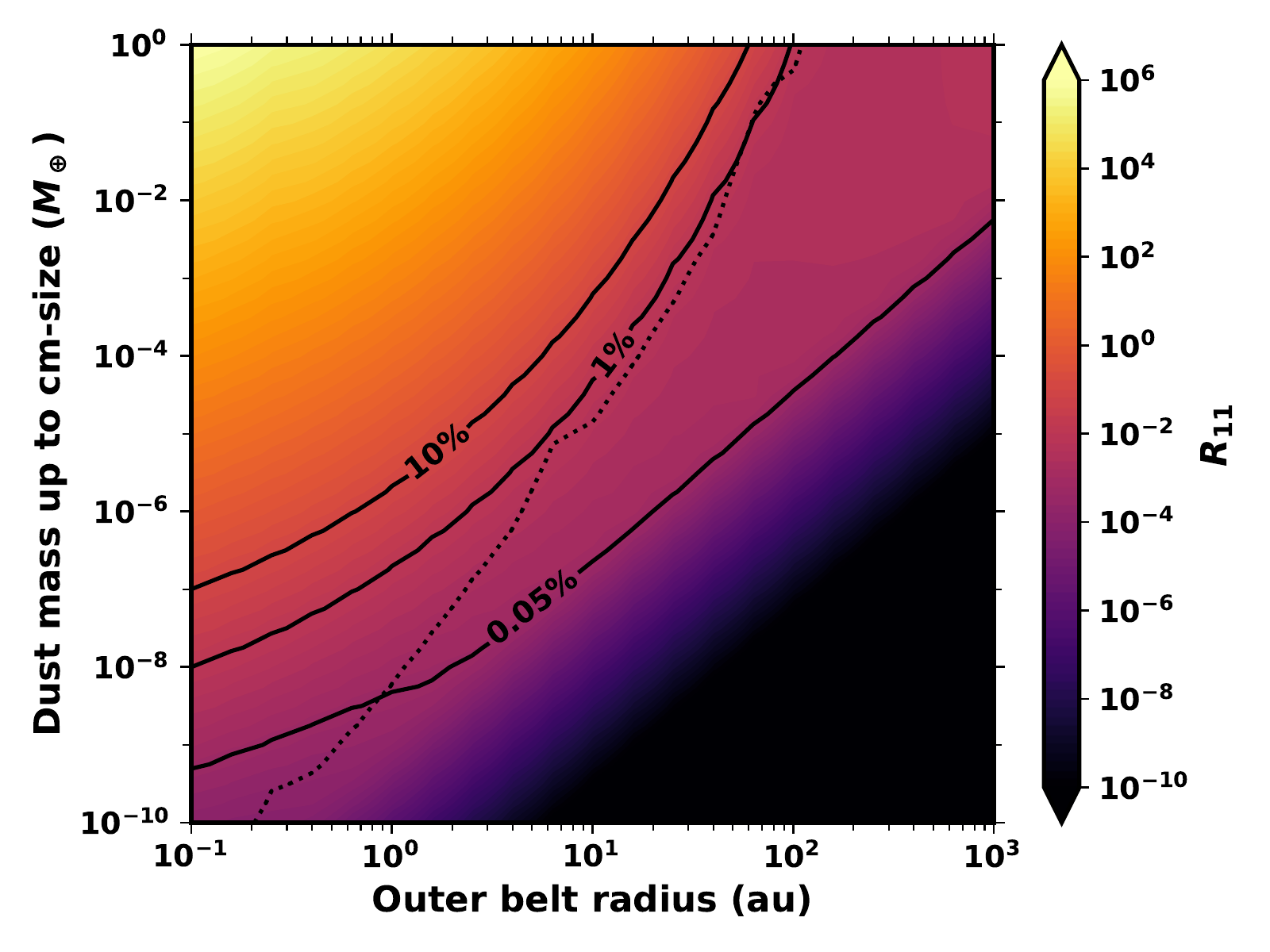}
		\caption{}
		\label{fig:R11}
	\end{subfigure}
	\hfill
	\begin{subfigure}[t]{0.49\textwidth}
		\centering
		\includegraphics[width=\textwidth]{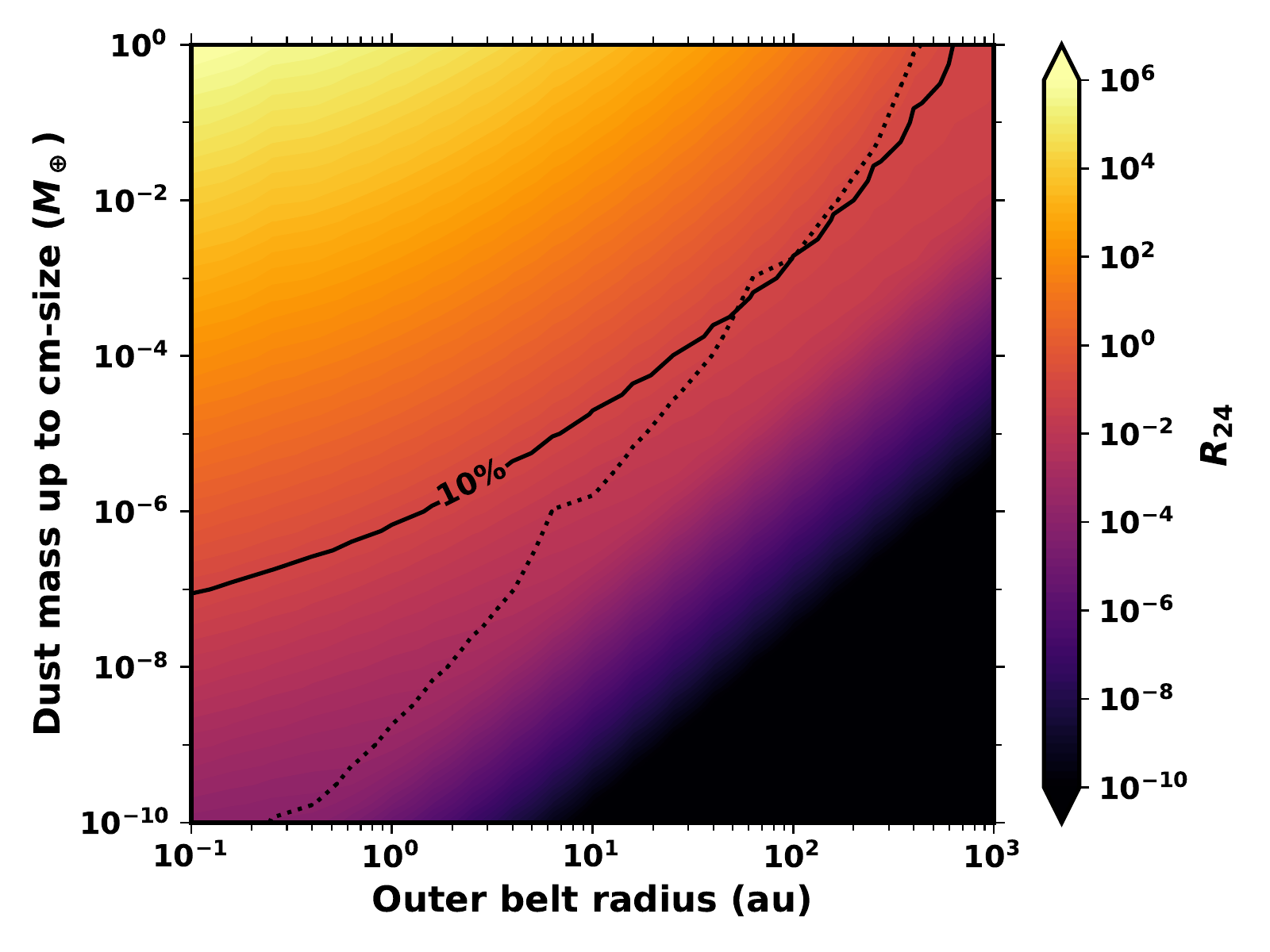}
		\caption{}
		\label{fig:R24}
	\end{subfigure}
	
	\begin{subfigure}[t]{0.49\textwidth}
		\centering
		\includegraphics[width=\textwidth]{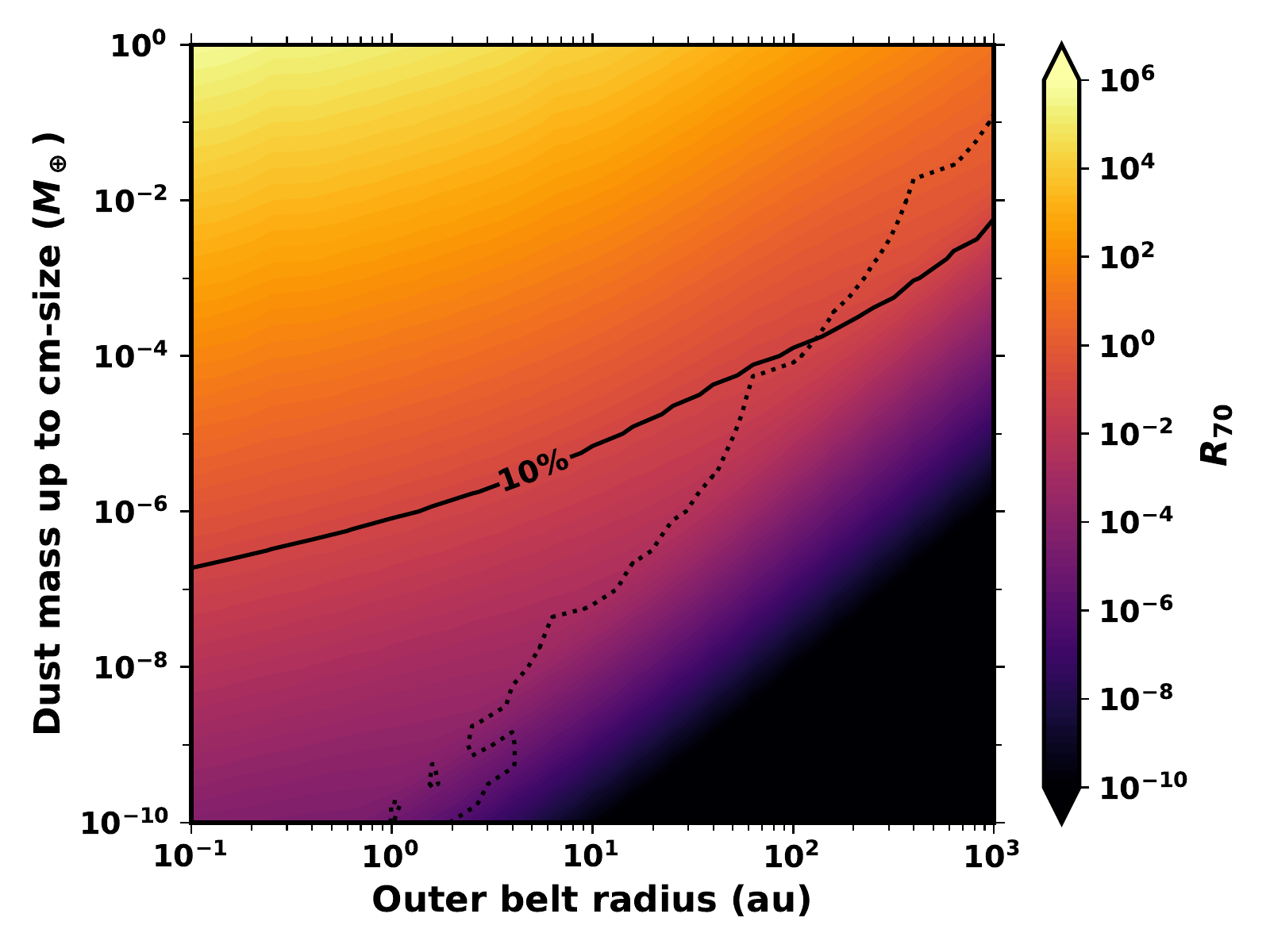}
		\caption{}
		\label{fig:R70}
	\end{subfigure}	
	\hfill
	\begin{subfigure}[t]{0.49\textwidth}
	\centering
	\includegraphics[width=\textwidth]{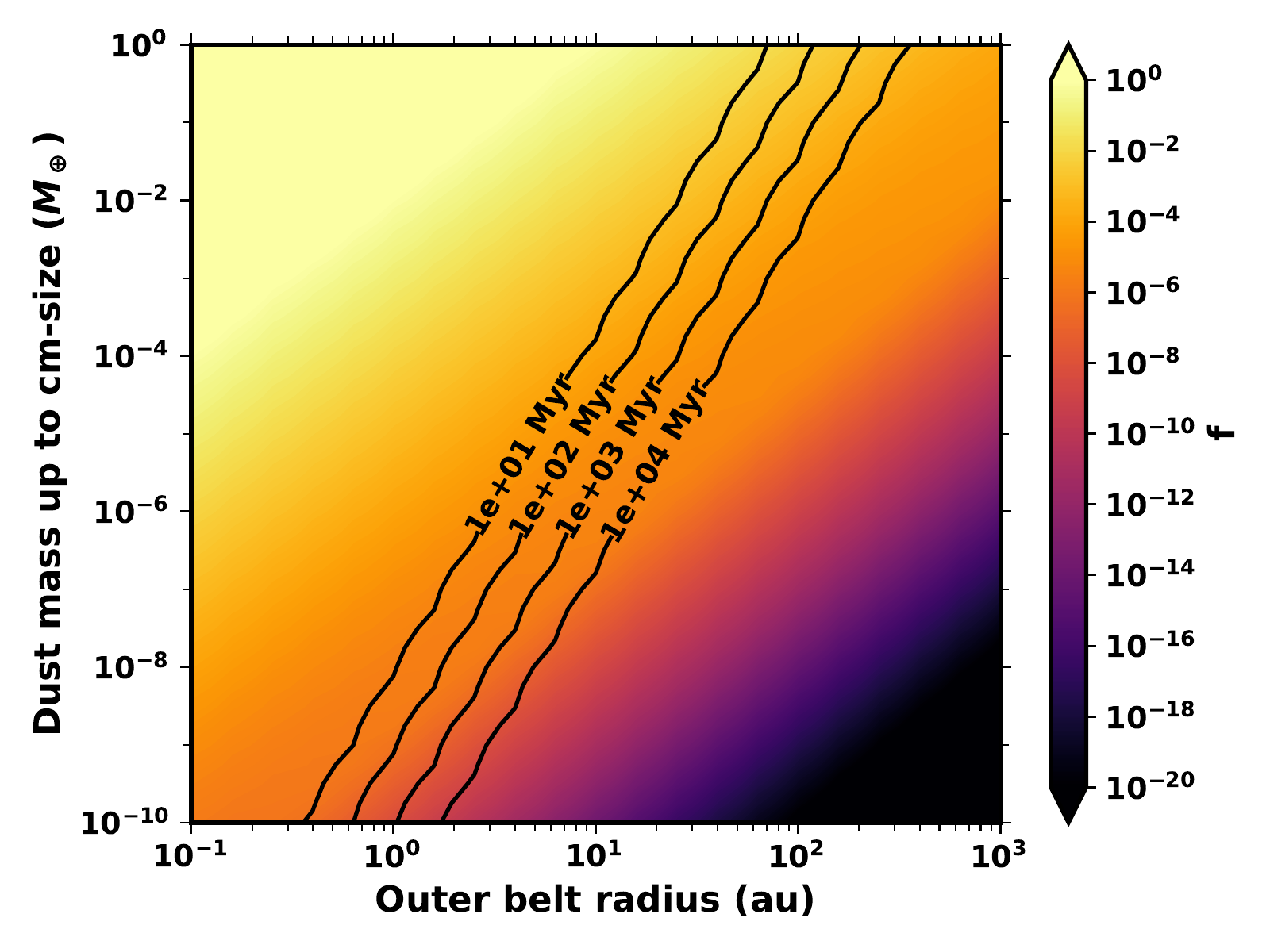}
	\caption{}
	\label{fig:f}
\end{subfigure}
	\caption{Fractional excesses and luminosities for discs of various dust masses $M_\mathrm{dust}$ and belt radii $r_0$ around a Sun-like star, as predicted by our analytical model for realistic asteroidal grains. Dotted contours show where 50\% of dust emission comes from the planetesimal belt, with the other half from dust interior to that belt, such that the thermal emission of discs to the left of these contours is dominated by the parent belt. (a) Fractional excess at $11~\mu$m, $R_{11}$. Contours show excesses of 10$\%$, 1$\%$, and 0.05$\%$, which correspond to the approximate sensitivities of WISE, KIN, and LBTI. (b) Fractional excess at $24~\mu$m, $R_{24}$. The sensitivity of Spitzer/MIPS photometry, $\sim10\%$, is indicated by a contour. (c) Fractional excess at $70~\mu$m, $R_{70}$. The sensitivity of Spitzer/MIPS or Herschel/PACS photometry, $\sim10\%$, is indicated by a contour. (d) Fractional luminosity $f$ of discs, as obtained by integrating the disc flux over the whole spectrum. Contours show upper limits on dust mass for discs around a Sun-like star at different ages, based on the model of \citet{Wyatt07_zodi}, using the parameters from \citet{Sibthorpe18}.}
	\label{fig:exozodi}
\end{figure*}

Figure~\ref{fig:exozodi} shows predictions of fractional excesses and fractional luminosity from the model for discs across the broader parameter space of dust mass and belt radius for a Sun-like star, where the dust mass is defined to be the mass in grains up to $D_\mathrm{max}~=~2~$cm in the belt. Figures~\ref{fig:R11}, \ref{fig:R24}, and \ref{fig:R70} show the predicted excesses at 11~$\mu$m, 24~$\mu$m, and 70~$\mu$m respectively for various discs. The highest excesses are seen for discs with small radii and high dust masses, but these will have short lifetimes, as they would rapidly grind down by collisions to a lower mass \citep{Wyatt07_zodi}. While a large range of belt radii are considered, most planetesimal belts would be expected to lie between $\sim1$~au and a few hundred au. Discs which are detectable with a given instrument should lie above a given excess level, which corresponds to the instrument sensitivity. The solid lines in Figures \ref{fig:R11}, \ref{fig:R24}, and \ref{fig:R70} give an indication of the regions of parameter space for which the discs would be detectable. For example, it is estimated that LBTI can detect mid-infrared ($11~\mu$m) null excesses down to 0.05$\%$ \citep{Hinz16}, and KIN had a sensitivity of $\sim1\%$. Thus Figure~\ref{fig:R11} shows how the improved detection capabilities of LBTI mean that the exozodi are detectable for a much larger range of outer belt properties than with previous instruments. Photometry, which is used at all wavelengths considered here, has a detection limit $\sim10\%$. For example, Spitzer/MIPS \citep[e.g.][]{Su06,Meyer06} and Herschel/PACS \citep[e.g.][]{Eiroa13,Sibthorpe18} have been used to detect debris discs at $24~\mu$m and $70~\mu$m. WISE \citep{Wright10} has been used at $12~\mu$m to observe bright exozodi \citep{Kennedy13}. \par
The dotted contours on the excess plots show the line where 50\% of the disc emission comes from the planetesimal belt. Discs to the left of this contour have most of their emission originating from dust in the parent belt, while discs to the right have emission which is dominated by dust in the inner regions of the system. At $24~\mu$m and $70~\mu$m, this means that the emission from most discs that can be detected must originate in the planetesimal belt, rather than closer in. However, at $11~\mu$m, for parent belts which are not very close to the star ($r_0 \gtrsim 10~$au), emission will be dominated by the warm dust which is dragged in to the inner regions. The fractional luminosity, $f$, of the discs is shown in Figure~\ref{fig:f}, as found by integrating the disc flux and stellar spectrum then finding the ratio. As expected, the fractional luminosity correlates with the fractional excesses. \par
As mentioned previously, in situ belts at small radii will rapidly grind down by collisions such that their mass is depleted. \citet{Wyatt07_zodi} showed that the mass of a planetesimal belt will decrease with time once the largest planetesimals in the belt are broken up by collisions, giving a time dependence of 
\begin{equation}
\label{eq:Mt}
M_\mathrm{tot}(t) = M_\mathrm{tot}(0) / (1 + t/t_\mathrm{c}(0)),
\end{equation}
where $M_\mathrm{tot}$ is the mass of the planetesimal belt, and $t_\mathrm{c}(0)$ is the collision timescale of the largest planetesimals at the initial time. Since the collision timescale depends on the total mass, at late times the mass of the belt will be independent of its initial mass. Based on equation~19 of \citet{Wyatt07_zodi}, the maximum dust mass at a given age, $t_\mathrm{age}$, is 
\begin{equation}
\label{eq:Mmax}
M_\mathrm{dust,max} =
\frac{2.3\times10^{-15} \rho r_\mathrm{BB}^{13/3}(dr/r)A}{M_\star^{4/3}t_\mathrm{age}}
\end{equation}
in $\mathrm{M}_{\earth}$, where $\rho$ is in kg~m$^{-3}$, $r_\mathrm{BB}$ is the radius which would be inferred from the dust temperature assuming black body emission in au, $A$ in $\text{km}^{0.5}\text{J}^{5/6}\text{kg}^{-5/6}$ is a parameter which can be found by fitting to observations, $M_\star$ is in $\mathrm{M}_{\sun}$, and $t_\mathrm{age}$ is in Myr. It has been assumed that the mean eccentricities and inclinations of planetesimals are equal. 
\par The \citet{Wyatt07_zodi} model gives the total planetesimal belt mass, which has been converted to mass in dust up to 2cm in diameter by scaling with a factor $M_\mathrm{dust, max}~=~\sqrt{\frac{2\times10^{-5}}{D_\mathrm{c}}}M_\mathrm{max}$, where $D_\mathrm{c}$ is the maximum planetesimal size in km. \citet{Sibthorpe18} fitted the model to observations  of Sun-like stars from the Herschel DEBRIS survey. The model was chosen to have the parameters $\rho~=~2700~\text{kg~m}^{-3}$ and $dr/r = 1/2$. The best fitting model also had $A = D_\mathrm{c}^{0.5}{Q_\mathrm{D}^\star}^{5/6}e^{-5/3} = 5.5 \times 10^5~\text{km}^{0.5}\text{J}^{5/6}\text{kg}^{-5/6}$. The model uses the black body radius of a planetesimal belt, while the resolved radius is typically 1-2.5$\times$ larger due to inefficient emission of dust grains \citep{Booth13,Pawellek14}. Therefore, to compare with our model we assume that the disc radius plotted in Figure~\ref{fig:exozodi} is $r_0 = 2 r_\mathrm{BB}$ as an approximation. The upper limits on dust mass up to cm-size grains from this model for a Sun-like star at different ages are shown in Figure~\ref{fig:f}. The \citet{Wyatt07_zodi} model shows that the brightest belts, which have very small radii and high dust masses, would not be in steady state even around very young stars of a few hundred Myr. Therefore, the region of parameter space we would expect to observe systems in is at lower dust masses and larger radii. \par

\begin{figure}
	\centering
	\begin{subfigure}[b]{\linewidth}
		\centering
		\includegraphics[width=\linewidth]{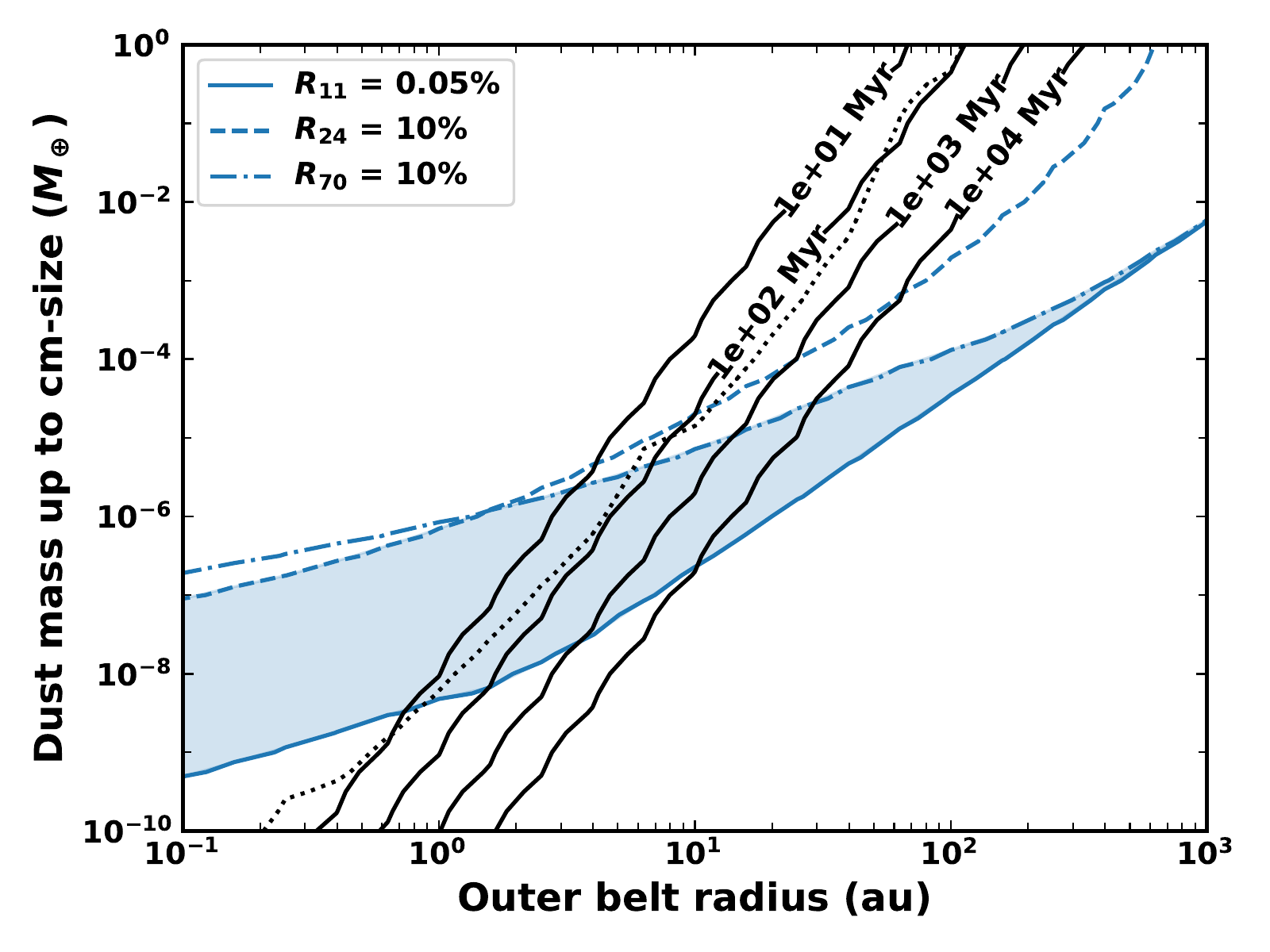}
	\end{subfigure}

	\begin{subfigure}[b]{\linewidth}
		\centering
		\includegraphics[width=\linewidth]{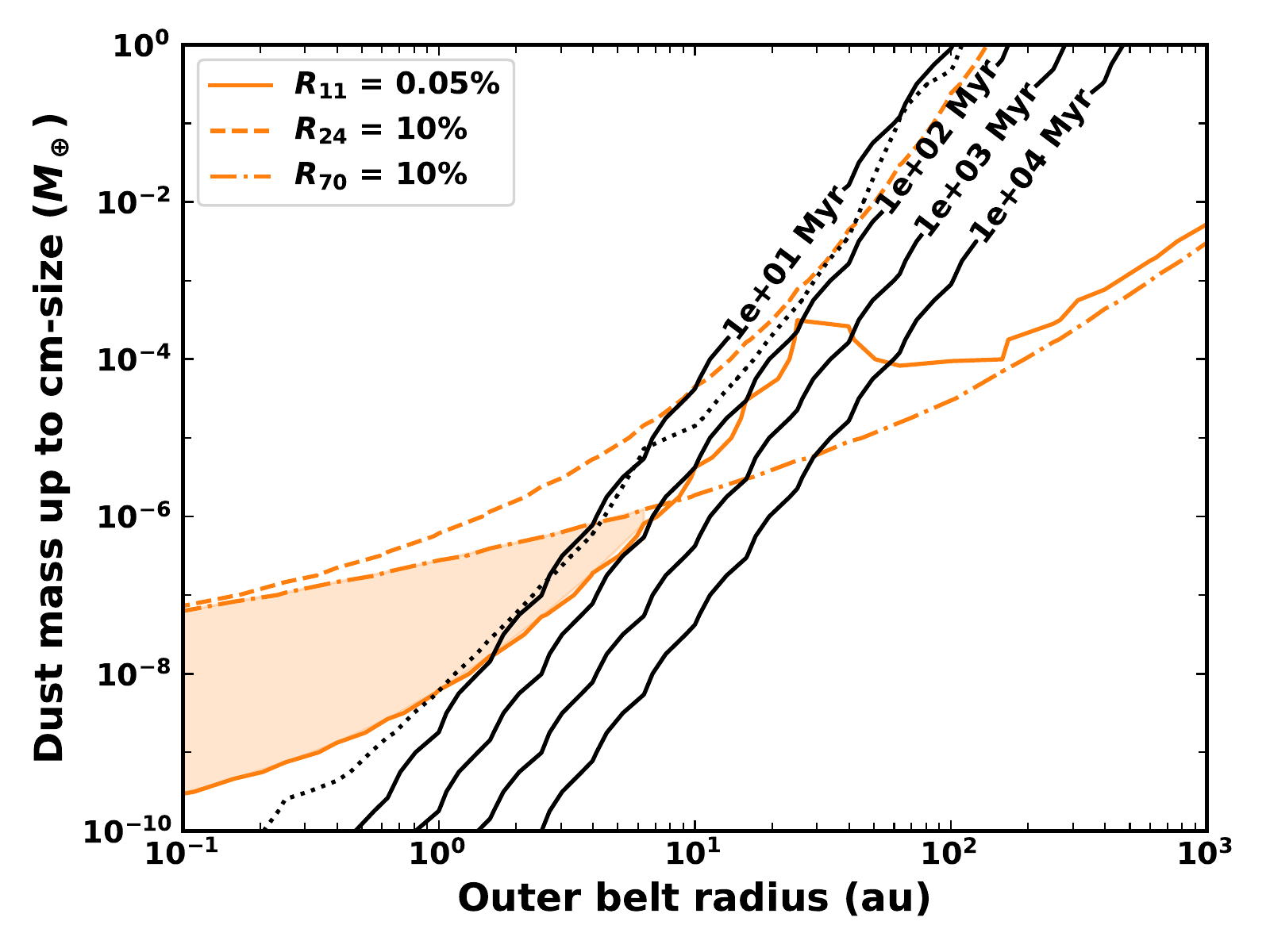}
	\end{subfigure}

	\caption{Detectability thresholds for different wavelengths for asteroidal grains (top) and cometary grains (bottom). The solid contours show $R_{11} = 0.05\%$, the level above which LBTI should be able to detect discs. Dashed and dashed-dotted contours show where a disc would have a 10\% excess at 24~$\mu$m and 70~$\mu$m around a Sun-like star, such that the parent belt is detectable in far-infrared photometry. The shaded region is the region of parameter space for which we predict that LBTI would be able to detect warm dust dragged in from a cold outer belt that has not been detected in the far-infrared. Solid lines show the upper limits on dust mass at given ages based on the model of \citet{Wyatt07_zodi}, as fitted to Sun-like stars in \citet{Sibthorpe18}. Dotted contours show where 50\% of dust emission comes from the planetesimal belt, with the other half from dust interior to the belt, such that the thermal emission of discs to the left of these contours is dominated by the parent belt.}
	\label{fig:excesses_det}
\end{figure}

A direct comparison of the mid-infrared excesses which can be detected by LBTI with the $10\%$ limits at $24~\mu$m and $70~\mu$m is shown in Figure~\ref{fig:excesses_det}. The model predicts that stars which have excesses detected at longer wavelengths with photometry should have exozodiacal dust levels due to P-R drag from the outer belt which are detectable by LBTI. Therefore, non-detections around stars with known cold dust could imply other mechanisms are depleting habitable zone dust. For example, planets could deplete exozodi levels by accreting dust or ejecting it from the planetary system, such that this could be a way to infer the presence of planets \citep[see][]{Bonsor18}. \par The shading in Figure~\ref{fig:excesses_det} highlights the region of parameter space for which it may be possible to detect warm exozodiacal dust that has been dragged inwards from an outer belt which is not currently detectable in far-infrared photometry. However, the limits on dust mass  based on the model of \citet{Sibthorpe18} shown in Figure~\ref{fig:f} rule out a lot of this shaded region around stars older than $\sim 100$~Myr. For example, for a Gyr Sun-like star, it remains possible for LBTI to detect warm dust dragged in from planetesimal belts without a far-infrared detection, however the region of parameter space in which such a disc might be present is smaller than the shaded region shown in Figure~\ref{fig:excesses_det}, e.g. requiring a planetesimal belt $\gtrsim 3$~au in radius for asteroidal grains. Consequently, it may be the case that LBTI detections without far-infrared excesses are the result of dust being dragged inwards from a planetesimal belt which is too faint to detect at longer wavelengths, but there are limits on the possible disc radius and dust mass of such systems, depending on the age of the star. There were three such detections in the HOSTS survey \citep{Ertel20}. Also of note in Figure~\ref{fig:excesses_det} is the dotted contour, which shows where half of the emission comes from the planetesimal belt. Discs for which emission is dominated by the belt will lie to the left of this line, but comparison with the age limits shows that such discs will typically collisionally deplete on a $\sim$10~Myr timescale.
Therefore if warm dust is detected by LBTI in a system without a far-infrared detection, it would not be dust in the planetesimal belt that is being detected, but rather the dust dragged into the inner region. \par

\begin{figure}
	\centering
	\begin{subfigure}[b]{\linewidth}
		\centering
		\includegraphics[width=\linewidth]{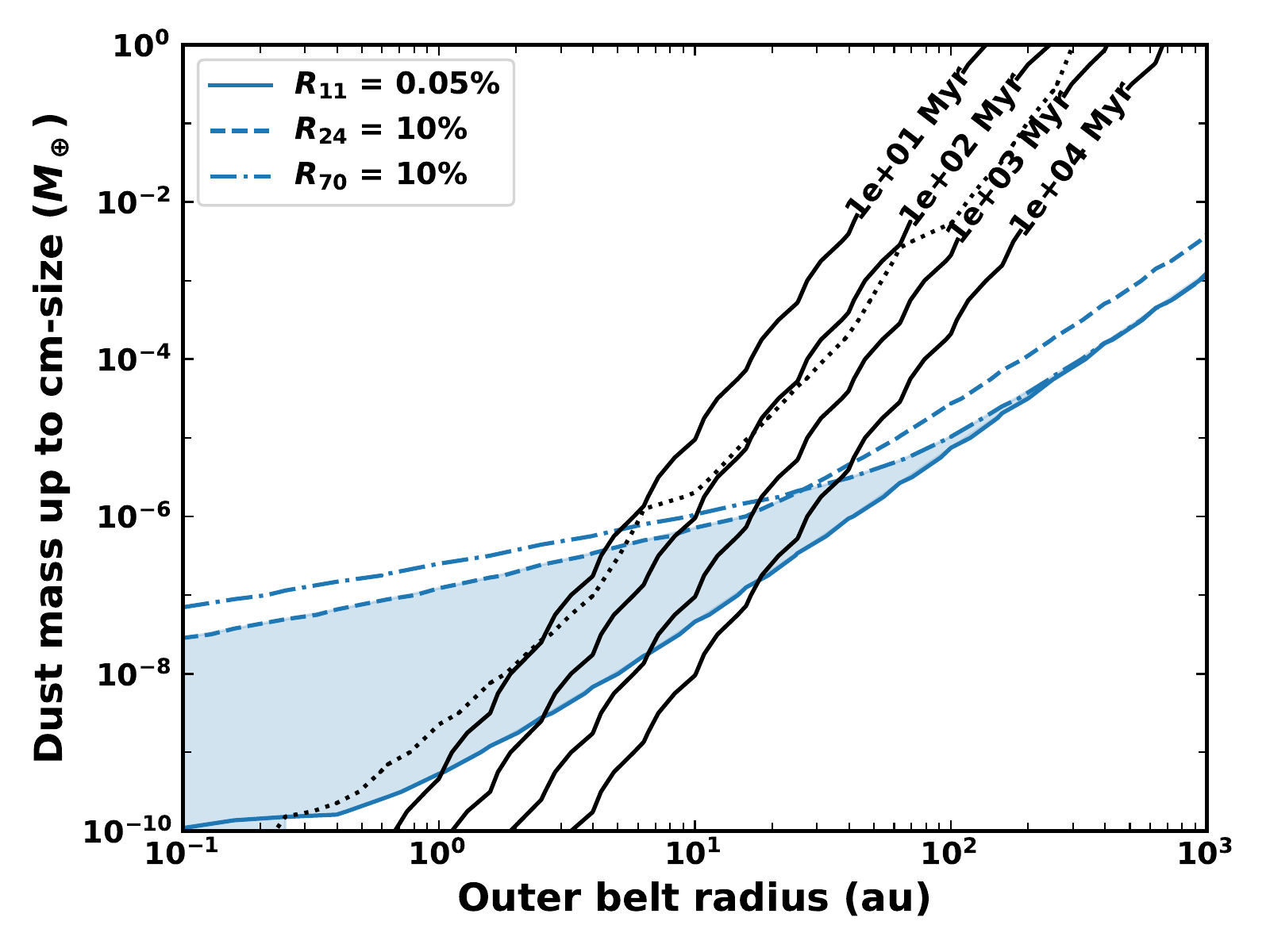}
	\end{subfigure}
	
	\begin{subfigure}[b]{\linewidth}
		\centering
		\includegraphics[width=\linewidth]{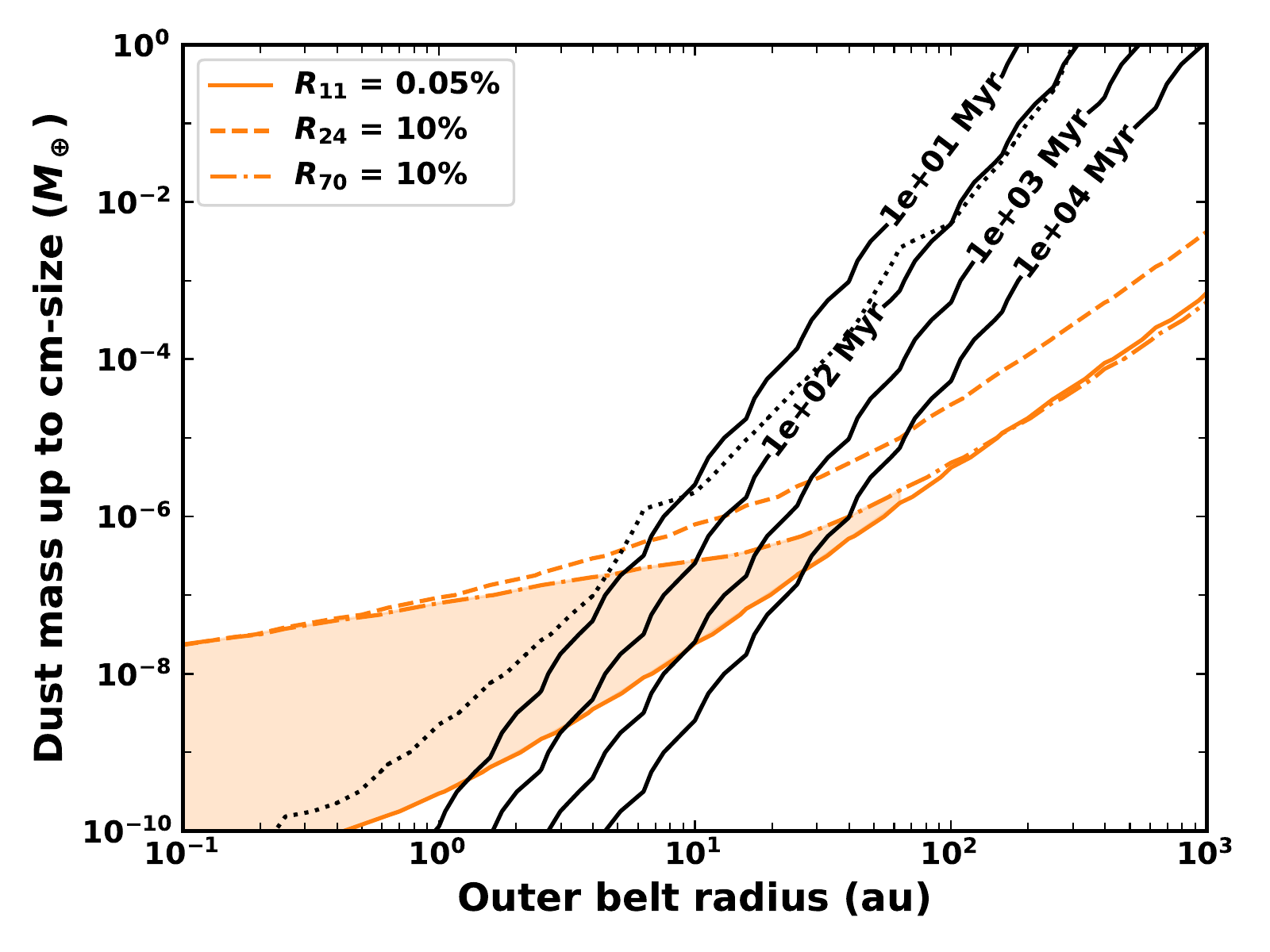}
	\end{subfigure}
	
	\caption{The same as Figure~\ref{fig:excesses_det} but for A type stars. Solid lines show the upper limits on dust mass at given ages based on the model of \citet{Wyatt07_A}.}
	\label{fig:excesses_Astar}
\end{figure}

Throughout this study grains have been assumed to be asteroidal, with no porosity, and a core-mantle model with 1/3 silicates and 2/3 organic material. Grain composition may differ from that assumed in Section~\ref{subsec:optprops}, so the effect of using cometary grains was investigated. These grains came from the core-mantle model of \citet{Li98}, with a porosity $p = 0.95$; half of the vacuum is filled with water ice. The matrix remains 1/3 silicates and 2/3 organic material, as for the asteroidal composition. This gives grains with a much lower density of $\rho = 0.688~\text{g cm}^{-3}$. Overall there is no qualitative change to the conclusions with this alternative composition, but there are relatively minor quantitative changes. The excess plots shown in Figure~\ref{fig:exozodi} were not significantly affected, with the same overall trends, but a slight change in contour shape. Figure~\ref{fig:excesses_det} shows how a cometary composition affects the region of parameter space for which dust is detectable. The parameter space for which LBTI can detect exozodiacal dust dragged inwards from an undetected planetesimal belt decreases slightly with a cometary composition. Only belts which are relatively close to the star, $r_0 \lesssim 5~$au, can be detected uniquely by LBTI, and the relevant region of parameter space is greatly reduced by collisional evolution. Stars known to host cold debris belts should still be detectable by LBTI in the mid-infrared. \par
The detectability of discs around A stars is demonstrated in Figure~\ref{fig:excesses_Astar}, which shows the detection thresholds for realistic asteroidal and cometary grains. Again the model predicts that stars with excesses detectable at longer wavelengths should have detectable levels of exozodiacal dust. Overall the conclusions are similar to those for Sun-like stars, though LBTI could detect slightly lower dust masses around an A star.

\subsection{Application to the HOSTS survey}
\label{subsec:HOSTS}

\begin{figure}
	\centering
	\begin{subfigure}[b]{\linewidth}
		\centering
		\includegraphics[width=\linewidth]{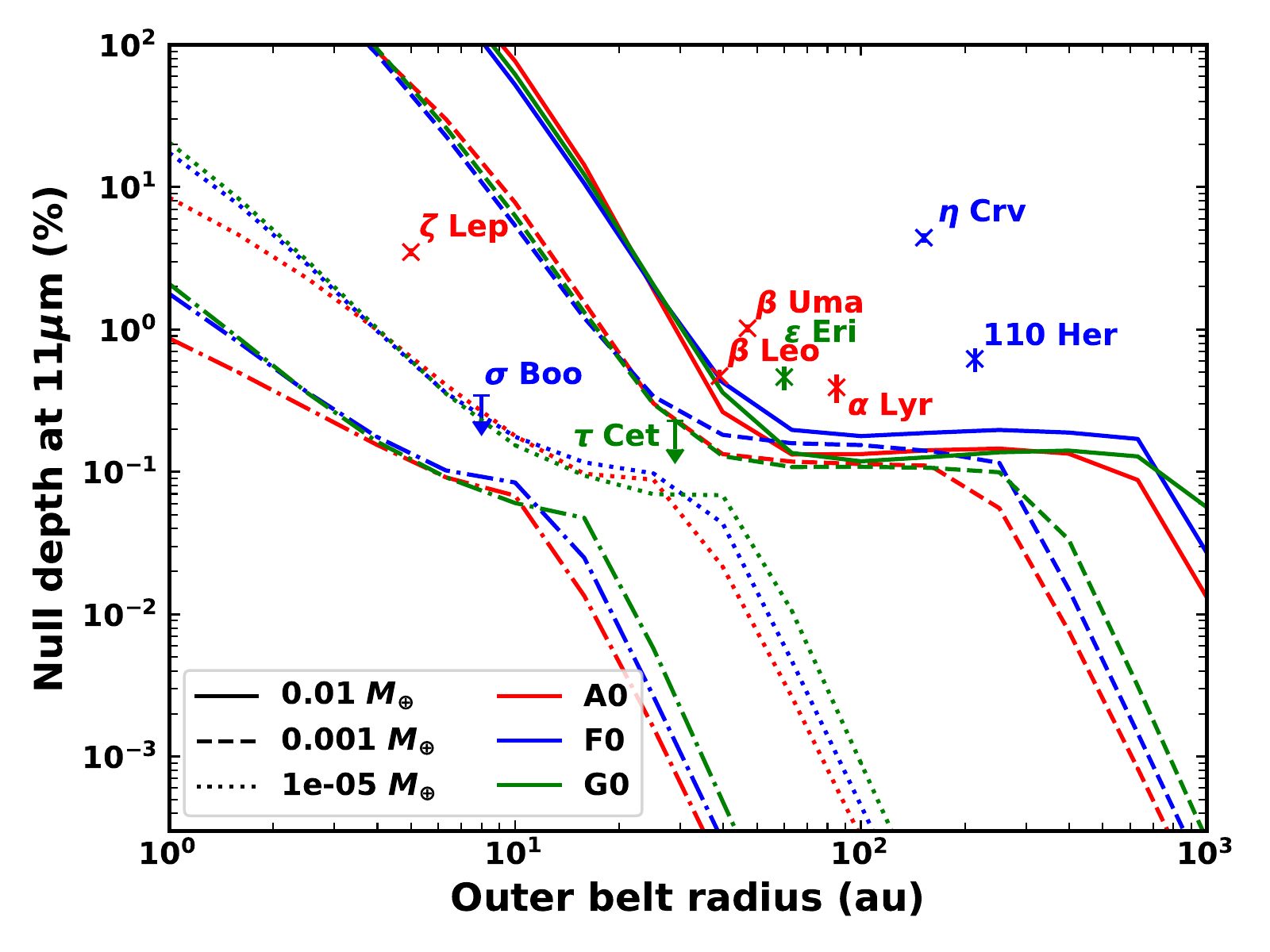}
	\end{subfigure}

	\begin{subfigure}[b]{\linewidth}
		\centering
		\includegraphics[width=\linewidth]{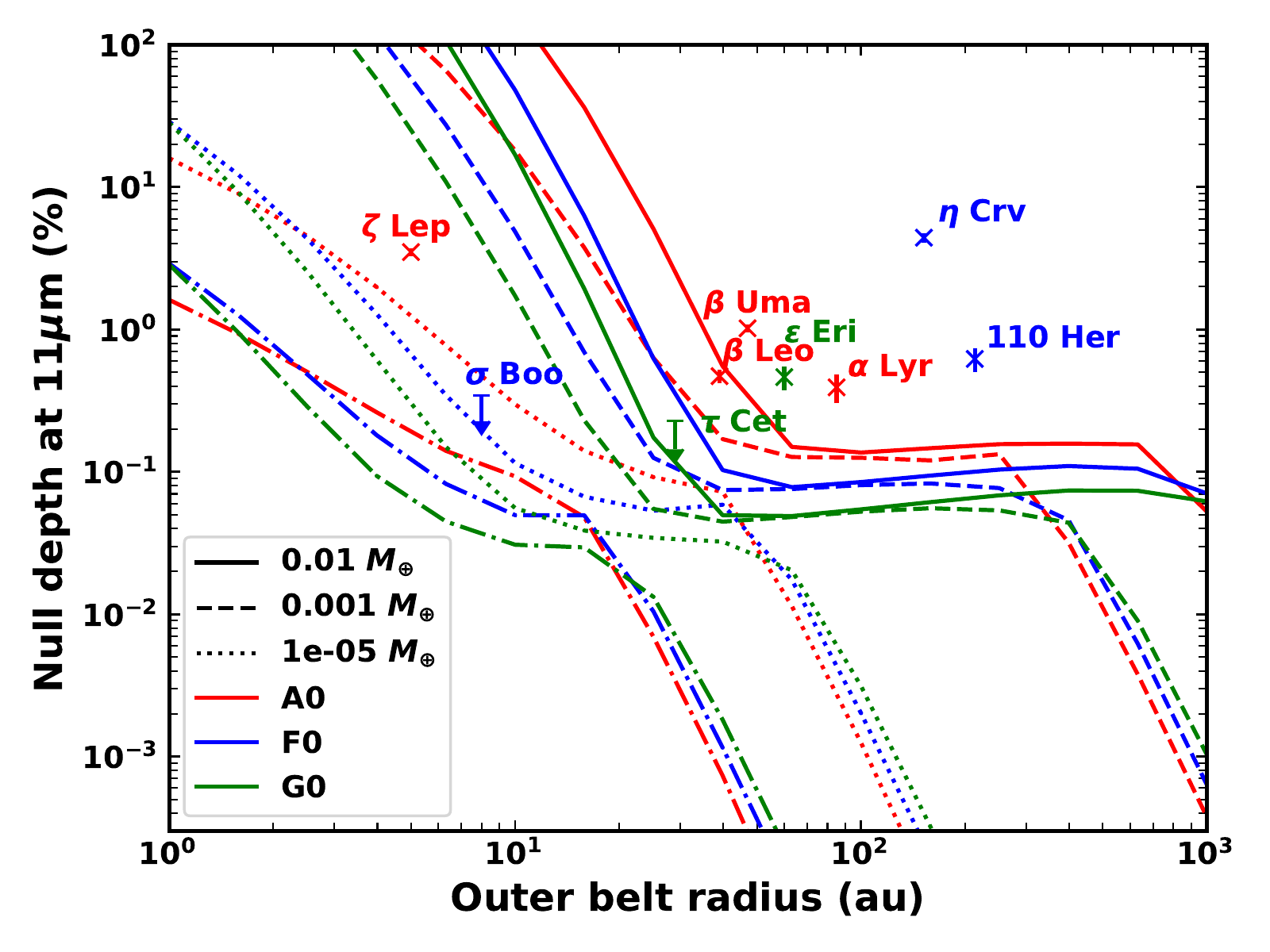}
	\end{subfigure}
	\caption{Null excess predictions at $11~\mu$m for planetesimal belts of different radii, dust masses and stellar spectral types with asteroidal grains (top) and cometary grains (bottom). The HOSTS results for the nine stars with detected debris discs are also shown. Arrows show $3\sigma$ upper limits for stars which had no detection.}
	\label{fig:nulls_rM}
\end{figure}

\begin{figure}
	\centering
	\begin{subfigure}[b]{\linewidth}
		\centering
		\includegraphics[width=\linewidth]{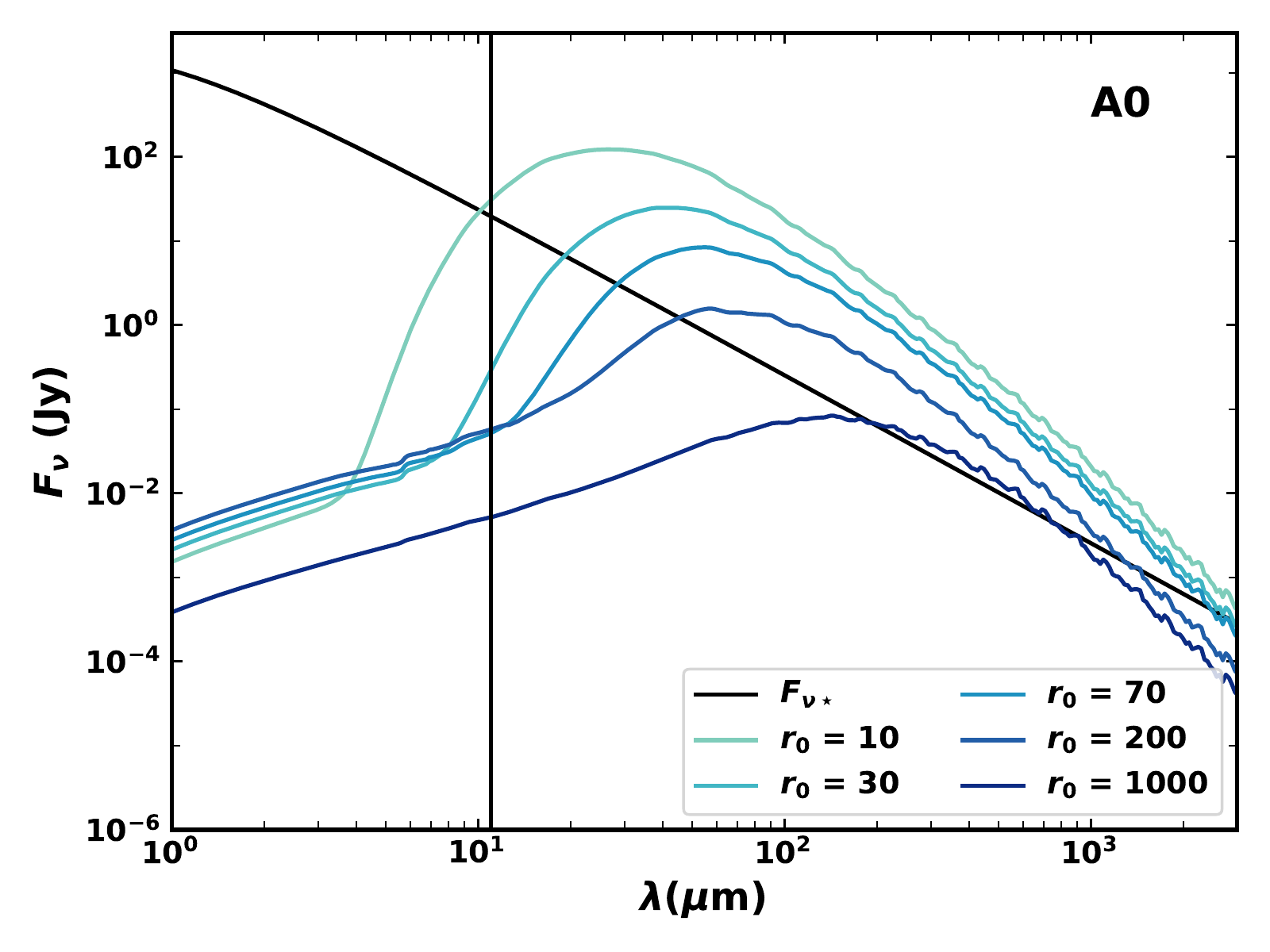}
	\end{subfigure}
	
	\begin{subfigure}[b]{\linewidth}
		\centering
		\includegraphics[width=\linewidth]{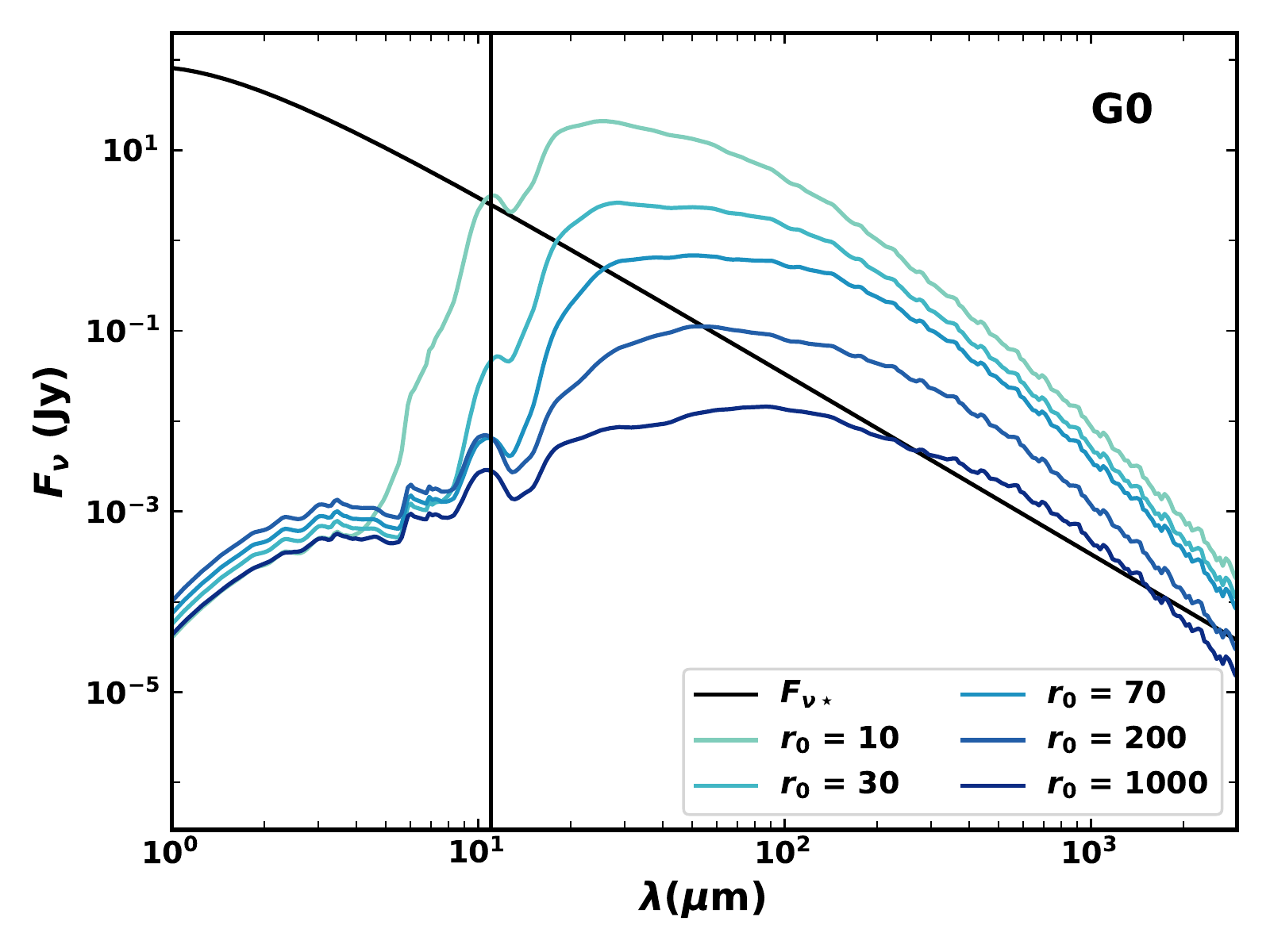}
	\end{subfigure}	
	
	\caption{SEDs for discs of mass $M_\mathrm{dust} = 0.01~\mathrm{M}_{\earth}$ around stars of spectral type A0 (top) and G0 (bottom), with planetesimal belts of different radii. In each case the stellar spectrum is shown in black. The vertical line highlights $11~\mu$m, the wavelength at which LBTI measurements are made. It is assumed that the systems are at a distance of 10~pc from the Sun.}
	\label{fig:SEDs}
\end{figure}

\begin{figure}
	\centering
	\includegraphics[width=\linewidth]{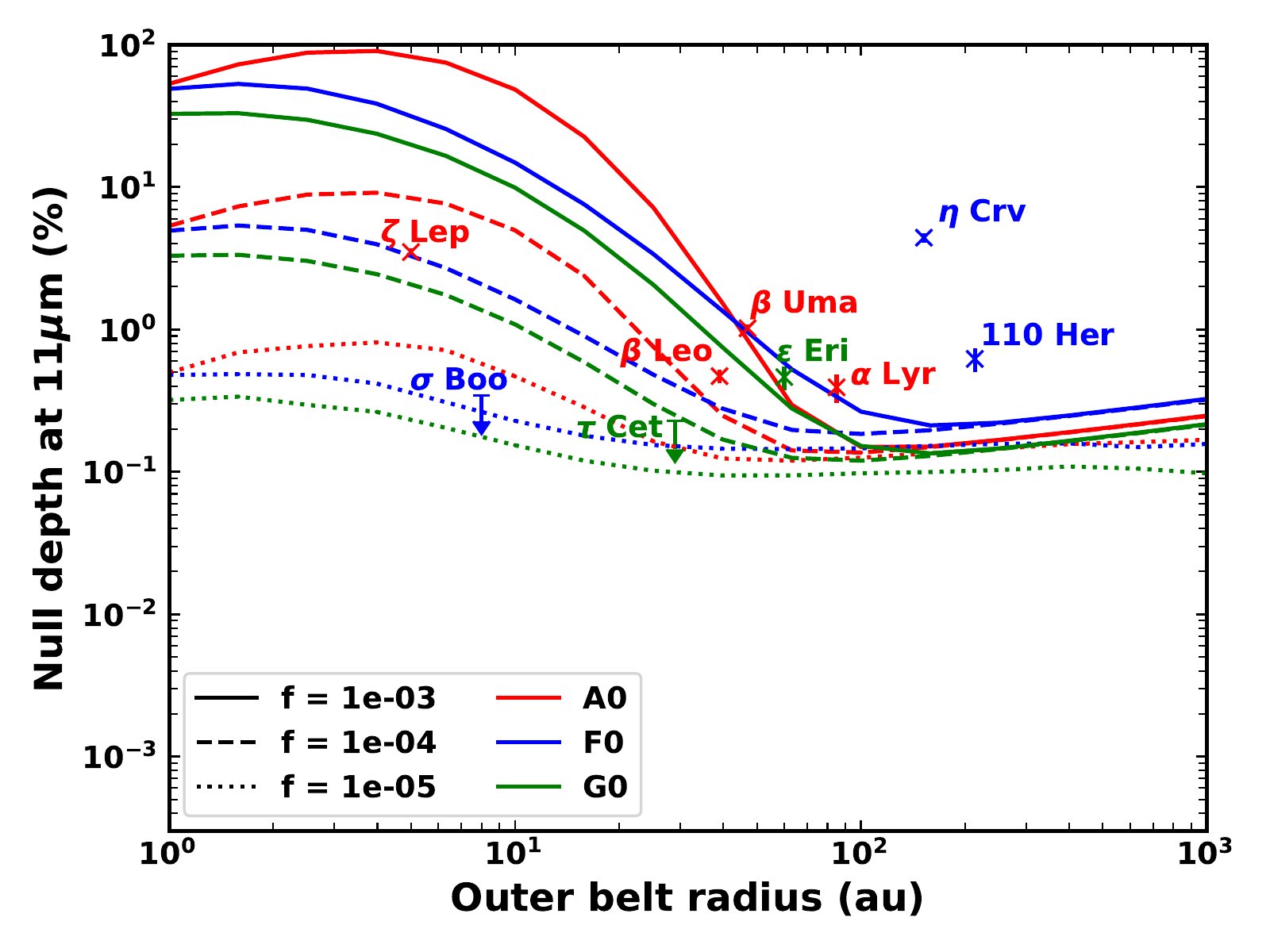}
	\caption{Null excess predictions at $11~\mu$m for planetesimal belts of different radii, fractional luminosities and stellar spectral types with asteroidal grains. The HOSTS results for the nine stars with detected debris discs are also shown. Arrows show $3\sigma$ upper limits for stars which had no detection. }
	\label{fig:nulls_rf}
\end{figure}

\begin{figure}
	\centering
	\includegraphics[width=\linewidth]{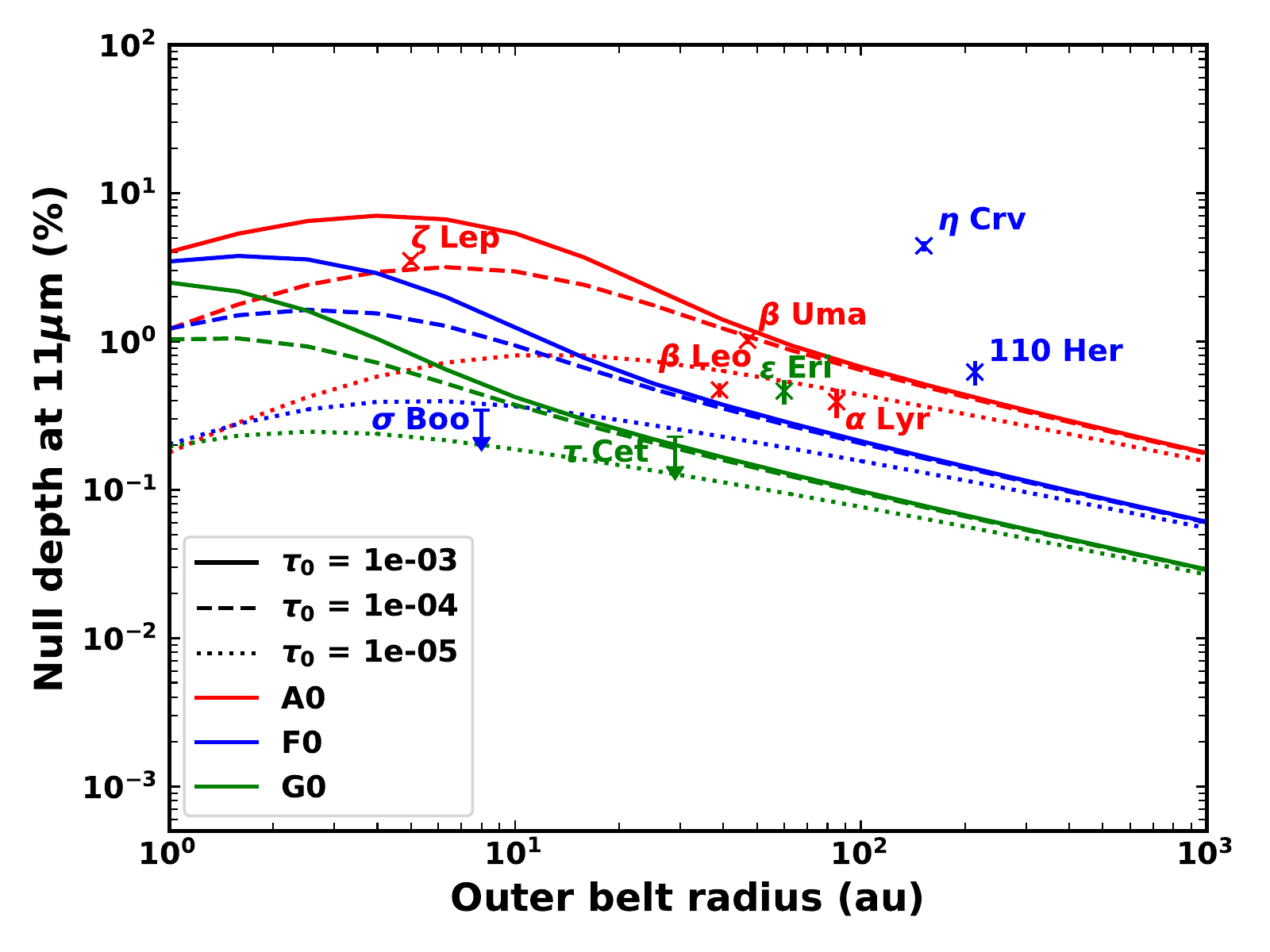}
	\caption{Null excess predictions at $11~\mu$m for planetesimal belts of different radii and stellar spectral types using the model of \citet{Wyatt05}, for different values of the belt optical depth $\tau_0$. The HOSTS results for the nine stars with detected debris discs are also shown. Arrows show $3\sigma$ upper limits for stars which had no detection. }
	\label{fig:nulls_W05}
\end{figure}

The HOSTS survey \citep{Ertel18_prelim,Ertel20}, searched for exozodiacal dust around 38 nearby main sequence stars, of which 9 have previously detected cold outer debris belts. LBTI uses nulling interferometry to subtract the stellar emission, resulting in a measurement of null depth. Predictions can be made using the analytical model presented in this paper for the excess around a certain star given the radius of the planetesimal belt and the mass of millimetre to centimetre-sized grains. To take into account the fact that total flux (which is that reported in Figure~\ref{fig:exozodi}) will be attenuated by the LBTI transmission pattern, our predicted fluxes are divided by a factor of 2 to better correspond to null depths. This is a very rough approximation, as the transmission pattern is highly dependent on the distance to the star and disc orientation. It is likely that the observed null depth will be less than half of the fractional excess, but the precise factor relating fractional excesses to null depths will depend on the geometry and vary between systems, such that individual systems need to be modelled \citep[see][]{Kennedy15_LBTI}. \par
Predictions are presented in Figure~\ref{fig:nulls_rM} (top) for null depths around stars of different spectral types for discs of certain dust masses and planetesimal belt radii, assuming asteroidal grains. Also shown are results from HOSTS survey stars with known debris discs. While the sensitivity of LBTI is around $0.1\%$, predictions are shown down to $10^{-3}\%$, i.e. just below the null depth corresponding to 1 zodi. While the exact null depth corresponding to 1 zodi will vary between systems, in Section~\ref{subsec:applySS} we found a fractional excess of $7.6\times10^{-3}\%$ for our toy model of the zodiacal cloud, giving a null depth in good agreement with the value of $\sim2\times10^{-3}\%$ found by \citet{Kennedy15_LBTI} for the null depth around a Sun-like star corresponding to 1 zodi. Predictions down to this level are important because detection of Earth-like planets would be hampered by dust at the level of 10-20 zodi \citep{Beichman06,Defrere10,Defrere12,Roberge12} for both visible coronagraphs and nulling interferometers. \par
When dust mass is kept fixed, Figure~\ref{fig:nulls_rM} shows that the predicted null as a function of planetesimal belt radius shows the same behaviour for all disc masses and spectral types. The null starts off high for the smallest radii, and drops sharply as radius is increased, before reaching a plateau for intermediate radii, then drops off rapidly again at the largest radii. The first transition (i.e. the beginning of the plateau) occurs when the $11~\mu$m emission changes from being dominated by dust in the planetesimal belt for smaller radii, to being dominated by warm dust interior to the belt. For example, for a disc with mass $0.01~\mathrm{M}_{\earth}$, half of the flux comes from the planetesimal belt for $r_0 \sim 50~$au (see dotted line in Figure~\ref{fig:R11}), and the plateau begins at around 50~au. The origin of the plateau is evident from Figure~\ref{fig:tau_r0}, which showed that at fixed dust mass there is little change in the levels of dust dragged into the innermost regions when varying the belt radius. Thus the null does not vary significantly with planetesimal belt radius when emission is due to dust interior to the belt. However, at the largest belt radii there is a sharp decrease in null with belt radius once more. This is due to the discs becoming drag-dominated, such that increasing radius decreases the density, reducing the levels of warm dust. Less of a plateau is seen for lower mass discs, as they become drag-dominated at much smaller radii. Many of the observed HOSTS stars appear to cluster around the region where the high dust mass curves plateau. \par
It should be noted that the field of view of LBTI is 2.3" in diameter, such that emission from a planetesimal belt at tens of au would be outside the aperture used to observe stars in the HOSTS survey. The radii from which emission can be detected will be dependent upon distance to the star, and depend on other factors such as the disc orientation, which will affect the LBTI transmission pattern. For example, if a star is very far away, all of the dust emission could be within the first null of the LBTI transmission pattern such that no emission is seen. A more detailed discussion of modelling the transmission pattern can be found in \citet{Kennedy15_LBTI}, and it would be necessary to consider the specific parameters of individual systems to fully understand the effect. Discs with outer belts at more than $\sim40$~au will have their $11~\mu$m emission dominated by warm dust in the inner regions, however the limitations of the field of view will still reduce the observed null depth from our predictions here. \par
Interestingly, stellar spectral type does not significantly affect the predicted nulls (for fixed dust mass). To investigate this further, SEDs are shown in Figure~\ref{fig:SEDs} for discs around both A (top) and G (bottom) stars, with a dust mass of $0.01~\mathrm{M}_{\earth}$. The main effect of stellar spectral type is the SED shape: SEDs are mostly smooth for discs around A stars, while some features are seen at shorter wavelengths around F and G stars. This is because the larger blowout size for higher-mass stars (equation~\ref{eq:Dbl}) prevents the appearance of the silicate feature. The 11~$\mu$m silicate feature will impact the null depth predictions with asteroidal grains. These SEDs also show how as radius is increased, the $11~\mu$m flux decreases, then reaches a point where it becomes constant, before decreasing again. The peak of the SED moves to larger wavelengths as the planetesimal belt becomes colder. \par
Grain composition plays an important role, as the optical properties of grains impose a lot of structure on the SED. The null depths for grains which are cometary are presented in Figure~\ref{fig:nulls_rM} (bottom). Broadly the shape of the null curves is the same for a cometary composition, with plateaus which occur at a similar level to the asteroidal case. Asteroidal grains exhibit a feature at $11~\mu$m due to the presence of silicates, which impacts predictions of mid-infrared excesses. SEDs for cometary grains lack the $11~\mu$m feature, but have other features at different wavelengths. Changing the composition therefore means that different stellar spectral types no longer give the same null for a disc of certain parameters, however the difference between spectral types is small, and generally less than a factor $\sim 2$. \par
Whereas Figure~\ref{fig:nulls_rM} showed null predictions for discs when dust mass was kept constant, Figure~\ref{fig:nulls_rf} shows the same prediction, but keeping the disc's fractional luminosity constant. The shape of the curves in Figure~\ref{fig:nulls_rf} are similar to those for fixed dust mass. However, at fixed fractional luminosity, the disc mass increases with belt radius (Figure~\ref{fig:f}). Therefore, rather than flattening out the way the constant dust mass curves do, there is an upturn at larger belt radius due to increasing dust mass causing an increased null. Which of Figures~\ref{fig:nulls_rM} and \ref{fig:nulls_rf} is appropriate depends on what is known about the disc it is being applied to. If the disc has been observed at sub-mm wavelengths, the dust mass can be derived, whereas if the disc has been observed at far-infrared wavelengths, its fractional luminosity may be known. In either case, the disc should ideally be resolved, such that its radius is known, rather than having to infer this from the spectrum, given the uncertainties in such an inference \citep{Booth13,Pawellek14}. In general, the reader should bear in mind that implicit with the $11~\mu$m null predictions is the full SED at all wavelengths (Figure~\ref{fig:SEDs}). \par
Figure~\ref{fig:nulls_W05} shows the null excesses which are predicted using the simpler analytical model of \citet{Wyatt05}, assuming a single grain size of $\beta = 0.5$ and black body grains, for different values of belt optical depth $\tau_0$. More variation is seen between the spectral types, due to the black body assumption. At fixed optical depth $\tau_0$ there is also a weaker dependence on the belt radius than for fixed dust mass (Figure~\ref{fig:nulls_rM}) or fractional luminosity (Figure~\ref{fig:nulls_rf}). While this model gives broad trends, and can be used for an order of magnitude estimate \citep[e.g.][]{Mennesson14}, the conclusions are significantly different to the more accurate model of this paper. Therefore, the two-dimensional model of this paper is necessary for more detailed analysis of exozodi. \par
In Table~\ref{tab:HOSTS} our model is applied to the HOSTS stars with far-infrared excesses, based on observed dust masses $M_\mathrm{dust}$ and fractional luminosities $f$. While detailed analysis of the SEDs of individual systems is necessary to make precise predictions, this application of the model gives a first approximation. Where given, dust masses are derived from sub-mm observations of SCUBA-2 \citep{Holland17}, using equation 5 of \citet{Wyatt08}, assuming an absorption opacity of $\kappa_\nu~=~45~\text{au}^2~\mathrm{M}_{\earth}^{-1} = 1.7~\text{cm}^2~\text{g}^{-1}$. The emission for $\sigma~$Boo had a large offset from the star, such that it was likely from a background source, and a $3\sigma$ upper limit $F_{850} < 2.7~$mJy is used. Two systems, $\sigma$~Boo and 110~Her, have not been resolved, so instead their black body radii are used. To convert to real radius, we use the power law from \citet{Pawellek15}, assuming a composition of 50\% astrosilicate and 50\% ice, such that $\Gamma = r_0 / r_\mathrm{BB} = 5.42(L_\star/\mathrm{L}_{\sun})^{-0.35}$. We can categorise the HOSTS detections as follows:
\begin{table*}
	\centering
	\caption{Comparison of predictions for null depths from the model based on dust mass $M_\mathrm{d}$ and fractional luminosity $f$ with measurements from the HOSTS survey for stars with cold dust, assuming asteroidal grains.}
	\label{tab:HOSTS}
	\begin{threeparttable}
		\renewcommand\TPTminimum{\linewidth}
		\centering
		\makebox[\linewidth]{
			\begin{tabular*}{0.71\linewidth}{ccccccc}
				\hline
				Star & $M_{\mathrm{dust}}$ & $f$ & $r_0$ & Observed Null & Predicted Null $(M_{\mathrm{d}})$ & Predicted Null ($f$) \\
				& $\mathrm{M}_{\earth}$ & $10^{-5}$ & au & \% & \% & \% \\
				\hline
				$\tau$ Cet\tnote{a} & $2.3\times10^{-4}$ & 1.2\tnote{c} & 29\tnote{h}  & < 0.228 & 0.095  & 0.073 \\
				$\epsilon$ Eri & $2.2\times10^{-3}$  & 8\tnote{c} & 60\tnote{i} & 0.463 & 0.080 & 0.079 \\
				$\zeta$ Lep &  & 8.9\tnote{d} & 5\tnote{j} & 3.50 &  & 5.74 \\
				$\beta$ Uma & $<6.4\times10^{-3}$ & 1.4\tnote{d} & 47\tnote{k} & 1.02 & < 0.157 & 0.119 \\
				$\beta$ Leo & $<1.3\times10^{-3}$ & 2.2\tnote{d} & 39\tnote{k} & 0.470 & < 0.152 & 0.148 \\
				$\eta$ Crv & 0.038 & 18.6\tnote{e} & 152\tnote{l} & 4.41 & 0.133 & 0.135 \\
				$\sigma$ Boo\tnote{a} & $<5.1\times10^{-4}$ & 1.4\tnote{f} & 8\tnote{b,m} & < 0.344 & < 0.388 & 0.139 \\
				$\alpha$ Lyr & 0.013 & 1.9\tnote{d} & 85\tnote{n} & 0.392 & 0.164 & 0.159 \\
				110 Her &  & 0.07\tnote{g} & 213\tnote{b,o} & 0.621 &  & 0.0056  \\
				\hline
		\end{tabular*}}
		\begin{tablenotes}
			\item[a] Stars with non-detections have $3\sigma$ upper limits given.
			\item[b] For discs which have not been resolved, the black body radius is used, corrected by a factor from \citet{Pawellek15} to convert to real radius.
			\item[c] \citet{DiFolco04}
			\item[d] \citet{Thureau14}
			\item[e] \citet{Lebreton16}
			\item[f] \citet{Sibthorpe18}
			\item[g] \citet{Eiroa13}
			\item[h] Mean of $R_\mathrm{in}$ and $R_\mathrm{out}$ from \cite{Macgregor16}.
			\item[i] \citet{Greaves98}
			\item[j] Mean of inner and outer radii from \citet{Moerchen07}.
			\item[k] \citet{Matthews10}
			\item[l] \citet{Marino17_eta}
			\item[m] \citet{Sibthorpe18}
			\item[n] \citet{Sibthorpe10}
			\item[o] \citet{Eiroa13}.
		\end{tablenotes}
	\end{threeparttable}
\end{table*}

\begin{itemize}
	\item Despite large uncertainties, for example given the breadth of the $\tau$~Ceti disc (6 - 52~au) and the fact that the $\sigma$~Boo disc does not have a resolved radius, our model predicts levels for the two systems with non-detections, $\tau$~Ceti and $\sigma$~Boo, which are consistent with their $3\sigma$ upper limits. Our model suggests that they have exozodi, but these are below the detection limits (unless there is something preventing dust from reaching the inner system).
	\item 3/7 detections could potentially be explained by our P-R drag model: Vega ($\alpha$~Lyr), $\beta$~Leo, and $\zeta$~Lep, taking into account that we have assumed the null depth to be half of the fractional excess, but this  will depend on the geometry of the system. Two of these systems are believed to have additional, warm planetesimal belts closer to the star, which could provide an additional source of exozodiacal dust. Modelling of $\beta$~Leo suggests the presence of warm dust which is inside the outer belt but outside of the habitable zone \citep{Stock10,Churcher11}, and Vega is thought to have a warm belt close to the star \citep{Su13}.  While not considered by our model, an additional inner belt (and the dust dragged inwards from it) would provide another contribution to the null depth. Thus this strengthens the conclusion that these LBTI detections can be explained by dust dragged in from known planetesimal belts, though if the warm emission contributed by dust dragged in from these inner belts is already large enough to explain the observations then additional processes may be needed to prevent too much dust from reaching the inner regions. $\zeta$~Lep has not been observed in the sub-mm, such that there is no reliable estimate of its dust mass, and has only been resolved in the mid-infrared. However, based on its fractional luminosity and the mid-infrared resolved size, it is plausible that the P-R drag scenario explains the observed null.
	\item 3/7 detections are much higher than expected: $\eta$~Corvi, $\beta$~Uma, and $\epsilon$~Eri. $\eta$~Corvi has a null depth which is a factor $\sim10$ higher than predicted, but \citet{Marino17_eta} showed that its exozodi could be explained by inward scattering of exocomets from its cold planetesimal belt. $\beta$~Uma may also be explained by the exocomet scenario, though it cannot be ruled out that more accurate SED fitting would allow the observed null depth to be explained by dust dragged in from the outer belt by P-R drag. Mid-infrared observations with Spitzer of $\epsilon$~Eri imply two warm inner belts at 3~au and 20~au \citep{Backman09}, such that there could be another contribution to its exozodi from a second belt. Indeed, \citet{Su17} already suggested that the $35~\mu$m SOFIA/FORCAST detection towards $\epsilon$~Eri is incompatible with all of its warm dust originating in the outer belt. More detailed modelling of this system than that presented here would be needed to assess this, as well as to consider the role of stellar winds on the amount of dust dragged in \citep[e.g.][]{Reidemeister11_drag}.
	\item 110~Her has an observed null depth much higher than its predicted excess based on fractional luminosity. However, this system has only marginal excesses from Spitzer at 70~$\mu$m \citep{Trilling08} and Herschel at 70 and 100~$\mu$m \citep{Eiroa13,Marshall13}, and is poorly constrained both in terms of its fractional luminosity and its radius. As such we cannot make strong statements about the consistency of the observed null with P-R drag from the known outer belt. However, our model could be used to provide further constraints on the properties of the outer belt on the assumption that the null arises from dust dragged inwards from that belt (e.g. using Figure~\ref{fig:R11}). 
	\item There are three stars in the HOSTS survey (~$\delta$~Uma, $\theta$~Boo, and 72~Her) which had detections of exozodiacal dust, but no known cold planetesimal belt. Based on our model, we suggest that they may have planetesimal belts that lie in the shaded region of Figure~\ref{fig:excesses_det}, such that they have cold planetesimal belts which are too faint to be detected at longer wavelengths, but produce observable levels of exozodiacal dust via P-R drag (see Section~\ref{subsec:applyexo}).
\end{itemize}

While more comprehensive modelling of individual systems is needed, overall the model provides a good explanation for the majority of systems observed by HOSTS with known planetesimal belts. The levels of dust dragged in from the planetesimal belts is expected to result in exozodiacal dust levels similar to those observed, or compatible with the upper limits. The exceptions to this are two systems which may have an additional contribution from exocomets, one system which may have an additional, warmer belt, and one system for which the outer belt is poorly constrained by observations. \par
If the three HOSTS detections with no far-infrared excesses are due to P-R drag from planetesimal belts not yet detected in the far-infrared, this suggests that the outer belt population continues to lower far-infrared flux levels. Based on our model (Figure~\ref{fig:exozodi}), discs which are just below the far-infrared detection threshold would have $11~\mu$m excesses of $0.1-1\%$. While the planetesimal belts of the $80\%$ of stars without far-infrared detections are not yet known, these three HOSTS detections suggest the existence of belts below the detection threshold, with 3/38 stars potentially having faint far-infrared planetesimal belts that result in mid-infrared excesses of $0.2-0.7\%$. It is reasonable to assume that the distribution of outer belts continues to even lower far-infrared flux levels, and so that mid-infrared excesses can be expected to be present at levels below $0.1\%$ for some stars. This means they could have exozodi at levels above the limit tolerable by exo-Earth imaging of 10-20 zodi \citep[e.g.][]{Defrere10,Roberge12}, which would be equivalent to a null depth of $\sim0.02-0.04\%$. Therefore, even systems where no planetesimal belt is detected may have exozodi levels which are problematic for exo-Earth characterisation.

\section{Conclusions}
\label{sec:conc}
We have developed an analytical model which can predict two-dimensional size distributions in debris discs for particle size and radial distance, taking into account the effects of both collisional evolution and P-R drag. This builds on previous, simpler analytical models which only considered a single dimension. Our model provides a reasonable approximation to results from detailed numerical models whilst being much faster, and can be used with realistic grain properties to predict the thermal emission resulting from planetesimal belts of different parameters around Sun-like and A stars. Applying the model to stars where exozodiacal dust has been detected allows us to determine whether their exozodi originate from dust being dragged inwards from an outer planetesimal belt while undergoing collisions, or whether an alternative scenario is needed. \par
We have shown that the effect of P-R drag transporting dust inwards from an outer belt means that systems with known planetesimal belts should have sufficient levels of exozodiacal dust to be detectable with LBTI. Non-detections could imply the presence of unseen planets which are accreting or ejecting dust from the habitable zone. Further, we have shown that LBTI may be able to detect exozodiacal dust which has been dragged inwards from outer belts which are too faint to detect in the far-infrared, particularly for belts with lower dust masses and small radii. Grain composition has only a minor effect on the results, such that our conclusions remain unchanged. \par
Application of our model to systems observed by HOSTS shows that our model can provide a good explanation for the majority of the detections, with the exception of two systems which are particularly bright, potentially due to exocomets, one system which is believed to have a warm inner belt, and one system which is poorly constrained. This means that the scenario of P-R drag transporting dust inward from an outer belt may be a viable source of exozodiacal dust. Further, for the three exozodi detections with no known planetesimal belt, we suggest that the source of the exozodiacal dust could be a faint outer belt which is not yet detectable in the far-infrared. In the future it may be possible to use models such as the one presented in this paper to determine whether particular exozodi originate from P-R drag or an alternative scenario. \par
Future attempts to detect and characterise exo-Earths will be impeded by levels of exozodiacal dust even ten times that of the solar system. We have shown that even planetesimal belts much less massive than the bright Kuiper belt analogues which have already been detected could produce mid-infrared excesses a few times greater than the zodiacal cloud. While systems with known belts are expected to have exozodiacal dust, even those where no belt has been detected could therefore be problematic for exo-Earth imaging. Understanding the occurrence of exozodiacal dust will therefore be crucial to the design of these exo-Earth missions. \par

\section*{Data Availability}
The data underlying this article will be shared on reasonable request to the corresponding author.

\section*{Acknowledgements}

JKR would like to acknowledge support from the Science and Technology Facilities Council (STFC) towards her doctoral research.




\bibliographystyle{mnras}
\bibliography{debris} 


%
%
%


\bsp	
\label{lastpage}
\end{document}